%% file: main.tex
\documentclass[leqno,12pt,a4paper]{article}
\usepackage{setspace}
\usepackage{latexsym}
\usepackage{amsmath}
\usepackage{amssymb}
\usepackage{mathtools}
\usepackage{bbm}
\usepackage{amsthm}
\usepackage{marvosym}
\usepackage{tensor}
\usepackage{cases}

\usepackage{algorithm}
\usepackage[noEnd=true]{algpseudocodex}
\algnewcommand\algorithmicforeach{\textbf{for each}}
\algdef{S}[FOR]{ForEach}[1]{\algorithmicforeach\ #1\ \algorithmicdo}

\usepackage{graphicx}
\usepackage{epsfig}
\usepackage{subcaption}
\usepackage{tabularx}
\newcolumntype{L}[1]{>{\raggedright\arraybackslash}p{#1}} % linksbÃ¼ndig mit Breitenangabe
\newcolumntype{C}[1]{>{\centering\arraybackslash}p{#1}} % zentriert mit Breitenangabe
\newcolumntype{R}[1]{>{\raggedleft\arraybackslash}p{#1}} % rechtsbÃ¼ndig mit Breitenangabe

\usepackage{array}

\allowdisplaybreaks
\usepackage{tikz}
\usetikzlibrary{shapes,arrows,patterns,backgrounds,fit,arrows.meta}
\usepackage{pgfplots}
\pgfplotsset{
  compat=newest,
  width=0.65\textwidth,
  height=0.4\textwidth,
  tick align=outside,
  tick pos=left,
  scaled ticks=false,
  grid=major,
  grid style={dotted},
}
\pgfkeys{/pgf/number format/.cd,
  1000 sep={\,}, % 1000 separation with distance instead of comma
  min exponent for 1000 sep=4,
}

\usepackage{siunitx}
\usepackage{makecell}

\usepackage{comment}

\newcommand{\boldface}[1]{\boldsymbol{#1}}
\newcommand{\bfa}{\boldface{a}}
\newcommand{\bfb}{\boldface{b}}

\newcommand{\bfe}{\boldface{e}}

\newcommand{\bfA}{\boldface{A}}
\newcommand{\bfB}{\boldface{B}}

\newcommand{\bfI}{\boldface{I}}

\newcommand{\bfT}{\boldface{T}}

\newcommand{\bfvarepsilon}{\boldsymbol{\varepsilon}}

\newcommand{\bfeta}{\boldsymbol{\eta}}

\newcommand{\bfsigma}{\boldsymbol{\sigma}}

\usepackage{bbold}
\newcommand{\dsA}{\mathbb{A}}

\newcommand{\dsD}{\mathbb{D}}
\newcommand{\dsE}{\mathbb{E}}

\newcommand{\dsI}{\mathbb{I}}

\newcommand{\dsP}{\mathbb{P}}

\newcommand{\dsR}{\mathbb{R}}
\newcommand{\dsS}{\mathbb{S}}

\newcommand{\be}{\begin{equation}}
\newcommand{\ee}{\end{equation}}
\newcommand{\bea}{\begin{eqnarray}}
\newcommand{\eea}{\end{eqnarray}}
\newcommand{\bes}{\begin{equation*}}
\newcommand{\ees}{\end{equation*}}
\newcommand{\beas}{\begin{eqnarray*}}
\newcommand{\eeas}{\end{eqnarray*}}

\newcommand{\pf}[2]{\frac{\partial #1}{\partial #2}}

\newcommand{\dd}{\ \mathrm{d}}

\usepackage{lipsum}
\usepackage{placeins}

\definecolor{RUBblue}{rgb}{0.0470588,0.262745,0.411765}
\definecolor{lightgray}{gray}{0.92}
\definecolor{blau}{rgb}{0,0.25,1.0}
\definecolor{RUBCDblue}{rgb}{0.062745,0.305882,0.545098}
\definecolor{RUBCDgreen}{rgb}{0.517647,0.741176,0.0}
\definecolor{RUBCDgray}{rgb}{0.815686,0.815686,0.807843}
\definecolor{RUBCDgrayDark}{rgb}{0.615686,0.615686,0.607843}

\makeatletter

\def\env@cases#1{%
  \let\@ifnextchar\new@ifnextchar
  \left\lbrace\def\arraystretch{1.2}%
  \array{@{}#1@{\quad}l@{}}}
\makeatother

\usepackage{booktabs}

\RequirePackage{relsize}
\newfont{\Sf}{cmssbx10 scaled 2074}

\usepackage[
  colorinlistoftodos,
  backgroundcolor=green,
  linecolor=green,
  textsize=footnotesize
]{todonotes}

\usepackage{empheq}

\newcommand{\rev}[1]{\textcolor{black}{#1}}
\newcommand{\revv}[1]{\textcolor{black}{#1}}

\begin{document}
\title{A new paradigm for the efficient inclusion of stochasticity in engineering simulations}
\date{} 
\maketitle
{\large
\noindent{Hendrik}  Geisler$^{1,2}$, Cem Erdogan$^1$, Jan Nagel$^3$, Philipp Junker$^{1,2}$\\[0.5mm]
}
1: Leibniz University Hannover, Institute of Continuum Mechanics, Hannover, Germany\\
2: IRTG 2657: Computational Mechanics Techniques in High Dimensions \\
3: TU Dortmund, Department of Mathematics, Dortmund, Germany\\[2mm]
\Letter \hspace{0.1cm} junker@ikm.uni-hannover.de

\section*{Abstract}
\rev{As a physical fact, randomness is an inherent and ineliminable aspect in all physical measurements and engineering production.}
\revv{As a consequence, material parameters, serving as input data, are only known in a stochastic sense and thus, also output parameters, e.g., stresses, fluctuate. For the estimation of those fluctuations it is imperative to incoporate randomness into engineering simulations. Unfortunately,} incorporating uncertain parameters into the modeling and simulation of inelastic materials is often computationally expensive, as many individual simulations may have to be performed. \\
The promise of the proposed method is simple: using extended material models to include stochasticity reduces the number of needed simulations to one. \rev{This single computation is cheap, i.e., it has a comparable numerical effort as a single standard simulation.} The extended material models are easily derived from standard \revv{deterministic} material models and account for the effect of uncertainty by an extended set of deterministic material parameters. The time-dependent and stochastic material behavior are separated, such that only the deterministic time-dependent behavior of the extended material model needs to be simulated. The effect of stochasticity is then included during post-processing. \\
\rev{The feasibility of this approach is demonstrated for three different and highly non-linear material models}: viscous damage, viscous phase transformations and elasto-viscoplasticity. 
A comparison to the Monte Carlo method showcases that the method is indeed able to provide reliable estimates of the expectation and variance of internal variables and stress at a \rev{minimal} fraction of the computation cost. 
\\ \\
Keywords: time-separated stochastic mechanics, inelasticity, damage, elasto-viscoplasticity, phase transformation, stochastic material, uncertainty 

\section{Introduction}
Predicting the effect of uncertainty in engineering simulations has grown in importance over the last decades.
In order to obtain high fidelity results, the inherent uncertainty of systems can not be ignored in many practical engineering problems \cite{stefanou_stochastic_2008, oden_research_2003}.
Therefore, a large number of methods have been developed for uncertainty quantification and propagation in the context of finite element simulations.
Most of these techniques have been first proposed for linear elastic materials and are known under the framework of stochastic finite element methods: Monte Carlo sampling method \cite{liu_monte_2004, hurtado_monte_1998, ghanem_hybrid_1998}, perturbation method \cite{liu_random_1986, kleiber_stochastic_1992, kaminski_generalized_2005}, spectral stochastic finite element method \cite{ghanem_stochastic_1991, xiu_wiener--askey_2002} and stochastic collocation method \cite{ghanem_sparse_2017, xiu_stochastic_2016, xiu_high-order_2005}. Every technique comes with its own unique strengths and weaknesses. These differences stem from the balance of approximation accuracy with the number of needed deterministic simulations or equivalently the size of the system to solve.

\begin{figure}
    \centering
    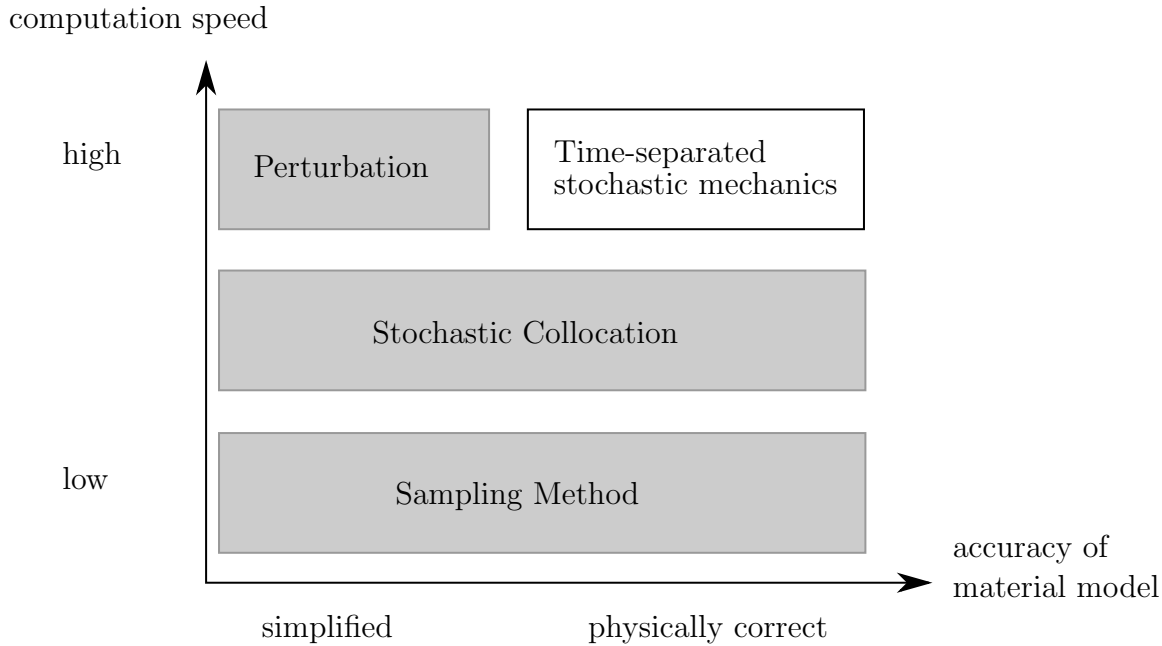
    \caption{Literature Overview. Grey boxes denote existing methods.}
    \label{fig:LiteratureOverview}
\end{figure}

In contrast \rev{to methods for elastic material behavior}, research of applicable methods to simulate nonlinear and/or inelastic materials is rather sparse.
In particular, the topics of hyperelasticity \cite{melink_automation_2012, kaminski_introduction_2016} and plasticity \cite{anders_threedimensional_2001, liu_applications_1987, rosic_stochastic_2011, basmaji_anisotropic_2022, zheng_nonlinear_2023} have received the majority of attention in recent years. These material models can be set up in a fashion that resembles linear elasticity up to some degree.
Then, the perturbation method and stochastic collocation method could be used with no or little modification  \cite{feng_performance_2021, dannert_investigations_2022}.
The same is true for the references \cite{acharjee_non-intrusive_2007, doltsinis_inelastic_2003, doltsinis_perturbation-based_2006}, dealing with uncertainty quantification in metal forming processes for the large deformation setting. Here, simplified material models allow a relative simple uncertainty propagation at the expense of physical accuracy.

More accurate material models are using internal variables \cite{carnot_reflections_1988,clausius_ueber_1850,horstemeyer_historical_2010} to track the evolution of the microstructure over time. The change of the microstructure is the reason for the time and history-dependent mechanical properties of such materials. 
Internal variables are used in nearly all advanced material models. Despite the importance of internal variables for accurate material modeling, there are only very few references combining uncertainty quantification and inelastic material models with internal variables to the knowledge of the authors, e.g., \cite{feng_performance_2021, zuo_sensitivity_2021, ren_rate-dependent_2015}. 

\rev{
An overview of the existing methods is presented in Figure~\ref{fig:LiteratureOverview}. The methods are sorted by their computation speed and the complexity of the possible material models to be investigated.
\begin{description}
    \item[Sampling method] The classical Monte Carlo scheme \cite{liu_monte_2004, ghanem_hybrid_1998} is one of the best known sampling methods. Sampling methods are based on a repeated evaluation, each time with a different realization of the random parameters. For a high enough number, of iterations the response characteristics can be approximated. These methods are typically very robust albeit have a significant computational effort.
    \item[Stochastic collocation] The number of individual simulations can be reduced by choosing special sampling points, e.g., Smolyak points \cite{xiu_high-order_2005}, and interpolating in between by a reduced-order polynomial surface \cite{xiu_stochastic_2016, ghanem_sparse_2017}. However, the number of sampling points can be prohibitively large in the high-dimensional case. Additionally, instability is a reoccoring problem \cite{dannert_investigations_2022}.
    \item[Perturbation] Perturbation methods \cite{kleiber_stochastic_1992, kaminski_stochastic_2013} are the most computationally efficient strategies. They make use of a Taylor series expansion of the response around the expectation of the random variables. Unfortunately, this method can only handle simplified polynomial material models.
\end{description}
It is obvious that a method \revv{which combines} high computation speed and the ability to handle physically correct material models was missing up to now.  
This research gap motivates the development of a new paradigm for the inclusion of stochasticity in engineering simulations. 
}

Our approach \rev{roots back to} time-separated stochastic mechanics (TSM) \rev{which was derived for linear viscoelasticity for which an analytic solution of the evolution equation could be found.} 
The method is based on the separation of the evolution equation of the internal variable into two parts: random but time-invariant terms and time-dependent but deterministic terms. This in return enables to only simulate the time-dependent deterministic evolution of the \revv{material} and to incorporate all sources of uncertainty during post-processing. Thus, the computational effort compared to Monte Carlo simulations is \rev{orders of magnitude smaller}. However, all relevant stochastic information such as expectation, standard deviation \rev{higher stochastic moments and even the probability density function} can be extracted. 
It was first implemented for a viscoelastic material model with random material parameters in \cite{junker_analytical_2018, junker_modeling_2020} to approximate the expectation and variance of stress and reaction force. An excellent agreement with the results of Monte Carlo simulations could be observed. Consequently, the TSM was used in other works with similar convincing results. In particular, it was used for dynamic simulation of viscoelastic materials in \cite{geisler_simulation_2022} and for local material fluctuations in \cite{geisler_time-separated_2023}. However, the application of the TSM to a larger class of material models was not investigated yet.

\subsection{Contribution}
\begin{figure}
	\centering
	\includegraphics[trim=1.6cm 19cm 0cm 2.5cm, clip, scale=1.]{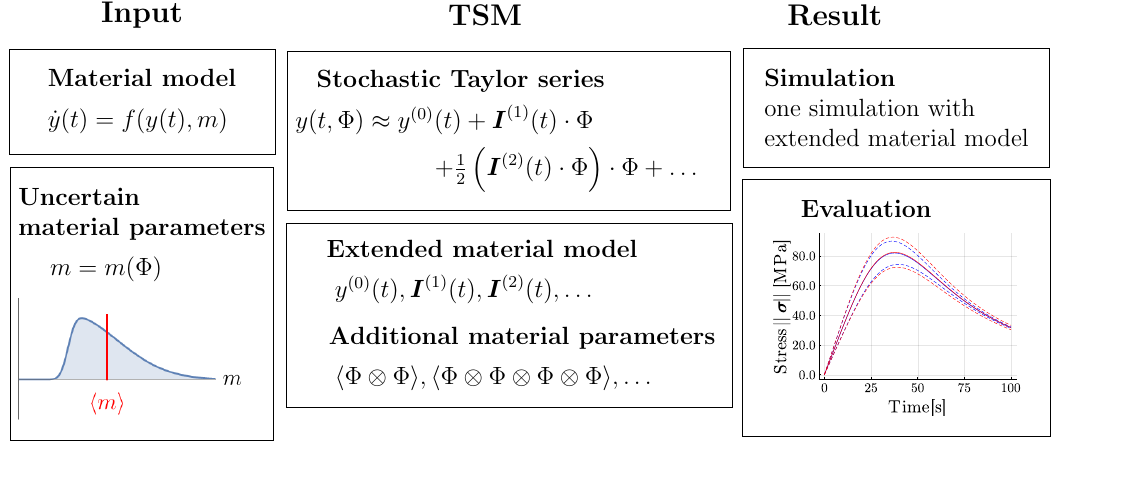}
	\caption{Graphical overview of the approach based on time-separated stochastic mechanics}
	\label{fig:TSMapproach}
\end{figure}
%\begin{figure}
%    \centering
%    \includegraphics[scale=0.6]{Pictures/Gen-TSM.png}
%    \caption{TSM approach}
%    \label{fig:TSMapproach}
%\end{figure}

\rev{In this contribution, we present a generalization of time-separated stochastic mechanics (TSM) when no analytical solution of the evolution equation is known. This enables TSM to be applicable to material models of any kind, paving the way for a paradigm shift in material modeling.}

The TSM requires only one deterministic simulation to estimate the effect of uncertain material parameters on all quantities \revv{such} as internal variables and stress. \rev{\revv{The} point of departure} is a standard deterministic material model which is augmented by additional internal variables to track the effect of fluctuations of the material parameters.
Effectively, time-dependent and stochastic behavior are separated. Thus, only the time-dependent deterministic behavior of the extended material model needs to be simulated. The effect of the uncertain material parameters is then included during post-processing. \rev{In particular, expectation and standard deviation of all engineering quantities as internal variables and stress are calculated.}
The method can be used for any material model without any restrictions on the equation structure of the material model or the 
type and quantity of uncertain material parameters.
A graphical overview of the proposed method is shown in Figure~\ref{fig:TSMapproach}.

The versatility of the \revv{generalized} time-separated stochastic mechanics is demonstrated on three different material models: viscous damage, viscous phase transformations and elasto-viscoplasticity.

\subsection{Notation and mathematical preliminaries}
Scalar variables are indicated by a small letter, e.g. $a$. For vectors and tensors bold letters, e.g., $\bfb = (b_1, b_2, b_3)^T$ and $\bfA = (A_{ij})$, are used. Some special tensors as the elasticity tensor $\dsE$ are indicated by calligraphic letters.
The basis of the vectors and tensors is the standard Cartesian coordinate system given by the orthogonal base vectors $\bfe_1, \bfe_2, \bfe_3$. 

We make use of Einsteins's summation convention which implies summation over repeated indices. Fixed indices over which no summation is carried out are indicated by greek letters. Repeatedly, we switch between notation styles whenever the readability could be enhanced.
Furthermore, the basis is not explicitly written \rev{in index notation} but indicated by free indices. 
We present shortly the different styles for common operations.
The inner product is defined as
\begin{equation*}
    \bfa \cdot \bfb = a_i \bfe_i \cdot b_j \bfe_j = a_i b_j \delta_{ij} = a_i b_i
\end{equation*}
and similar\revv{ly}
\begin{equation*}
    \bfA \cdot \bfB = A_{ij} \bfe_i \bfe_j \cdot B_{kl} \bfe_k \bfe_l = A_{ij} B_{jl} \bfe_i \bfe_l = A_{ij} B_{jl} \,. 
\end{equation*}
The tensor product is given by
\begin{equation*}
    \bfa \otimes \bfb = a_i \bfe_i \otimes b_j \bfe_j = a_i b_j \bfe_i \bfe_j = \left( a_i b_j \right) \,.
\end{equation*}
The identity matrix is indicated by $\dsI$. A matrix with all entries equal zero is indicated by $\bf0$.

The Voigt notation is used for symmetric tensors whenever feasible, e.g., the stress tensor is represented by the 6-dimensional vector $(\sigma_{x}, \sigma_{y}, \sigma_{z}, \sigma_{yz}, \sigma_{xz}, \sigma_{xy})$. Consequently, the elasticity tensor is a $6\rev{\times}6$ tensor in Voigt notation.

\rev{
We use random variables to represent the uncertainties. A random variable $X: \Omega \to \dsR$ is a function that assigns a numerical value to each outcome in a random experiment. Thus, it maps outcomes from a sample space $\Omega$ to the real numbers $\dsR$. The likelihood of different values is encoded in the probability density function $\dsP$. 
}

\rev{
For random scalar quantities, we denote the expectation as
\begin{equation}
    \textrm{Exp}(a) = \langle a \rangle := \int_{-\infty}^\infty a \ \dsP(\mathrm{d}\theta).
\end{equation}
The variance of a scalar random variable is given by
\begin{equation}
    \textrm{Var}(a) := \langle (a-\langle a \rangle)^2 \rangle = \langle a^2 \rangle - \langle a \rangle^2. \label{eq:variance}
\end{equation}
The standard deviation is then given as the square root of the variance, i.e., $\textrm{Std}(\cdot) = \sqrt{\textrm{Var}(\cdot)}$.
}
\rev{
For vectors or tensor valued random variables, the expectation is defined componentwise, e.g.,
\begin{equation*}
    \langle \bfA\rangle := \langle A_{ij}\rangle \bfe_i \bfe_j ,
\end{equation*}
and the variance is the tensor of covariances,
\begin{equation*}
    \text{Var}(\dsA) := \big\langle (\dsA - \langle \dsA\rangle )\otimes (\dsA - \langle \dsA\rangle)\big\rangle .
\end{equation*}
}

\subsection{Structure}
The work is structured as follows:
\rev{the general approach is derived} in Section~\ref{sec:TSM}. \rev{An overview of the method is presented in Section~\ref{sec:MethodOverview}.} The application examples are presented in Section~\ref{sec:ApplicationExamples}. The computational effort of the method is discussed in Section~\ref{sec:ComputationalEffort}. A conclusion is found in Section~\ref{sec:Conclusion}.
{
\color{black}
\section{Time-separated stochastic mechanics}
\label{sec:TSM}
In order to accurately predict how a structure behaves under various loads and conditions, one has to model the material behavior. Typically, some set of equations relating stress $\bfsigma$ and strain $\bfvarepsilon$ is defined. Of course, the specific equations depend on the type of material being modeled and the physical phenomena under consideration such that a large number of material models is \revv{presented in} literature.
Essential characteristic values of each material are the material parameters $m$.
Some common material parameters include the Lam\'e parameters, the yield stress, viscosity, etc. 
Depending on the material model, different material parameters have to be determined, typically through experiments. Unfortunately, the material parameters are not fixed constants but vary for several reasons. Some common reasons include: microstructural variations, impurities, the testing procedure or environmental factors. Of course, these variations can have significant implications for the behavior and performance of the materials.

Mathematically, the fluctuations can be modeled as a dependence of a material parameter on a random variable $\varphi$ as $m = m(\varphi)$. For mathematical simplicity, we use zero-centered random variables, i.e., $\langle \varphi \rangle = 0$. Any fluctuating material parameter $m$ can be split up into a deterministic and a fluctuating part with zero-centered random variable, e.g., $m = \langle m \rangle + \varphi$, where $\varphi$ results trivially as $\varphi = m - \langle m \rangle$. It may be remarked that some material parameters enter the equations in a structured way, e.g., in form of the elasticity tensor~$\dsE$. The elasticity tensor can be parameterized through the Lam\'e parameter $(\lambda, \mu)$ or Young's modulus and Poisson ratio $(E, \nu)$\revv{, equivalently}. If these quantities are fluctuating we may similarly split up the elasticity tensor into a deterministic and a fluctuating tensor-valued part $\dsE = \langle \dsE \rangle + \dsD(\varphi_1, \varphi_2) = \revv{\dsE^{(0)} + \dsD}$. Consequently, the fluctuating part $\dsD$ depends on the random variables $\varphi_1, \varphi_2$ of the parameters \revv{$(\lambda, \mu)$ or $(E, \nu)$}. 

To accurately predict the effect of the fluctuations, one has to trace the effect on the equations relating stress, strain and internal variables. In particular, we are interested in calculating expectation and variance of the engineering quantities.
In general, one can differentiate two different kinds of equations relating stress, strain and internal variables: algebraic equations and differential equations. A classical example for an algebraic equation is Hooke's law $\bfsigma = \dsE \cdot \bfvarepsilon$. It relates stress and strain by algebraic expressions. In contrast, differential equations relate unknown functions with their time-derivative and thus describe the time evolution. They are often used to describe the inner state of the material by means of internal variables. Such, they account for dissipative microstructure evolution \cite{junker_extended_2021,horstemeyer_historical_2010}. Of course, the different nature of these types of equation require a different approach to predict the change of the output variable. The more simple case of algebraic equations is presented in Section~\ref{sec:AlgEq}. Up to now, no approach for differential equations existed. This severly hindered uncertainty quantification for advanced material models. The time-separated stochastic mechanics offers a solution which is presented in Section~\ref{sec:DiffEq}.

\subsection{Algebraic equations}
\label{sec:AlgEq}
Let us denote an algebraic equation linking the quantities $y$ and $x$ over time $t$ as
\begin{equation}
 y(t) = f(t, x, m(\varphi)).    
\end{equation}
The quantities $y$ and $x$ can be, e.g., components of stress and strain.
For simplicity, we first discuss the approach for scalar quantities.
The material parameters $m$ of course influence any quantity $y$. Consequently, as the material parameters can be random, i.e., depend on a random variable $\varphi$, also $y$ is random.
A feasible approach to track the change in $y$ on the fluctuations $\varphi$ is to use a Taylor series.
This is a reasonable approach as fluctuations are typically small compared to the expectation and the investigated functions are smooth.

A Taylor series for one random variable $\varphi$ reads
\begin{equation}
    y(t, \varphi) \approx  f(t, \varphi) \vert_{\varphi = 0} + \sum_{k=1}^\infty \frac{1}{k!} \varphi^k \frac{\mathrm{d}^k f}{\mathrm{d}\varphi^k} \Big\vert_{\varphi = 0} = f(t, 0) + \sum_{k=1}^\infty \frac{1}{k!} \varphi^k T^{(k)}(t), \label{eq:Taylor1}
\end{equation}
\revv{where
\begin{equation}
	T^{(k)}(t) := \frac{\mathrm{d}^k f}{\mathrm{d} \varphi^k} \bigg\vert_{\varphi = 0}.
\end{equation}
}
Of course, the approximation quality increases with a higher order of the \revv{series}. As the influence of the higher-order terms decreases for a converging series, the Taylor series is typically broken after $n$ terms. The order of the approximation is then $n$. 
In fact, in the simulations in Section~\ref{sec:ApplicationExamples}, we demonstrate that even a linear approximation ($n = 1$) suffices for the investigated problems.

The Taylor series in Equation~\eqref{eq:Taylor1} effectively separates random terms $\varphi, \varphi^2, \dots, \varphi^k$ from time-dependent terms $T^{(k)}(t)$. In fact, \revv{since} $f(t, 0)$ and the tangents $T^{(k)}(t)$ are evaluated at $\varphi = 0$, they are deterministic. Thus, the Taylor series separates stochastic but time-independent terms from deterministic time-dependent terms. In return, this allows to first calculate $y(t, 0)$ and the tangents $T^{(k)}(t)$ and trace the effect of the random fluctuations by simple multiplication with the random variables $\varphi^k$. This approach can be justified mathematically by imposing moment conditions or boundedness assumptions on $\varphi$. This allows a wide class of distributions for $\varphi$. 
 
As the material parameters may fluctuate simultaneously and independently, they may depend on multiple random variables $\varphi_1, \dots, \varphi_p$.
Consequently, the Taylor series approach is extended to $p$ random variables as
\begin{align}
    y(t, \varphi_1 \dots \varphi_p) &\approx f(t, 0 \dots 0) + \sum_{i=1}^p \varphi_i \ \partial_i f(t)\vert_{\varphi_1, \dots, \varphi_p = 0} + \frac{1}{2} \sum_{j=1}^p \sum_{k=1}^p \varphi_j \varphi_k \ \partial_{jk} f \vert_{\varphi_1, \dots, \varphi_p = 0} \nonumber \\
    &\quad + \frac{1}{6} \sum_{q=1}^p \sum_{r=1}^p \sum_{s=1}^p \varphi_q \varphi_r \varphi_s \ \partial_{qrs} f(t)\vert_{\varphi_1, \dots, \varphi_p = 0} + \dots
\end{align}
where $\partial_i$ denotes the partial derivative with respect to random variable $\varphi_i$. Similarly, $\partial_{ij}$ denotes the second partial derivative with respect to $\varphi_i$ and $\varphi_j$ and $\partial_{ijk}$ is the third partial derivative. Here, the Taylor series is presented up to third degree. Higher order terms result similarly.
The general construction of the Taylor series stays the same for multiple random variables. The desirable property to separate stochastic from time-dependent behavior is unchanged.

For a more concise notation all random variables $\varphi_1 \dots \varphi_p$ can be collected in a vector $\Phi = [\varphi_1, \dots, \varphi_p]$. The Taylor series reads th\revv{e}n  
\begin{equation}
    y(t, \Phi) \approx f^{(0)}(t) + \bfT^{(1)}(t) \cdot \Phi + \frac{1}{2} \left( \bfT^{(2)}(t) \cdot \Phi \right) \cdot \Phi + \frac{1}{6} \left( \left( \bfT^{(3)}(t) \cdot \Phi \right) \cdot \Phi \right) \cdot \Phi + \dots \label{eq:TaylorMultiDim} 
\end{equation}
where $\revv{f^{(0)}}(t) = f(t, \textbf{0})$ and $\bfT^{(1)}(t)$ denotes the gradient of $f$, $\bfT^{(2)}(t)$ the Hessian and $\bfT^{(3)}(t)$ the derivative of the Hessian. 
All derivatives are evaluated at $\Phi = \textbf{0}$.
They are straightforward generalizations of the scalar case. Of course, higher-order terms are found similarly.

Two main observation can be made: 1) no assumption on the design of the random variables $\varphi_i$ needs to be done. In particular, they do not have to be stochastically independent \revv{n}or \revv{do they} have a specific probability density function. They just have to be zero-centered which can be easily fulfilled; 2) we are free in the organization of the random variables. This does not only allows for permutations but additionally allows for tensor-valued $\Phi$.
These observations have a simple practical implication: no assumption on the structure or organization $\Phi$ needs to be \revv{made}.
By consistent construction of the tangent, Hessian and higher-order derivatives a correct ordering automatically results.
This allows to, e.g., directly construct the Taylor series with respect to the random part of the elasticity tensor $\dsD$. In that case, $\Phi$ would be a matrix and gradient, Hessian and derivative of the Hessian would be one dimension larger.
 
With the Taylor series given in Equation~\eqref{eq:TaylorMultiDim}, it is now easy to determine expectation and variance.
As the random variables are zero-centered by construction, all odd moments equal zero, e.g., $\langle \phi \rangle = \langle \phi^3 \rangle = 0$.
Consequently, the expectation is found due to the linearity of the expectation operator in index notation as
\begin{equation}
    \langle y(t) \rangle = f^{(0)}(t) + \frac{1}{2} \bfT^{(2)}_{ij}(t) \langle \Phi_{i} \Phi_{j} \rangle + \dots. \label{eq:AlgExpectation} 
\end{equation}
For the variance, we make use of the relationship in Equation~\eqref{eq:variance} and find in index notation
\begin{align}
    \textrm{Var}(y(t)) &= \bfT^{(1)}_{a}(t) \langle \Phi_a \Phi_b \rangle \bfT^{(1)}_{b}(t) + \frac{1}{6} \bfT^{(1)}_{c}(t) \langle \Phi_c \Phi_d \Phi_e \Phi_f \rangle \bfT^{(3)}_{def}(t) \nonumber \\ &\quad + \frac{1}{36} \bfT^{(3)}_{ghi}(t) \langle \Phi_g \Phi_h \Phi_i \Phi_j \Phi_k \Phi_l \rangle \bfT^{(3)}_{jkl}(t) + \dots. \label{eq:AlgVariance} 
\end{align}
The standard deviation can be trivially found as the square-root of the variance.
For the calculation of expectation and standard deviation, the moments $\langle \Phi \otimes \Phi \rangle, \langle \Phi \otimes \Phi \otimes \Phi \otimes \Phi \rangle, \dots$ of the vector of random variables have to be determined.
As the vector of random variables $\Phi$ is time-independent, this can be done in advance of any simulation. In fact, the values of the moments can be calculated as soon as the probability density function of $\Phi$ is known. The determination of these higher moments can be performed, e.g., by means of sampling. This is a very fast method as the desired quantities directly result from realizations of $\Phi$. \revv{Furthermore,} this only has to be performed once \revv{per material}. The values of the moments of $\Phi$ can be saved and reused for any simulation. In that sense, the moments can be seen as additional material parameters encoding the stochastic fluctuations of the material.

Of course, the above result directly generalize to vector-valued and tensor-valued functions similarly. This is immediately clear if the components are investigated independently. 

The presented result, especially the scalar case in Equation~\eqref{eq:Taylor1} and the corresponding calculation of expectation and variance, is known in the literature as the perturbation method \cite{kleiber_stochastic_1992, kaminski_stochastic_2013}. It has shown reasonable success for simple algebraic material models as linear elasticity and hyperelasticity.
Nevertheless, this approach can be seen as the first instance of time-separated stochastic mechanics as it already shows the main property:
separating stochastic but time-independent from deterministic but time-dependent terms allows to approximate the stochastic behavior in one simulation. 

Unfortunately, the approach in the present form only allows for algebraic relationships between stress and strain.
Consequently, it falls short for many advanced material models who make repeatedly use of differential equations for the description of the microstructure evolution. For these material models, an algebraic equation as solution of the differential equation is not known or can not be recovered.

\subsection{Differential equations}
\label{sec:DiffEq}
Advanced material models make use of internal variables to describe the microstructure evolution. The evolution of the internal variable $y$ is typically given in form of a differential equation, e.g.,
\begin{equation}
    \dot{y}(t) = g(y, x, m(\varphi)) \label{eq:diffeq}
\end{equation}
Here, a scalar first order differential equation is presented. However, the arguments below generalize trivially to vector- and tensor-valued differential equations.
Furthermore, the approach generalizes to higher-order differential equations and even differential-algebraic systems as presented in the application example in Section~\ref{sec:ViscoPlasticity}.
Of course, the evolution of the internal variable may be related to the material parameters $m$ and the strain $\bfvarepsilon$.
If the material parameters are random, i.e., \revv{they} depend on random variables $\varphi_1, \dots, \varphi_p$, $y$ is \revv{itself} random, i.e., $y = f(\varphi_1, \dots, \varphi_p)$.
Therefore, the evolution equation \eqref{eq:diffeq} is a differential equation with stochastic coefficients, rendering the internal variable stochastic.
This increases the complexity of the mathematical and numerical treatment. In return, the estimation of the stochastic properties of the internal variable and of the derived quantitites as the stress is a highly challenging task.

As typically no analytical solution for the evolution laws is known, there exists no corresponding algebraic equation. Therefore, another approach to handle stochastic differential equations is required. For a specific realization of the random variables, a classical time integration scheme, e.g., the Euler method, allows the solution of the differential equation \eqref{eq:diffeq} at specific time points. The most simple approach to approximate the stochastic behavior would then be to evaluate the solution for a large number of realizations of the random variables. Obviously, this is computationally inefficient.

In contrast, most computationally efficient is the Taylor expansion approach of the preceding section. However, as it can only be applied to algebraic equations, one approach might be to transform the differential equation into an equivalent algebraic equation. As discussed before, this is only possible for a very limited subset of evolution equations, e.g., linear viscoelasticity \cite{junker_analytical_2018}.
However, if the differential equation is discretized in time and integrated, at each timestep $n_t+1$ an algebraic equation
\begin{equation}
    y_{n_t+1} = g(y_{n_t}, m) 
\end{equation}
results.
At first glance this allows the previously in Section~\ref{sec:AlgEq} described approach. However, the solution of the last timestep is $y_{n_t} = f(y_{n_t-1}, \phi)$ is itself random. A Taylor series approach requires the total derivative in the form
\begin{equation}
    \frac{\dd y_{n_t+1}}{\dd \varphi} = \frac{\partial y_{n_t+1}}{\partial \varphi} + \frac{\partial y_{n_t+1}}{\partial y_{n_t}} \frac{\mathrm{d} y_{n_t}}{\mathrm{d} \varphi}.
\end{equation}
Consequently, the total derivative requires to explicitly track the effect of the random variable on each timestep and to propagate it accordingly, \revv{a} task that quickly gets unbearable. Alternately, simply using the partial derivative quickly leads to large errors in the approximation, rendering the approach unusable. This \revv{behavior constitued} a serious bottleneck for uncertainty quantification for advanced material models. \revv{As mitigation strategy, one could} resort to slow sampling or collocation methods.

Here, we present a novel idea. As the time-integration destroys any hope to efficiently track the effect of stochasticity, the approximation has to be performed before time integration.
As choice \revv{for} the approximation, \revv{we choose a} Taylor series as in the preceding section. A Taylor series has the desired property of separating stochastic from time-dependent behavior and a polynomial approximation around the expectation is typically reasonable for smooth functions. Therefore, we adapt the Taylor series for multiple random variables as in Equation~\eqref{eq:TaylorMultiDim} to
\begin{equation}
y(t, \Phi) = y(t, \textbf{0}) + Dy(t) \Big\vert_{\Phi = \textbf{0}} \cdot \Phi + \frac{1}{2} \left( D^2 y(t) \Big\vert_{\Phi = \textbf{0}} \cdot \Phi \right) \cdot \Phi + \frac{1}{6} \left( \left( D^3 y(t) \Big\vert_{\Phi = \textbf{0}} \cdot \Phi \right) \cdot \Phi \right) \cdot \Phi + \dots. 
\end{equation}
The main difference to Equation~\eqref{eq:TaylorMultiDim} is that the Taylor series terms are given as derivatives of the unknown function $y(t)$.
As $Dy(t) \Big\vert_{\Phi = \textbf{0}}, D^2y(t) \Big\vert_{\Phi = \textbf{0}}, D^3y(t) \Big\vert_{\Phi = \textbf{0}}, \dots$ are unknown and can not be directly determined, we exchange them \revv{by} the variables $\bfI^{(1)}, \bfI^{(2)}, \bfI^{(3)}, \dots$ with appropriate dimensions. \revv{Let us further denote $y^{(0)}(t) = y(t, \bf0)$.} Accordingly, the Taylor series reads as
\begin{equation}
y(t, \Phi) \approx y^{(0)}(t) + \bfI^{(1)}(t) \cdot \Phi + \frac{1}{2} \left(\bfI^{(2)}(t) \cdot \Phi \right) \cdot \Phi + \frac{1}{6} \left( \left( \bfI^{(3)}(t) \cdot \Phi \right) \cdot \Phi \right) \cdot \Phi + \dots \label{eq:TaylorMultiDim2}
\end{equation}
where $y^{(0)}(t) = y(t, \textbf{0})$.
The time-derivative of Equation~\eqref{eq:TaylorMultiDim2} is given as
\begin{equation}
    \dot{y}(t, \Phi) \approx \dot{y}^{(0)}(t) + \dot{\bfI}^{(1)}(t) \cdot \Phi + \frac{1}{2} \left( \dot{\bfI}^{(2)}(t) \cdot \Phi \right) \cdot \Phi + \frac{1}{6} \left( \left( \dot{\bfI}^{(3)}(t) \cdot \Phi \right) \cdot \Phi \right) \cdot \Phi + \dots \label{eq:TaylorMultiDim2Time}
\end{equation}
\revv{since} the vector of random variables $\Phi$ is constant in time \revv{and} only the terms $\bfI^{(i)}(t)$ are time-dependent. 

The determination of the terms $\bfI^{(i)}(t)$ is now crucial. Equation~\eqref{eq:diffeq} links \revv{the} derivative $\dot{y}$ and \revv{the} function $y$. \revv{Thus}, it also has to hold for the approximation.
Accordingly, we can \revv{write for Equation~\eqref{eq:diffeq}}
\begin{align}
     \dot{y}^{(0)}(t) + \dot{\bfI}^{(1)}(t) \cdot \Phi &+ \frac{1}{2} \left( \dot{\bfI}^{(2)}(t) \cdot \Phi \right) \cdot \Phi + \frac{1}{6} \left( \left( \dot{\bfI}^{(3)}(t) \cdot \Phi \right) \cdot \Phi \right) \cdot \Phi + \dots = \nonumber \\
     &g\left(y^{(0)}(t) + \bfI^{(1)}(t) \cdot \Phi + \frac{1}{2} \left( \bfI^{(2)}(t) \cdot \Phi \right) \cdot \Phi + \frac{1}{6} \left( \left( \bfI^{(3)}(t) \cdot \Phi \right) \cdot \Phi \right) \cdot \Phi + \dots, \Phi\right).
\end{align}
As the equation has to hold for any specific realization of $\Phi$ and even more any possible value for $\Phi$, the equation is separable in multiple equations which have to hold individually.
For each polynomial degree of the random variable, one equation arises which has to be fulfilled.
For a polynomial degree of zero the classical deterministic equation
\begin{equation}
	\dot{y}^{(0)}(t) = g(y^{(0)}(t), \textbf{0})) \label{eq:y0}
\end{equation}
can be easily identified.
To identify equations corresponding to the terms $\bfI^{(1)}, \bfI^{(2)}, \bfI^{(3)}, \dots$ we make use of the Gâteaux derivative and evaluate for $\Phi = \textbf{0}$.
For the term $\bfI^{(1)}$, we calculate
\begin{align}
	\frac{\dd}{\dd \Phi} \Big[ \dot{y}^{(0)}(t) &+ \dot{\bfI}^{(1)}(t) \cdot \Phi + \frac{1}{2} \left( \dot{\bfI}^{(2)}(t) \cdot \Phi \right) \cdot \Phi + \frac{1}{6} \left( \left( \dot{\bfI}^{(3)}(t) \cdot \Phi \right) \cdot \Phi \right) \cdot \Phi + \dots \Big]\bigg\vert_{\Phi = \textbf{0}} = \nonumber \\
	&\frac{\dd}{\dd \Phi} g\left(y^{(0)}(t) + \bfI^{(1)}(t) \cdot \Phi + \frac{1}{2} \left( \bfI^{(2)}(t) \cdot \Phi \right) \cdot \Phi + \frac{1}{6} \left( \left( \bfI^{(3)}(t) \cdot \Phi \right) \cdot \Phi \right) \cdot \Phi + \dots, \Phi\right) \bigg\vert_{\Phi = \textbf{0}} 
\end{align}
and thus, \revv{we} arrive at \revv{the} differential equation 
\begin{equation}
	\dot{\bfI}^{(1)}(t) = \frac{\dd}{\dd \Phi} g\left(y(t, \textbf{0}) + \bfI^{(1)}(t) \cdot \Phi + \frac{1}{2} \left( \bfI^{(2)}(t) \cdot \Phi \right) \cdot \Phi + \frac{1}{6} \left( \left( \bfI^{(3)}(t) \cdot \Phi \right) \cdot \Phi \right) \cdot \Phi, \Phi\right) \bigg\vert_{\Phi = \textbf{0}}. \label{eq:I1}
\end{equation}
It is important to notice that $\dot{\bfI}^{(1)}$ can only depend on terms of equal or lower degree of the random variables. Thus, Equation~\eqref{eq:I1} is a deterministic evolution equation which can be solved for as soon as $y^{(0)}(t)$ is calculated. 
Similarly, we find
\begin{equation}
\dot{\bfI}^{(2)}(t) = \frac{\dd^2}{\dd \Phi \dd \Phi} g\left(y(t, \textbf{0}) + \bfI^{(1)}(t) \cdot \Phi + \frac{1}{2} \left( \bfI^{(2)}(t) \cdot \Phi \right) \cdot \Phi + \frac{1}{6} \left( \left( \bfI^{(3)}(t) \cdot \Phi \right) \cdot \Phi \right) \cdot \Phi, \Phi \right) \bigg\vert_{\Phi = \textbf{0}} \label{eq:I2}
\end{equation}
and 
\begin{equation}
\dot{\bfI}^{(3)}(t) = \frac{\dd^3}{\dd \Phi \dd \Phi \dd \Phi} g\left(y(t, \textbf{0}) + \bfI^{(1)}(t) \cdot \Phi + \frac{1}{2} \left( \bfI^{(2)}(t) \cdot \Phi \right) \cdot \Phi + \frac{1}{6} \left( \left( \bfI^{(3)}(t) \cdot \Phi \right) \cdot \Phi \right) \cdot \Phi, \Phi\right)  \bigg\vert_{\Phi = \textbf{0}}. \label{eq:I3}
\end{equation}
Higher-order terms can be found similarly.
The Equations~\eqref{eq:y0}, \eqref{eq:I1} - \eqref{eq:I3} can be solved by a staggered scheme as they are only coupled in one direction.
Therefore, a standard time-integration scheme can be applied.
The terms $\bfI^{(1)}(t), \bfI^{(2)}(t), \bfI^{(3)}(t), \dots$ increase the number of equations describing the material model to incorporate information about the effect of the random fluctuations.
Thus, the material model consisting of $y^{(0)}(t), \bfI^{(1)}(t), \bfI^{(2)}(t), \bfI^{(3)}(t), \dots$ can be seen as an extended material model.

As \revv{soon as} the time-integration of Equations~\eqref{eq:y0}, \eqref{eq:I1} - \eqref{eq:I3} has been performed, expectation and standard deviation can be calculated \revv{in a post-processing step.}
The expectation is found in index notation as
\begin{equation}
\langle y \rangle = y^{(0)}(t) + \frac{1}{2} \bfI^{(2)}_{ij}(t) \langle \Phi_{i} \Phi_{j} \rangle + \dots. \label{eq:DiffExpectation}
\end{equation}
For the variance we make use of the relationship in Equation~\eqref{eq:variance} \revv{and} we find in index notation
\begin{align}
\textrm{Var}(y) &= \bfI^{(1)}_{a}(t) \langle \Phi_a \Phi_b \rangle \bfI^{(1)}_{b}(t) + \frac{1}{6} \bfI^{(1)}_{c}(t) \langle \Phi_c \Phi_d \Phi_e \Phi_f \rangle \bfI^{(3)}_{def}(t) + \frac{1}{36} \bfI^{(3)}_{ghi}(t) \langle \Phi_g \Phi_h \Phi_i \Phi_j \Phi_k \Phi_l \rangle \bfI^{(3)}_{jkl}(t) + \dots. \label{eq:DiffVariance}
\end{align}
The values of $\langle \Phi \otimes \Phi \rangle, \langle \Phi \otimes \Phi \otimes \Phi \otimes \Phi \rangle, \dots $ can be determined \revv{in the same way as} for the case of algebraic equations, e.g., by means of sampling. The tensors $\langle \Phi \otimes \Phi \rangle, \langle \Phi \otimes \Phi \otimes \Phi \otimes \Phi \rangle, \dots $ can be seen as additional material parameters characterizing the stochastic fluctuations of the material parameters. We demonstrate in the application examples in Section~\ref{sec:ApplicationExamples} that even a linear approximation with the terms  $y^{(0)}(t)$ and $\bfI^{(1)}$ is sufficient for the investigated problem.
}

\section{Overview of the method}
\label{sec:MethodOverview}
The time-separated stochastic mechanics (TSM) is an approach to estimate the stochastic properties \rev{of engineering quantities} of inelastic material models with uncertain material parameters. 
The effect of uncertain material parameters on the evolution equation is estimated by means of an extended material model. \rev{The approach effectively separates stochastic but time-independent from deterministic but time-dependent terms. This makes it possible that the extended material model is deterministic and \revv{as} such, only has to be evaluated once.
In an post-processing step, the influence of the randomness is incorporated. Then, the calculation of expectation and variance of various properties only consists of several fast matrix multiplications.}

The general approach is presented in Algorithm~\ref{alg:TSM}. It can be divided in three distinct phases: pre-processing, simulation and post-processing.
\rev{The pre-processing is done only once for each material. It can be done as soon the material model and the uncertainty in the material \rev{parameters} are known.
Here, the evolution equations for the extended material model are determined by the approach as discussed in Section~\ref{sec:TSM}. Additionally, the value of all required moments of the random variables, e.g, $\langle \Phi \otimes \Phi \rangle, \langle \Phi \otimes \Phi \otimes \Phi \otimes \Phi \rangle, \dots$ is determined by means of sampling. The simulation phase consists of the evaluation of the extended material model consisting of $y^0(t), \bfI^{(1)}(t), \bfI^{(2)}(t), \dots$. For this, classical time-integration techniques as explicit/implicit Euler can be used.
During the post-processing, the expectation and variance of the internal variable and stress are calculated with the Equations~\eqref{eq:AlgExpectation} and \eqref{eq:AlgVariance}, \eqref{eq:DiffExpectation} and \eqref{eq:DiffVariance}.
}

\begin{algorithm}
	\caption{TSM approach}\label{alg:TSM}
	\begin{algorithmic}
		\color{black}
		\Procedure{Pre-processing}{}		\Comment{Once per material}
			\State derive extended material model using Equations~\eqref{eq:y0} and \eqref{eq:I1} 
			\State estimate moments $\langle \Phi \otimes \Phi \rangle, \dots$ by sampling
			\State \Return Evolution equations \rev{for} $y^{(0)}(t), \bfI^{(1)}(t), \dots$, $\langle \Phi \otimes \Phi \rangle, \dots$
		\EndProcedure
		\Procedure{Simulation}{}	\Comment{Once per load case}
			\State time-integration of extended material model with internal variables $\{y^{(0)}_n, \bfI^{(1)}_n, \dots \}$
			\State \Return $y^{(0)}_n(t)$ and $\bfI^{(1)}_n(t), \dots$ for all time steps \rev{$n \in \{1, \dots, N_t\}$}
		\EndProcedure
		\Procedure{Post-processing}{} \Comment{Once per load case}
			\State calculate expectation and variance with Equations~\eqref{eq:AlgExpectation} and \eqref{eq:AlgVariance}, \eqref{eq:DiffExpectation} and \eqref{eq:DiffVariance}
			\State \Return Eexpectation and variance of engineering quantities
		\EndProcedure
	\end{algorithmic}
\end{algorithm}

\section{Application examples}
\label{sec:ApplicationExamples}
In order to evaluate \rev{the time-separated stochastic mechanics (TSM)}, three different inelastic material models are investigated. For these material models, the approximation of expectation and standard deviation for the internal variable(s) and the stress are compared to reference Monte Carlo solutions.
The material models investigated are viscous damage in Section~\ref{sec:ViscousDamage}, viscous phase transformations in Section~\ref{sec:PhaseTransformation} and elasto-viscoplasticity in Section~\ref{sec:ViscoPlasticity}. These material models are fundamentally different as they are used to predict different kinds of material behavior. Viscous damage is used for modeling softening behavior, phase transformations can be used, e.g., to predict the behavior of shape memory alloys, and elasto-viscoplasticity is a fundamental material model used for example for metals. It may be remarked that each material model comes with its own intricacies: 
the damage model is nonlinear in the applied strains and internal variable; the material model of phase transformations is a \rev{coupled} system of nonlinear equations for each phase; elasto-viscoplasticity is described by a set of differential algebraic equations.
Thus, each material model presents a computational challenge for a fast and accurate evaluation under uncertainties.  
For each material model, we shortly present the standard material model, the derivation of the extended material model, the probabilistic analysis and show numerical results for selected loading scenarios. 
The sources of uncertainty are the Lam\'e parameter and additionally for elasto-viscoplasticity, the yield limit. A normal distribution of all sources of uncertainty is assumed. It may be remarked that the method is able to incorporate any material parameter with many standard distribution functions as source of uncertainty.
\rev{We use a linear Taylor series (n = 1) as approximation of algebraic and differential equations.}

\subsection{Viscous damage}
\label{sec:ViscousDamage}
Viscous damage models are a strategy to circumvent mesh dependency in damage simulations \cite{bazant_nonlocal_2002}. The internal variable is the scalar-valued damage variable $d$ with $d = 0$ indicating the undamaged state. The mechanical part of the Helmholtz free energy is given by
\begin{equation}
    \Psi(t) = \frac{1}{2} f(d(t)) \bfvarepsilon(t) \cdot \dsE \cdot \bfvarepsilon(t) = f(d(t)) \Psi_0(t)
\end{equation}
with the damage function $f(d(t)) = \text{exp}(-d(t)) \, \in ]0,1]$ 
and the undamaged Helmholtz energy
\begin{equation}
    \Psi_0(t) = \frac{1}{2} \bfvarepsilon(t) \cdot \dsE \cdot \bfvarepsilon(t).
\end{equation}

The viscous evolution of the damage variable is driven by the Helmholtz free energy as
\begin{equation}
    \dot{d}(t) = \frac{1}{\eta} \text{exp}(-d(t)) \Psi_0(t). \label{eq:DamageEvolEq}
\end{equation}
\rev{A detailed derivation of this material model is given in Section~\ref{sec:derivDamage}.}

\subsubsection{Extended material model}
The first step is the derivation of the extended material model. \rev{In total, three material parameters are used $m = \{\eta, \lambda, \mu\}$. We are assuming fluctuations in the Lam\'e parameters $\lambda$ and $\mu$. Consequently, the elasticity tensor is random. We are directly deriving the Taylor series as discussed in Section~\ref{sec:DiffEq} with respect to the uncertain part of the elasticity tensor \revv{$\Phi = \dsD = \dsE - \langle \dsE \rangle$}.}
\rev{Thus,} the linear Taylor series of the damage variable can be stated as
\begin{equation}
    d(t) = d^{(0)}(t) + \bfI(t) : \dsD. \label{eq:damageTSM}
\end{equation}
Consequently, the time derivative is given as
\begin{equation}
    \dot{d}(t) = \dot{d}^{(0)}(t) +  \dot{\bfI}(t) : \dsD,
\end{equation}
since the random part of the elasticity tensor $\dsD$ is time-independent (cf. Equation~\ref{eq:TaylorMultiDim2Time}).

With these approximations, the evolution equation \eqref{eq:DamageEvolEq} reads as
\begin{align}
    \dot{d}^{(0)}(t) + \dot{\bfI}(t) : \dsD &= \frac{1}{\eta} \exp(-(d^0(t) + \bfI(t) : \dsD) \Psi_0(t) \\
    &= \frac{1}{2 \eta} \exp(-d^0(t)) \exp(- \bfI(t) : \dsD) (\bfvarepsilon(t) \cdot (\dsE^{(0)} + \dsD) \cdot \bfvarepsilon(t)) \nonumber
\end{align}
with the expectation of the material tensor $\dsE^{(0)}$.

The standard deterministic material model is given by the evolution equation \eqref{eq:DamageEvolEq} evaluated at $\dsD = \bf0$ as
\begin{equation}
    \dot{d}^{(0)}(t)\big\vert_{\dsD=\bf0} = \frac{1}{2 \eta} \exp(-d^{(0)}(t)) (\bfvarepsilon(t) \cdot \dsE^{(0)} \cdot \bfvarepsilon(t)).
\end{equation}

The tangent $\dot{\bfI}(t)$ is calculated by taking the derivative of the evolution equation with respect to the uncertain material tensor $\dsD$ (cf. Equation~\eqref{eq:I1}), yielding
\begin{align}
    \dot{\bfI}(t) = \frac{\mathrm{d}\,\dot{d}}{\mathrm{d}\dsD_{st}} \Big\rvert_{\dsD = \bf0} &= \frac{1}{2 \eta} \exp(-d^{(0)}(t)) \Bigg( \frac{\mathrm{d}}{\mathrm{d}\dsD_{st}} \rev{\Big(} \exp(-I_{ab}(t)\dsD_{ab}) \rev{\Big)} (\varepsilon_c(t) (\dsE^{(0)}_{cd} + \dsD_{cd}) \varepsilon_d(t))  \\
    &\hspace{1cm} + \exp(- I_{ab}(t)\dsD_{ab}) \frac{\mathrm{d}}{\mathrm{d}\dsD_{st}} \rev{\Big(} \varepsilon_c(t) (\dsE^{(0)}_{cd} + \dsD_{cd}) \varepsilon_d(t) \rev{\Big)} \Bigg) \Bigg\rvert_{\dsD = \bf0} \nonumber \\
    &= \frac{1}{2\eta} \exp(-d^{(0)}(t)) \left( - I_{st}(t) (\varepsilon_c(t) \dsE_{cd}^0 \varepsilon_d(t)) + \varepsilon_s(t) \varepsilon_t(t) \right) \nonumber \\
    \dot{\bfI}(t) &= \frac{1}{2\eta} \exp(-d^{(0)}(t)) \left( - \bfI(t) (\bfvarepsilon(t) \cdot \dsE^{(0)} \cdot \bfvarepsilon(t)) + \bfvarepsilon(t) \otimes \bfvarepsilon(t) \right). \nonumber
\end{align}

\subsubsection{Probabilistic analysis}
After the time-dependent deterministic terms $d^{(0)}(t)$ and $\bfI(t)$ have been calculated the expectation and variance of internal variable and stress can be determined in a post-processing step.

\paragraph{Expectation and variance of the internal variable $d$}
The expectation of the internal variable $d(t)$ is given with Equation~\eqref{eq:DiffExpectation} by $d^{(0)}(t)$ as
\begin{equation}
    \langle d(t) \rangle = d^{(0)}(t)
\end{equation}
For the variance we find (cf. Equation~\eqref{eq:DiffVariance})
\begin{align}
    \textrm{Var}(d(t)) = \bfI(t) : \langle \dsD \otimes \dsD \rangle : \bfI(t).
\end{align}

\paragraph{Expectation and variance of the stress}
The stress of the damage model is given as
\begin{align}
    \bfsigma(t) &= \exp(-d(t))\dsE \cdot \bfvarepsilon(t) \nonumber \\
    &= \exp(-d^{(0)}(t)-\bfI(t) : \dsD) (\dsE^{(0)} + \dsD) \cdot \bfvarepsilon(t).
\end{align}
\rev{This is an algebraic equation, hence, we are using the approach of Section~\ref{sec:AlgEq} to calculate expectation and variance. Following Equation~\eqref{eq:TaylorMultiDim}, the linear approximation of the stress} is given for each component $\alpha$ as
\begin{align}
    \sigma_\alpha(t) = \sigma_\alpha^{(0)}(t) + T_{\alpha bc}(t) \dsD_{bc}. 
\end{align}
Here, $\alpha$ is a fixed index and no summation is carried out over this index.
The terms $\sigma_a^0(t)$ and $T_{\alpha bc}(t)$ are found by
\begin{equation}
    \sigma^{(0)}(t) = \exp(-d^{(0)}(t)) \dsE^{(0)} \cdot \bfvarepsilon(t)
\end{equation}
and 
\begin{align}
    T_{\alpha bc}(t) &= \pf{\sigma_\alpha(t)}{\dsD_{bc}} \Big\rvert_{\dsD = \bf0} \nonumber \\
    &= \exp(-d^{(0)}(t))\left(\dsI_{\alpha b} \varepsilon_c - I_{bc}(t) \dsE_{\alpha d} \varepsilon_d(t) \right)
\end{align}
\rev{with the identity matrix $\dsI$}.
The calculation of expectation and variance is now straightforward.
The expectation of the stress is given by Equation~\eqref{eq:AlgExpectation} as
\begin{equation}
    \langle \sigma(t) \rangle = \sigma^{(0)}(t) = \exp(-d^{(0)}(t)) \dsE \cdot \bfvarepsilon(t).
\end{equation}

The variance of each component $\alpha$ of the stress is calculated with Equation~\eqref{eq:AlgVariance} by
\begin{equation}
    \text{Var}(\sigma_\alpha(t)) = T_{\alpha bc}(t) \langle \dsD_{bc} \dsD_{de} \rangle T_{\alpha de}(t).
\end{equation}

\subsubsection{Numerical experiments}
The approximation obtained by the time-separated stochastic mechanics approach are compared against a reference Monte Carlo simulation. An explicit time-integration for the material model is used. For the Monte Carlo simulation, 1000 iterations \rev{are performed} each with a different realization of the Lam\'e parameters and consequently elasticity tensor $\dsE$. Expectation and standard deviation of internal variable and stress are calculated in an post-processing step.
The tensor $\langle \dsD \otimes \dsD \rangle$ is calculated by means of sampling before the simulation.

The material parameters $\lambda, \mu$ are chosen as independent normal random variables with $\langle \lambda \rangle = \SI{12}{GPa}, \langle \mu \rangle = \SI{8}{GPa}, \eta = \SI{10}{MPa.s}$. The standard deviation of $\lambda$ and $\mu$ are chosen as 15\% of their mean value. The system is simulated for a total of 100 seconds with a time increment of $\Delta t = \SI{5}{ms}$.

\begin{figure}
	\centering
	\begin{subfigure}[b]{0.49\textwidth}
		\centering
		\includegraphics[scale=0.3]{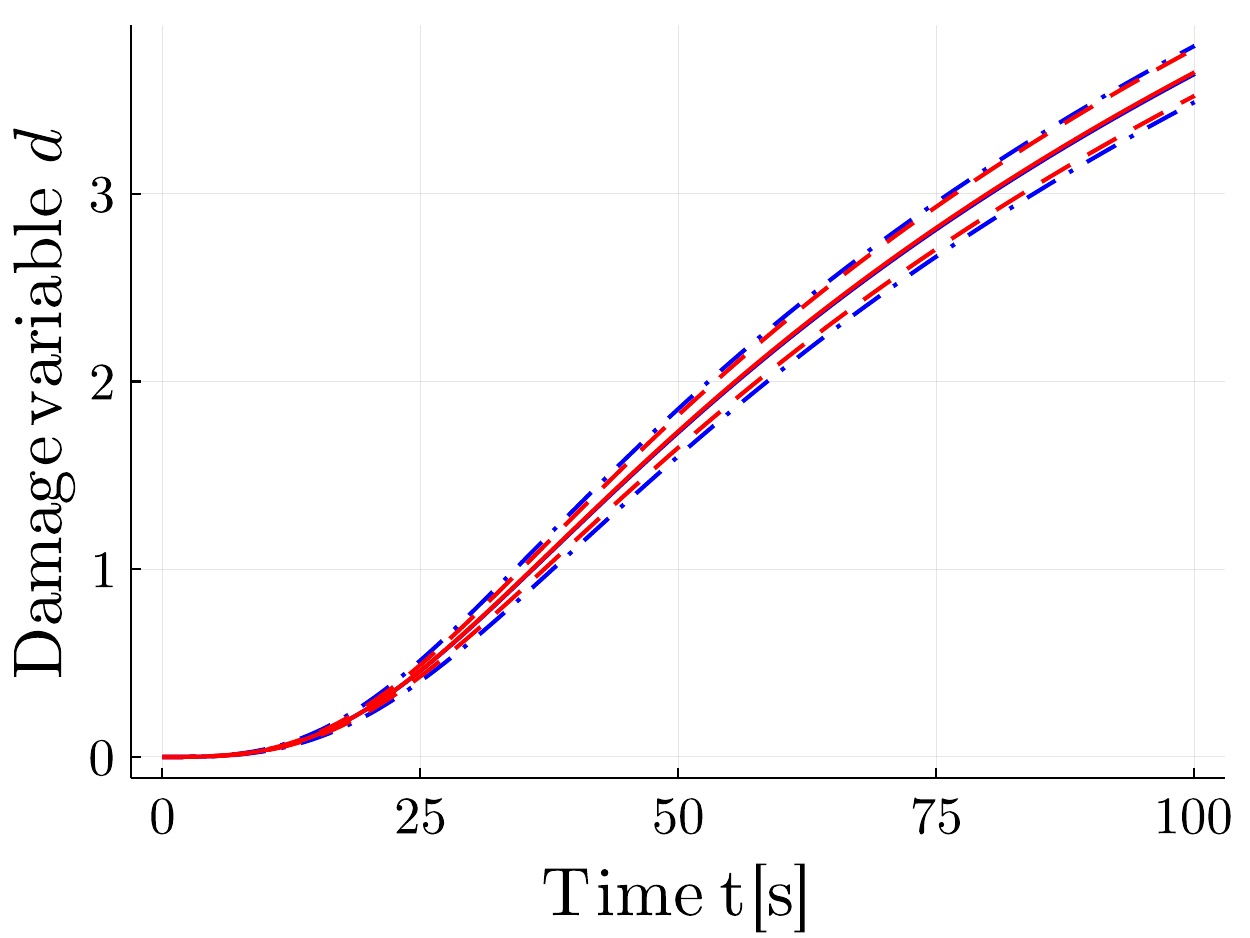}
		\caption{Internal variable $d$}
		\label{fig:DamageResFast1}
	\end{subfigure}
	\hfill
	\begin{subfigure}[b]{0.49\textwidth}
		\centering
		\includegraphics[scale=0.3]{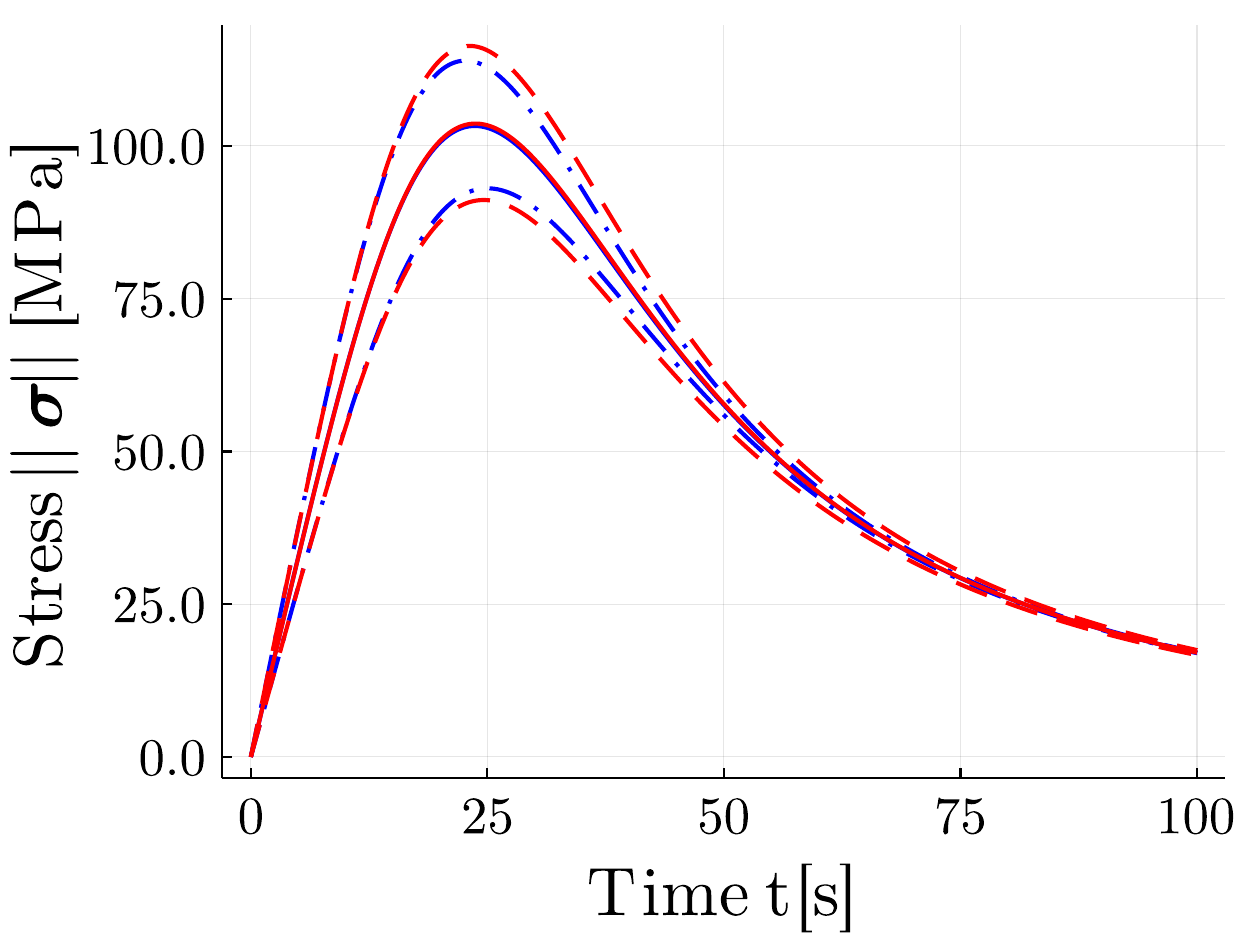}
		\caption{Stress $\bfsigma$}
		\label{fig:DamageResFast2}
	\end{subfigure}
	\caption{Comparison of the results of TSM and reference Monte Carlo for internal variable and stress for a doubled loading velocity for the damage model. Expectation (solid line) and expectation $\pm$ standard deviation (dashed lines) are presented. Results in blue are obtained by MC, \rev{results} in red by TSM.}
	\label{fig:DamageResFast}
\end{figure}

Two different load cases are investigated. The results are presented in Table~\ref{tab:DamageRes}. The loading scenario is presented in the first column, the expectation and standard deviation of internal variable and stress are shown in the second and third column respectively. The expectation is indicated by a solid line, the expectation plus/minus standard deviation by a dashed line. Results from Monte Carlo are colored blue; results from TSM are red. In the first loading scenario, a time-proportional increasing strain in \rev{$x$}-direction is applied. In the second loading scenario, a time-harmonic loading is applied for all strain components.
The typical properties of a damage model are visible: the damage variable is non-decreasing and a reduction in stiffness takes place.
There exists no simple correspondence between expectation and standard deviation of internal variable and stress. This becomes especially clear by the results for the stress for the first loading scenario. At the end of the simulation, the standard deviation is vanishingly small while the expectation is non-zero. Overall, the expectation and standard deviation are captured well by the TSM.
In addition, the effect of a change in loading velocity is investigated. For this, the first load case is simulated with a doubled loading velocity.
The result is presented in Figure~\ref{fig:DamageResFast}. At the beginning, a higher maximum stress is reached due to the viscous evolution of the damage. At the end of \rev{the simulation,} the stiffness of the material point is nearly completely reduced. Once again, expectation and standard deviation are remarkably well approximated by the TSM hinting at the robustness of the proposed method.  

\begin{table}[]
    \centering
    \begin{tabular}{l|cc}
        Loading & Internal variable & Stress  \\[2pt] \hline
        \includegraphics[scale=0.3]{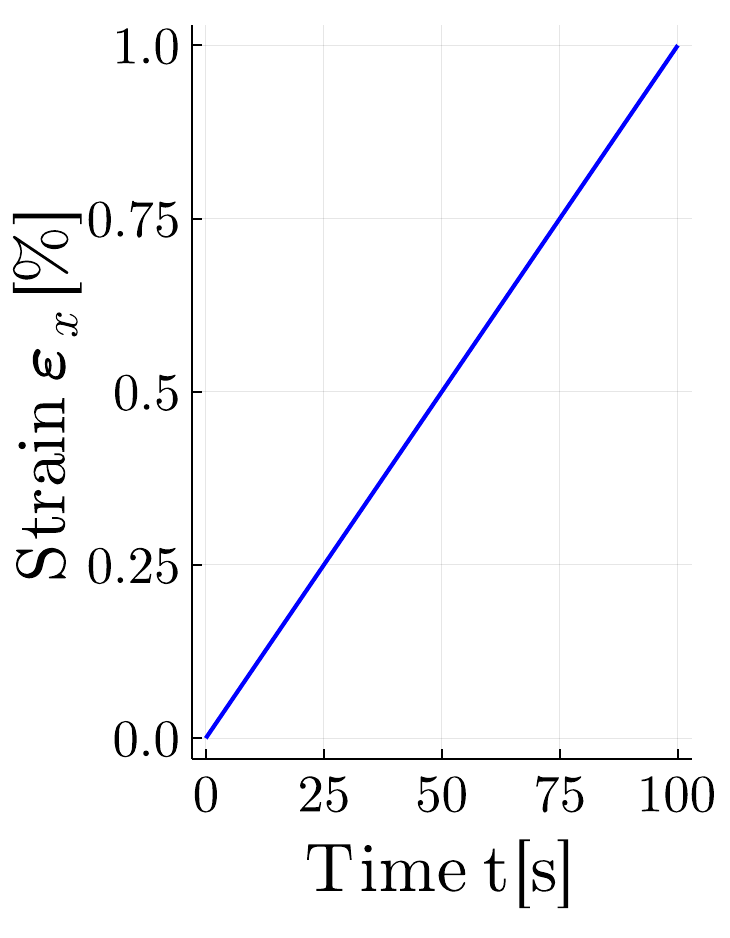}& \includegraphics[scale=0.3]{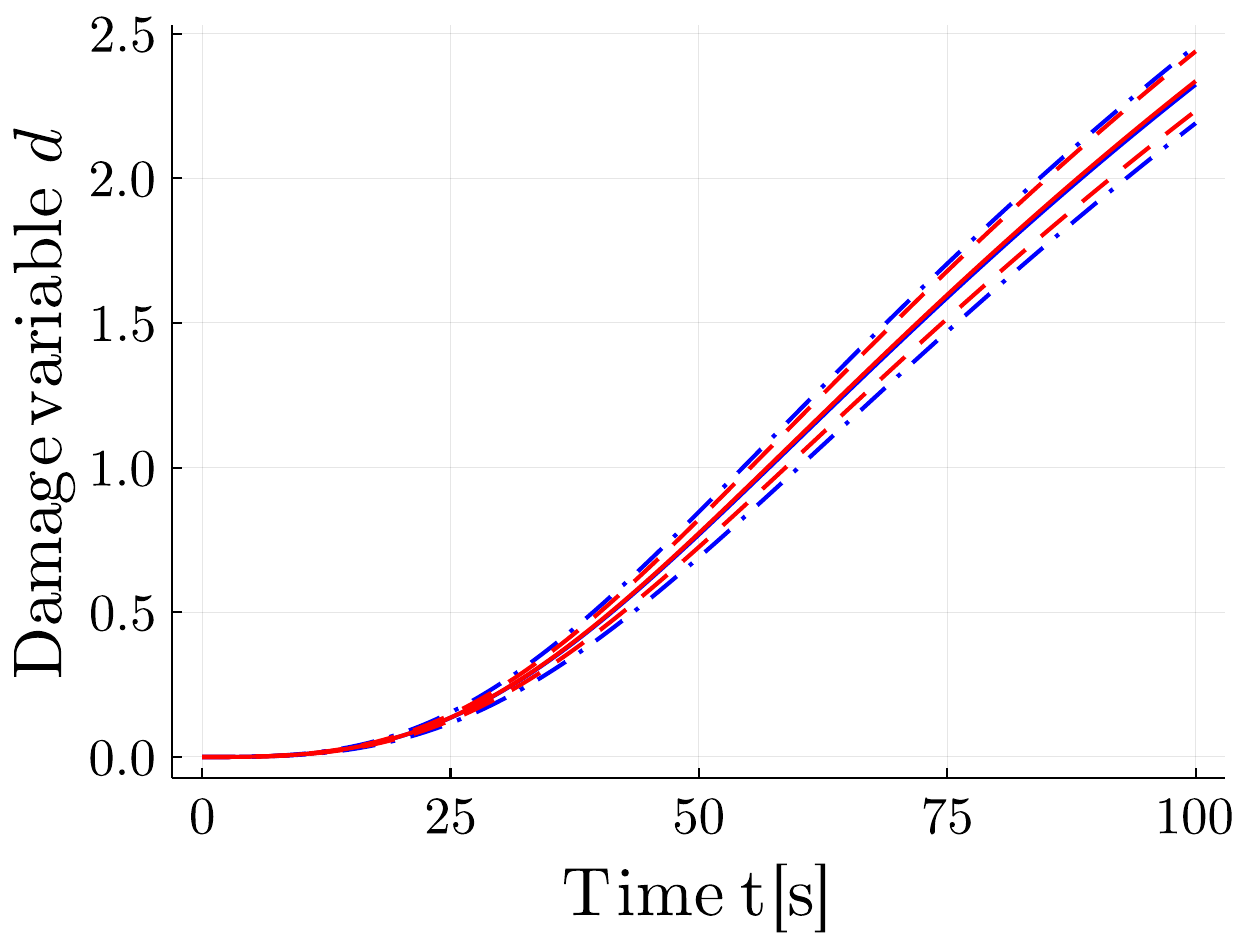} & \includegraphics[scale=0.3]{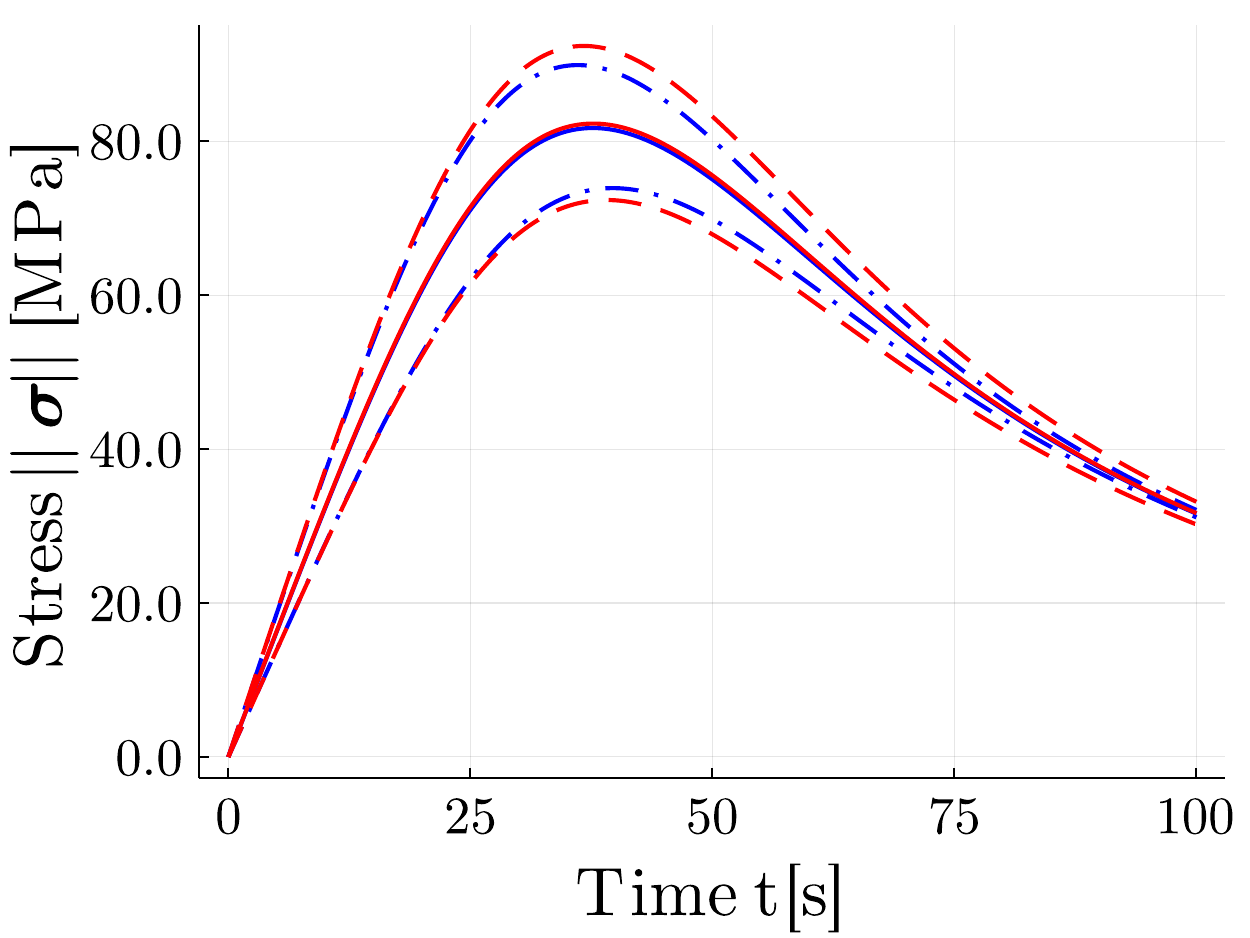}  \\
        
        \includegraphics[scale=0.3]{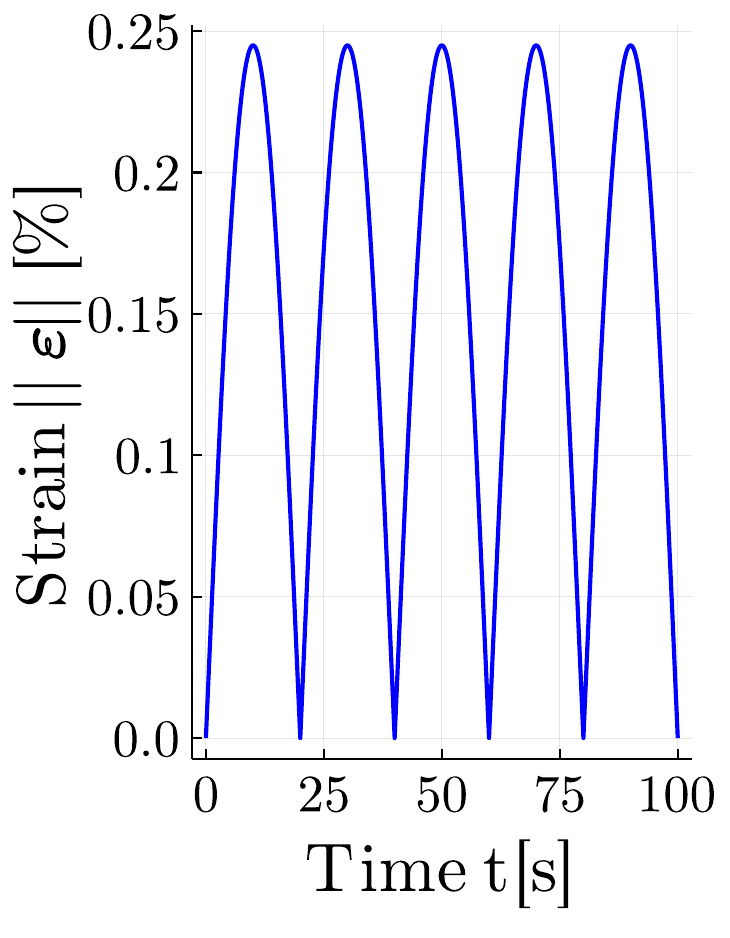}& \includegraphics[scale=0.3]{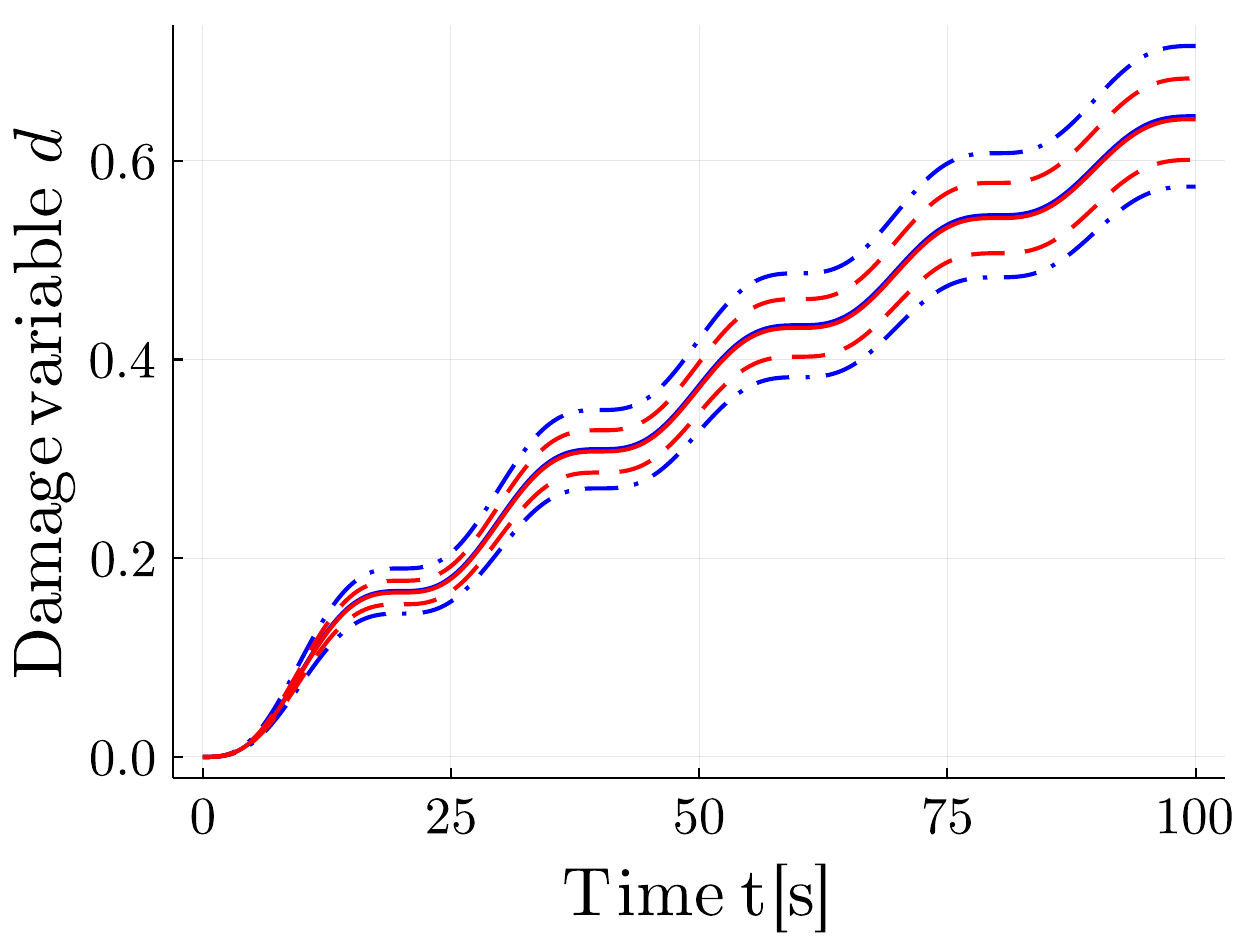} & \includegraphics[scale=0.3]{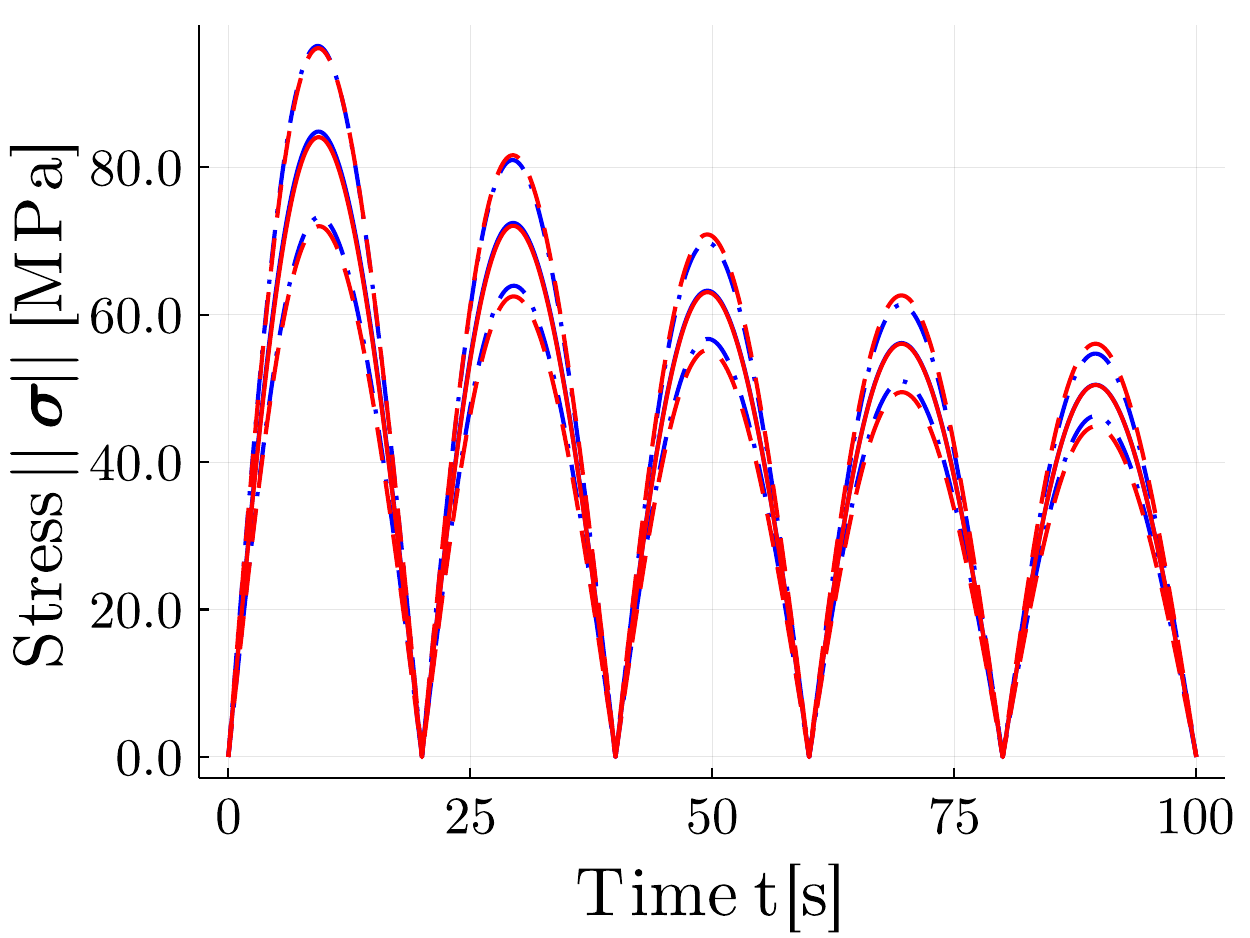}  \\
    \end{tabular}
    \caption{Comparison of the results of TSM and reference Monte Carlo for internal variable and stress of the damage model. Expectation (solid line) and expectation $\pm$ standard deviation (dashed lines) are presented. Results in blue are obtained by MC, \rev{results} in red by TSM.}
    \label{tab:DamageRes}
\end{table}

\subsection{Phase transformations}
\label{sec:PhaseTransformation}
The crystalline lattice structure within a solid, the so-called "phase", has a major influence on the behavior of the material. This transition from one crystalline lattice structure to another is called a "phase transformation". Phase transformations are influenced not only by thermal and mechanical loadings, but also by the loading history of the material. The \rev{mechanics} of phase transformations in metals depends on their alloy composition. A certain category of materials, \rev{which is referred to} as shape memory alloys, is particularly sensitive to the above factors during phase transformations. This property makes shape memory alloys suitable for a wide range of applications. However, this sensitivity is associated with complex material behavior, making research, development, and manufacturing of structures using shape memory alloys challenging \cite{otsuka1999shape}. \rev{The development processes are complicated due to} the need to account for the statistical distribution of characteristic values since the thermomechanical behavior varies as a result. Even with consistent manufacturing processes using bulk materials, this stochastic variation in material response remains significant \cite{lagoudas2008shape}.

\rev{
At least two different phases are involved in a phase transformation. Each phase $i$ has an associated volume fraction $\lambda_i \in [0,1]$. Of course, the sum of all volume fraction is $\sum_i \lambda_i = 1$ due to volume conversation. 
The evolution of the individual volume fractions over time due to external loads is described by the material model. }

\rev{
The material model is formulated as an evolution equation of the volume fraction variables $\chi_i \in [-\infty, \infty]$ as it yields a simpler structure.
The volume fraction variable $\chi_i$ and volume fraction $\lambda_i$ are linked by 
\begin{equation}
    \lambda_i = (1+\exp(-\chi_i))^{-1}. \label{eq:Lambda}
\end{equation}
}
\rev{
The material model for an \rev{isothermal} viscous phase transformation reads
\begin{equation}
    \dot{\chi}_i(t) = \left( \frac{\partial \lambda_i}{\partial \chi_i} \right)^{-1} \frac{1}{\eta} \left( - \frac{\partial \Psi}{\partial\lambda_i} +\frac{1}{n} \sum_j \frac{\partial \Psi}{\partial \lambda_j} \right), \label{eq:PhaseTransformDEq}
\end{equation}
where $\eta$ represents the viscosity parameter for the process of phase transformations. A full derivation of this material model is given in the Appendix~\ref{sec:derivPT}.
The Helmholtz free energy $\Psi$ is given by
\begin{equation}
    \Psi = \frac{1}{2} (\bfvarepsilon - \bar{\bfeta}) \cdot \bar{\dsE} \cdot (\bfvarepsilon - \bar{\bfeta})+ \Lambda \sum_i \left( \frac{1}{\lambda_i^2(\lambda_i-1)^2} \right) \label{eq:PTHelmholtz}
\end{equation}
with the strains $\bfvarepsilon$, the effective transformation strain $\bar{\bfeta}$ and the effective elasticity tensor $\bar{\dsE}$.
The last term introduces a potential wall to exclude states in which a phase approaches a volume fraction of 0 or 1. The parameter $\Lambda$ influences the magnitude of the potential wall.
The effective transformation strain and the effective elasticity tensor are given as geometric and harmonic mean, respectively.
In particular, they are defined as
\begin{equation}
    \bar{\bfeta} = \sum_i \lambda_i \bfeta_i  \quad\text{ and } \quad \bar{\dsE} = \left( \sum_i \lambda_i \dsE_i^{-1} \right)^{-1}
\end{equation}
where $\bfeta_i$ and $\dsE_i$ are material specific parameter for each phase \rev{\cite{junker_accurate_2014}}.
}
Equation~\eqref{eq:PhaseTransformDEq} is an evolution equation for the volume fraction variable $\chi_i$. In total, there are $n$ coupled \rev{and highly nonlinear} equations for $n$ phases.
The coupling stems from the volume conservation requirement ($\sum_i \lambda_i = 1$). It may be remarked that the evolution equation has no obvious analytical solution. 
Thus, this set of equations represents a challenge for an efficient probabilistic analysis \rev{of the high nonlinear behavior}.

\subsubsection{Extended material model}
One of the main sources of uncertainty stems from the Lam\'e parameter and thus the elasticity tensor $\dsE_i$ of each phase $i$. Here, we present how the effect of uncertainty of the elasticity tensors on the volume fractions and the stress can be approximated.
For simplicity, we assume $n = 2$ phases in the following. It may be remarked that a higher number of phases can be handled \rev{well without conceptual changes.}
\rev{As the Lam\'e parameters of both phases can fluctuate simultaneously and independently, the Taylor series has to incorporate all random variables. As before, we directly formulate the Taylor series in respect of the uncertain elasticity tensor. Therefore, in this case the Taylor series is formulated in respect to $\dsD_1$ and $\dsD_2$ \revv{for the two phases, i.e., $\Phi = [\dsD_1, \dsD_2]$}. For simplicity, we further split up the tangent $\bfI_i$ in a part $\bfI_{i1}$ which approximates the effect of fluctuations of $\dsD_1$ and a part $\bfI_{i2}$ for the fluctuations of $\dsD_2$. Thus, the Taylor series for the internal variable $\chi_i$ reads as}
\begin{align}
    \chi_i(t) = \chi_i^{(0)}(t) + \bfI_{i1}(t) : \dsD_1 + \bfI_{i2}(t) : \dsD_2. \label{eq:PhaseTransformTaylor}
\end{align}
\revv{Here, the tensor of random variables $\Phi$ is split up into the random part of both elasticity tensors $\dsD_1$ and $\dsD_2$ for readability.}
Due to the coupling of the phases to fulfill the volume constraint, each internal variable \revv{depends on the random fluctuations of} each elasticity tensor. 

With the evolution equation in \eqref{eq:PhaseTransformDEq}, we find
\begin{equation}
    \chi_i^{(0)}(t) = \left( \left( \frac{\partial \lambda_i}{\partial \chi_i} \right)^{-1} \frac{1}{\eta} \left( - \frac{\partial \Psi}{\partial\lambda_i} +\frac{1}{n} \sum_i \frac{\partial \Psi}{\partial \lambda_i} \right) \right) \Bigg{\rvert}_{\dsD_1 = \dsD_2 = \bf0}
\end{equation}
The tangents $\bfI_{i1}$ and $\bfI_{i2}$ are given by
\begin{align}
    \bfI_{ij}(t) = \frac{\mathrm{d}\chi_i(t)}{\mathrm{d}\dsD_j} \Big\vert_{\dsD_1 = \dsD_2 = \bf0}. \label{eq:PhaseTransformTangent}
\end{align}
A symbolic derivation of the tangent is given in Appendix~\ref{sec:tangentEMM}.
While the symbolic tangent offers important insights into the mathematical structure of the problem, it lacks computational efficiency. The identification of symmetry conditions and common computations offers a large potential for a reduce in computational complexity. Here, the software AceGen \cite{korelc_multi-language_2002, korelc_automation_2016} in Mathematica \cite{Mathematica} is used for the automatic derivation of the tangent in Equation~\eqref{eq:PhaseTransformTangent}. 

\subsubsection{Probabilistic analysis}
In the following, we assume that the fluctuations in $\dsD_1$ and $\dsD_2$ are independent. However, any stochastic dependence of the elasticity tensor fluctuations can be easily incorporated.
\paragraph{Expectation and variance of internal variables $\chi_i$}
The expectation of the volume fraction variable of each phase $i$ is given by 
\begin{equation}
	\langle \chi_i \rangle = \chi_i^{(0)}(t). 
\end{equation}
The variance is calculated by
\begin{align}
    \textrm{Var}(\chi_i) = \langle \left( \bfI_{i1}(t) : \dsD_1 \right)^{\otimes2} \rangle + \langle \left( \bfI_{i2}(t) : \dsD_2 \right)^{\otimes2} \rangle 
\end{align}
with
\begin{equation}
    \langle \left( \bfI_{ij}(t) : \dsD_j \right)^{\otimes2} \rangle = \bfI_{ij}(t) : \langle \dsD_j \otimes \dsD_j \rangle : \bfI_{ij}(t).
\end{equation}
If the material fluctuations in the two phases are not stochastically independent, an additional coupling term would result.

\paragraph{Expectation and variance of volume fraction $\lambda$}
The internal variable $\chi$ is not trivial to interpret. In contrast, the interpretation of the volume fraction $\lambda$ is straightforward. For this reason, we are interested in the expectation and variance of this variable.
\rev{However, $\lambda$ itself is nonlinear in the internal variable $\chi$ and additionally the uncertain material parameters.
	Equation~\eqref{eq:Lambda} is an algebraic equation if the approximation of Equation~\eqref{eq:PhaseTransformTaylor} is used. Thus, we make use of the approach of Section~\ref{sec:AlgEq}. The linearization of the volume fraction (cf. Equation~\eqref{eq:TaylorMultiDim}) is given as }
\begin{equation}
    \lambda_i(t) \approx \lambda_i^{(0)}(t) + \frac{\mathrm{d} \lambda_i(t)}{\mathrm{d} \dsD_1}\Big\vert_{\dsD_1 = \dsD_2 = \bf0} : \dsD_1 + \frac{\mathrm{d} \lambda_i(t)}{\mathrm{d} \dsD_2}\Big\vert_{\dsD_1 = \dsD_2 = \bf0} : \dsD_2.
\end{equation}
The expectation is then found as
\begin{equation}
    \langle \lambda_i(t) \rangle = \lambda_i^{(0)}(t).
\end{equation}
The derivatives in the linearization can be calculated as
\begin{equation}
    \frac{\mathrm{d}\lambda_i}{\mathrm{d}\dsD_j}\Big\vert_{\dsD_1 = \dsD_2 = \bf0} = \pf{\lambda_i}{\chi_i} \pf{\chi_i}{\dsD_j}\Big\vert_{\dsD_1 = \dsD_2 = \bf0} = (\lambda_i^{(0)})^2 \exp(-\chi_i^{(0)}) \bfI_{ij} =: T^\lambda_{ij}. \label{eq:dlambda_dD}
\end{equation}
Therefore, the variance of the volume fraction is given as 
\begin{equation}
    \textrm{Var}(\lambda_i) = \sum_j T^\lambda_{ij} : \langle \dsD_j \otimes \dsD_j \rangle : T^\lambda_{ij}
\end{equation}
for stochastically independent fluctuations in $\dsD_j$.

\paragraph{Expectation and variance of stress}
The stress can be computed using the relation
\begin{align}
    \bfsigma(t) = \bar{\dsE} \cdot (\bfvarepsilon - \bar{\bfeta}) = \left( \sum_i \lambda_i \dsE_i^{-1} \right)^{-1} \cdot (\bfvarepsilon - \bar{\bfeta}) \label{eq:PhaseTransformStress}
\end{align}
As presented in \rev{Section~\eqref{sec:AlgEq}} we linearize the stress for each component $\alpha$ as
\begin{equation}
    \sigma_\alpha(t) \approx \sigma_\alpha^{(0)}(t) + \frac{\mathrm{d}\sigma_\alpha(t)}{\mathrm{d}\dsD_1}\Big\rvert_{\dsD_1=\dsD_2=\bf0} : \dsD_1 + \frac{\mathrm{d}\sigma_\alpha(t)}{\mathrm{d}\dsD_2}\Big\rvert_{\dsD_1=\dsD_2=\bf0} : \dsD_2. \label{eq:PhaseTransformLinearStress}
\end{equation}
The expectation is found for each component $\alpha$ \rev{following Equation~\eqref{eq:AlgExpectation}} as
\begin{equation}
    \langle \sigma_\alpha(t) \rangle = \sigma_\alpha^{(0)}(t) = \bar{\dsE}_{\rev{\alpha b}} \Big\rvert_{\dsD_1 = \dsD_2 = 0} (\bfvarepsilon_b - \bar{\bfeta}_b) 
\end{equation}

The derivatives in Equation~\eqref{eq:PhaseTransformLinearStress} can be found as
\begin{equation}
    \frac{\mathrm{d}\sigma_\alpha(t)}{\mathrm{d}\dsD_j}\Big\rvert_{\dsD_1 = \dsD_2 = \boldface{0}} =: T^\sigma_{j, \alpha}(t).
\end{equation}
The detailed derivation is presented in Appendix~\ref{sec:tangentStress}.
The variance of the stress for each component~$\alpha$ is given \rev{by Equation~\eqref{eq:AlgVariance}} as
\begin{equation}
    \textrm{Var}(\sigma_a(t)) =  \sum_j (T_{j, \alpha}^\sigma)^T(t) : \langle \dsD_j \otimes \dsD_j \rangle : T_{j, \alpha}^\sigma(t) 
\end{equation}
in the case of stochastically independent material fluctuations. Otherwise, an additional coupling term would arise.

\subsubsection{Numerical experiments}
\begin{figure}
	\centering
	\begin{subfigure}[b]{0.3\textwidth}
		\centering
		\includegraphics[trim=0cm 12.5cm 0cm 0cm,clip,scale=.25]{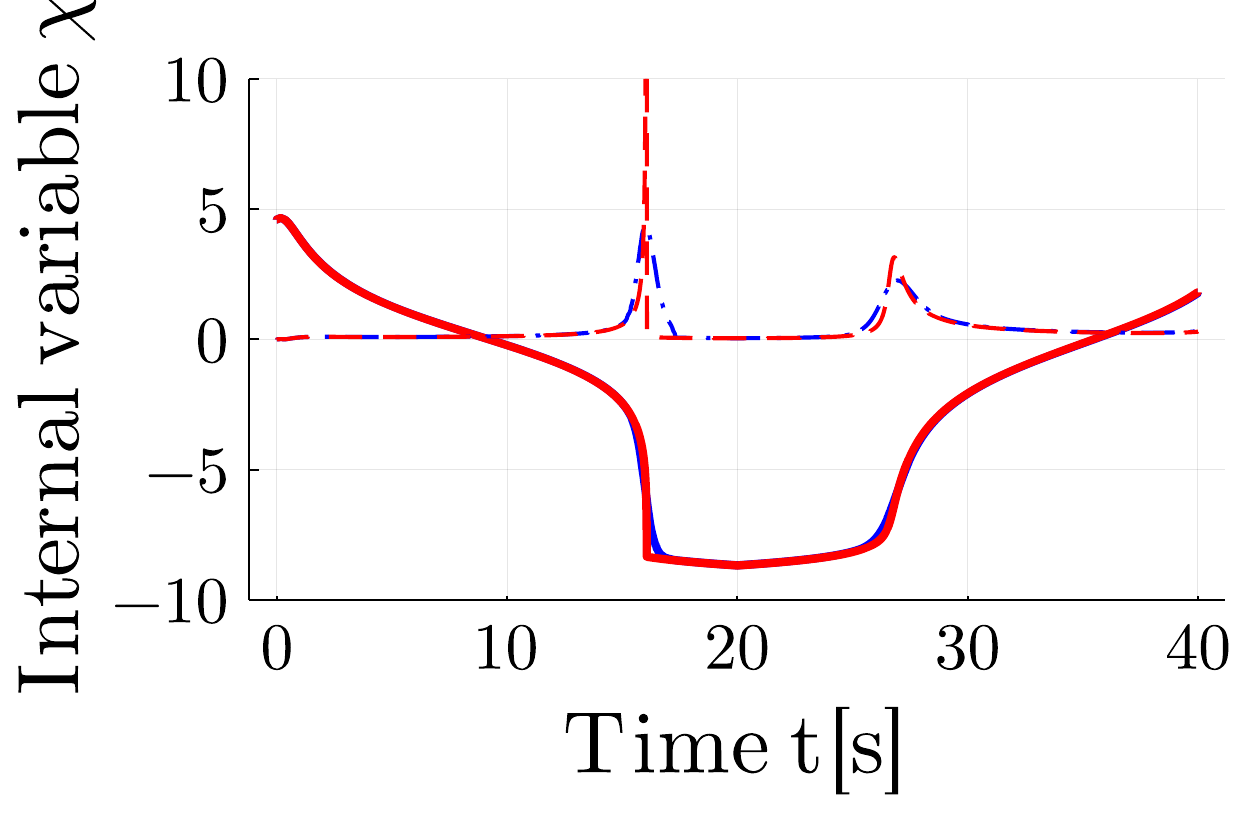}
	\end{subfigure}
	\begin{subfigure}[b]{0.3\textwidth}
		\centering
		\includegraphics[trim=0cm 12.5cm 0cm 0cm,clip,scale=.25]{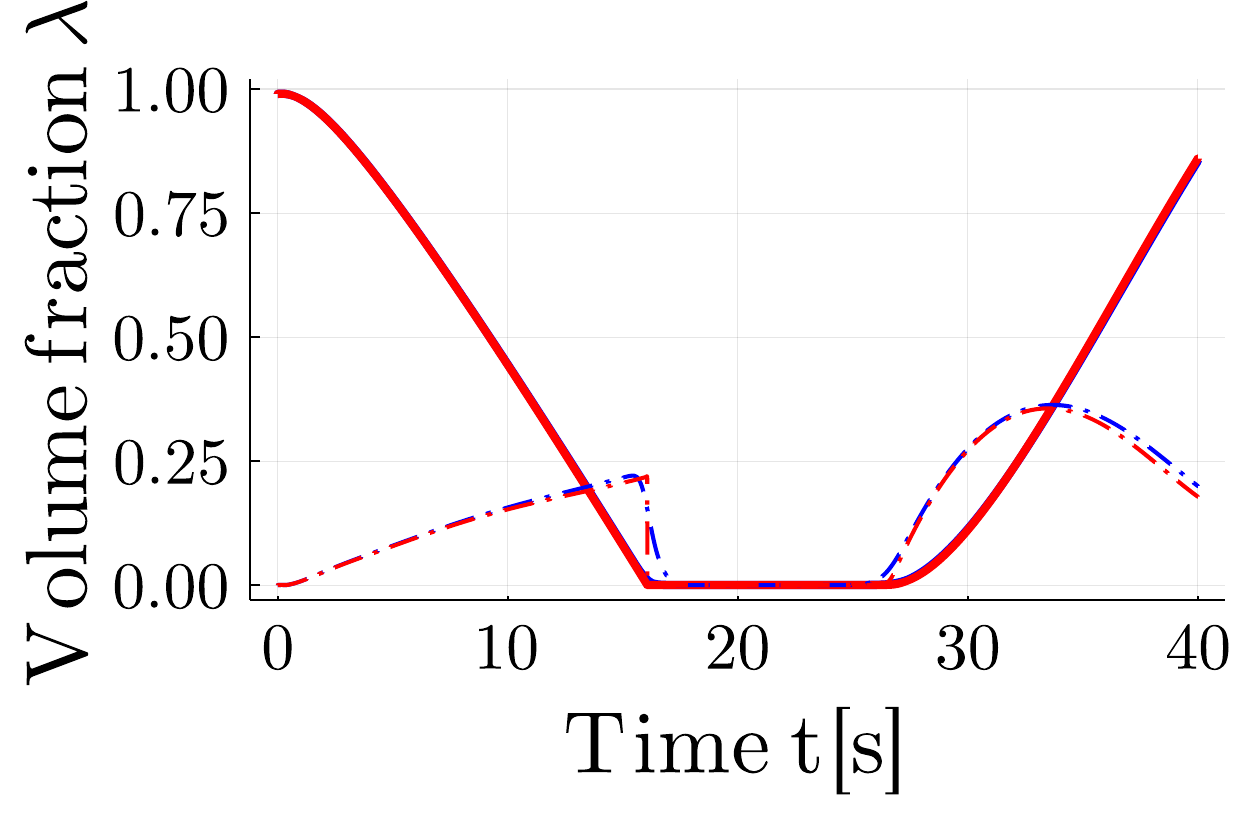}
	\end{subfigure}
	\begin{subfigure}[b]{0.3\textwidth}
		\centering
		\includegraphics[trim=0cm 12.5cm 0cm 0cm,clip,scale=.25]{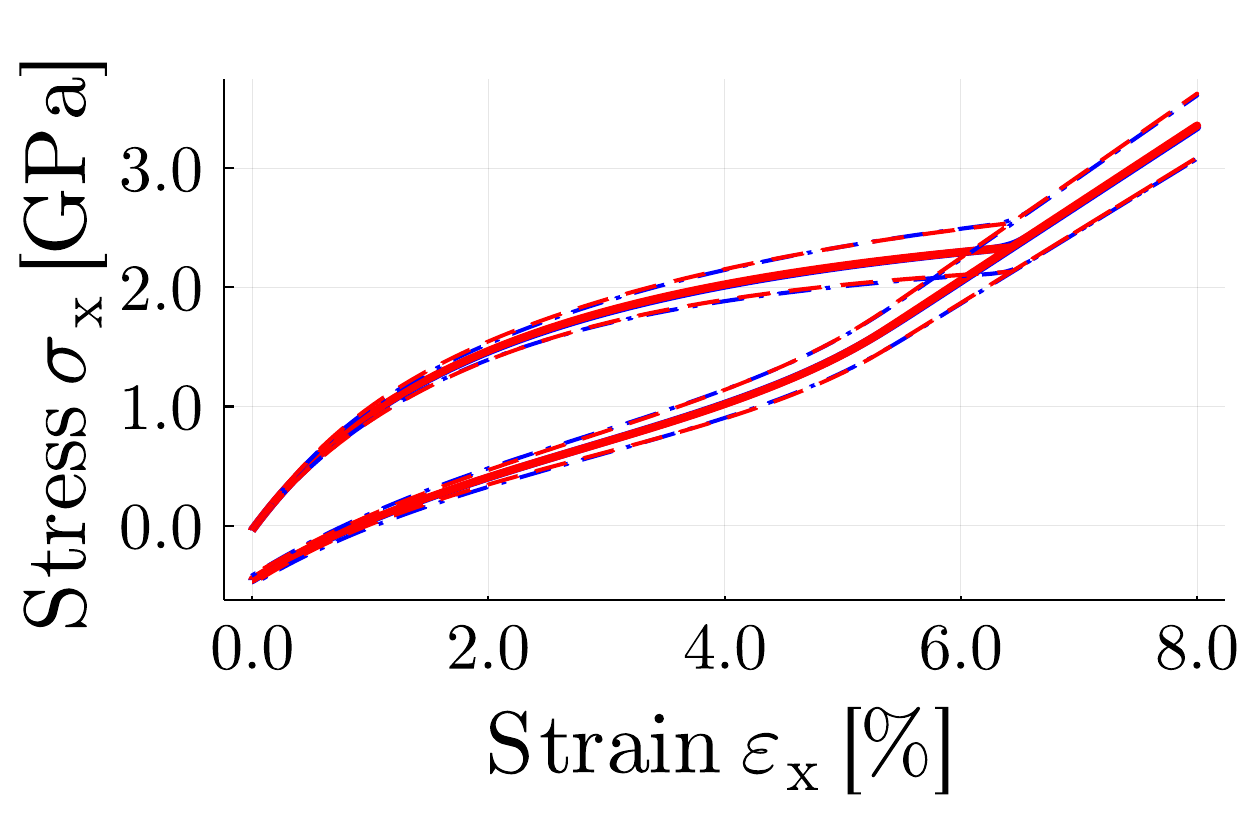}
	\end{subfigure}
	\begin{subfigure}[b]{0.3\textwidth}
		\centering
		\includegraphics[trim=0cm 4cm 0cm 12.5cm,clip,scale=.25]{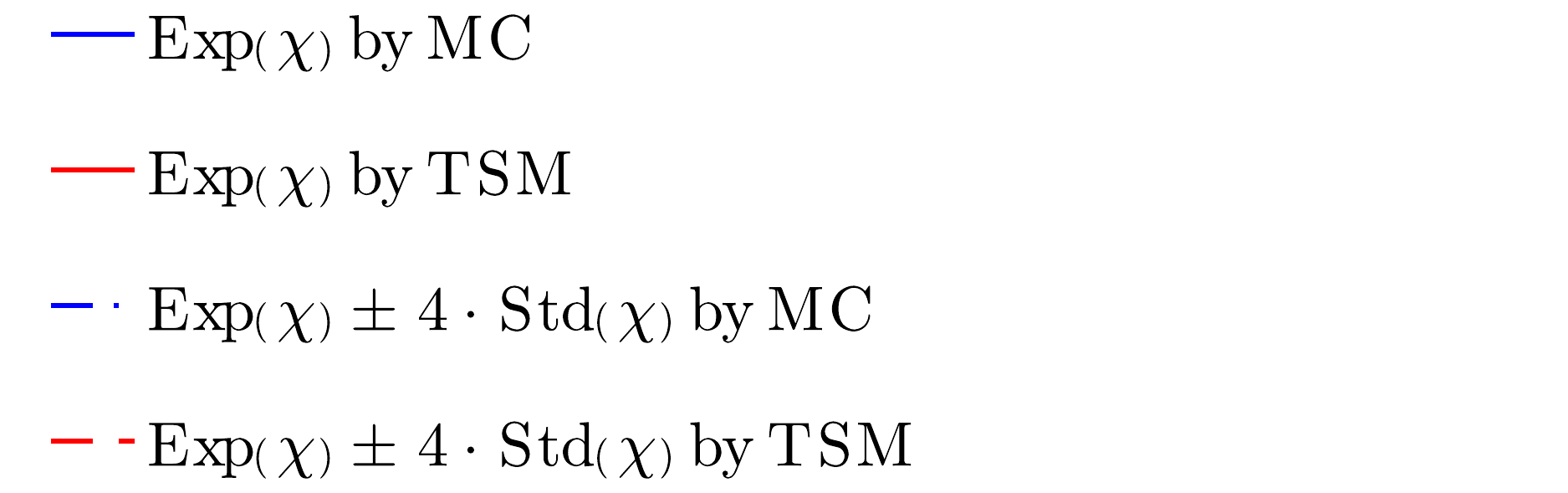}
		\caption{Internal variable $\chi_1$}
		\label{fig:resPhaseTransform2}
	\end{subfigure}
	\begin{subfigure}[b]{0.3\textwidth}
		\centering
		\includegraphics[trim=4cm 4cm 0cm 12.5cm,clip,scale=.25]{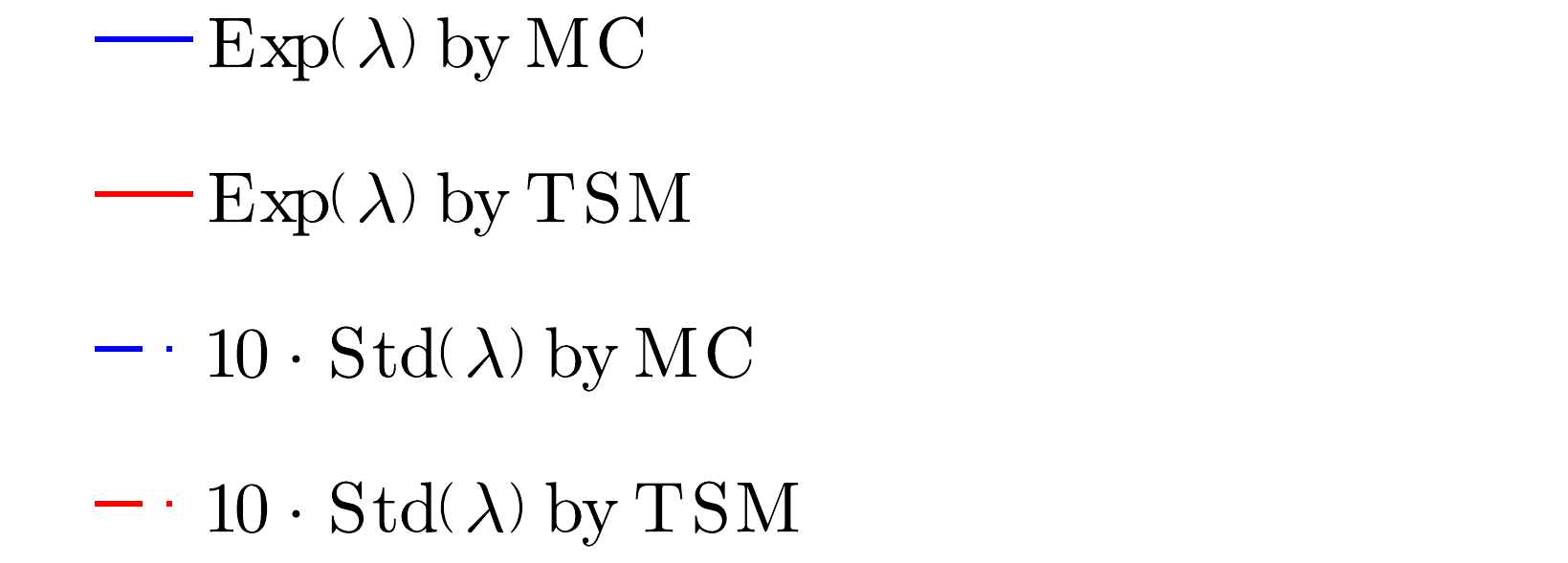}
		\caption{Volume fraction $\lambda_1$}
		\label{fig:resPhaseTransform3}
	\end{subfigure}
	\begin{subfigure}[b]{0.3\textwidth}
		\centering
		\includegraphics[trim=0cm 4cm 0cm 12.5cm,clip,scale=.25]{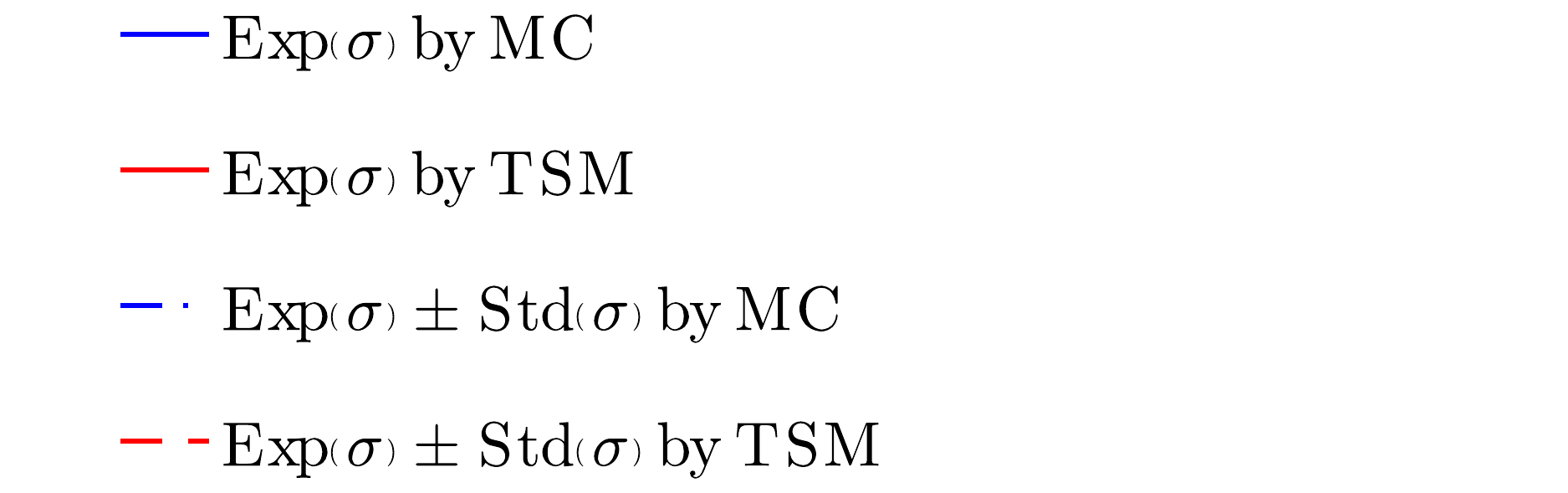}
		\caption{Stress $\sigma_\mathrm{x}$}
		\label{fig:resPhaseTransform1}
	\end{subfigure}
	\caption{Results for internal variable $\chi_{1}$ over \rev{time} $t$, volume fraction $\lambda_{1}$ over \rev{time} $t$ and stress $\sigma_{x}$ over strain $\epsilon_{x}$ for viscosity parameter $\eta=\SI{0.2}{GPas}$. Results in blue are obtained by MC, \revv{results} in red by TSM.}
	\label{fig:resPhaseTransform}
\end{figure}

\begin{figure}
	\centering
	\begin{subfigure}[b]{0.3\textwidth}
		\centering
		\includegraphics[trim=0cm 12.5cm 0cm 0cm,clip,scale=.25]{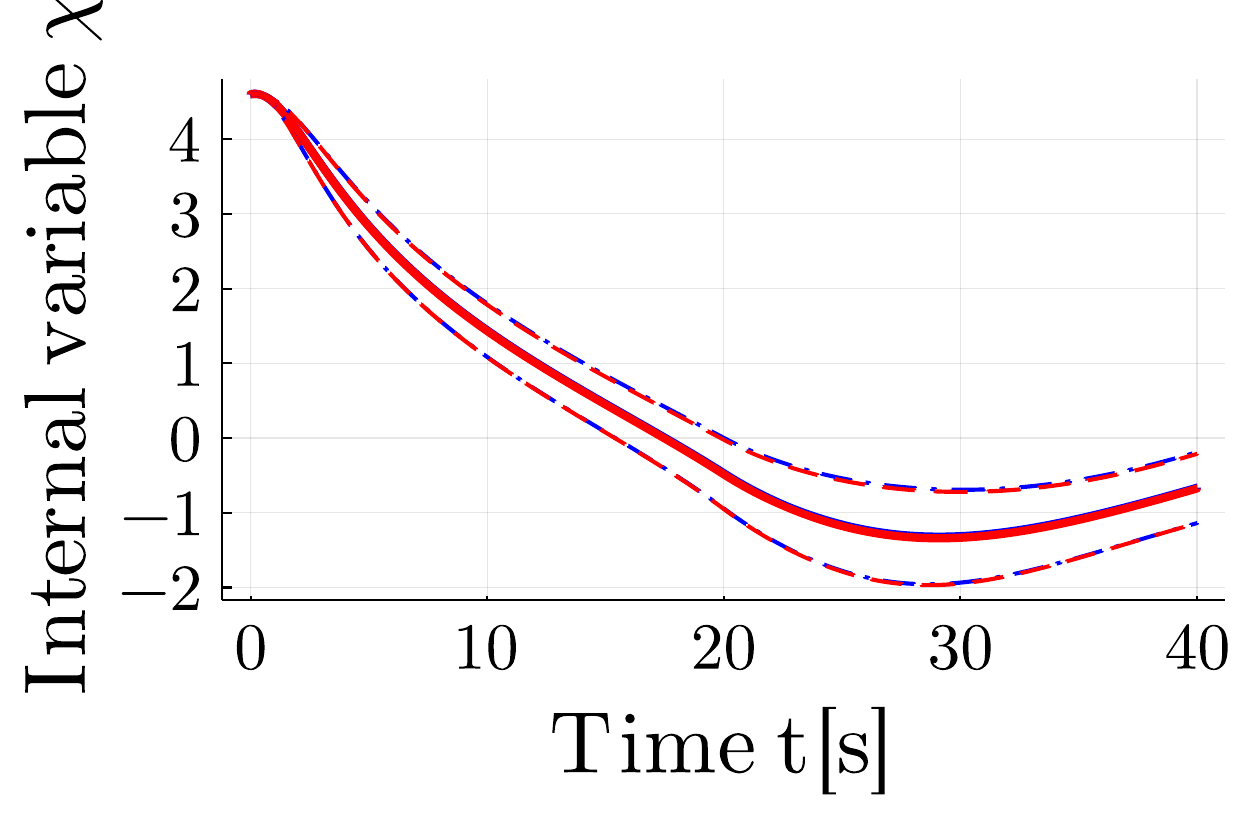}
	\end{subfigure}
	\begin{subfigure}[b]{0.3\textwidth}
		\centering
		\includegraphics[trim=0cm 12.5cm 0cm 0cm,clip,scale=.25]{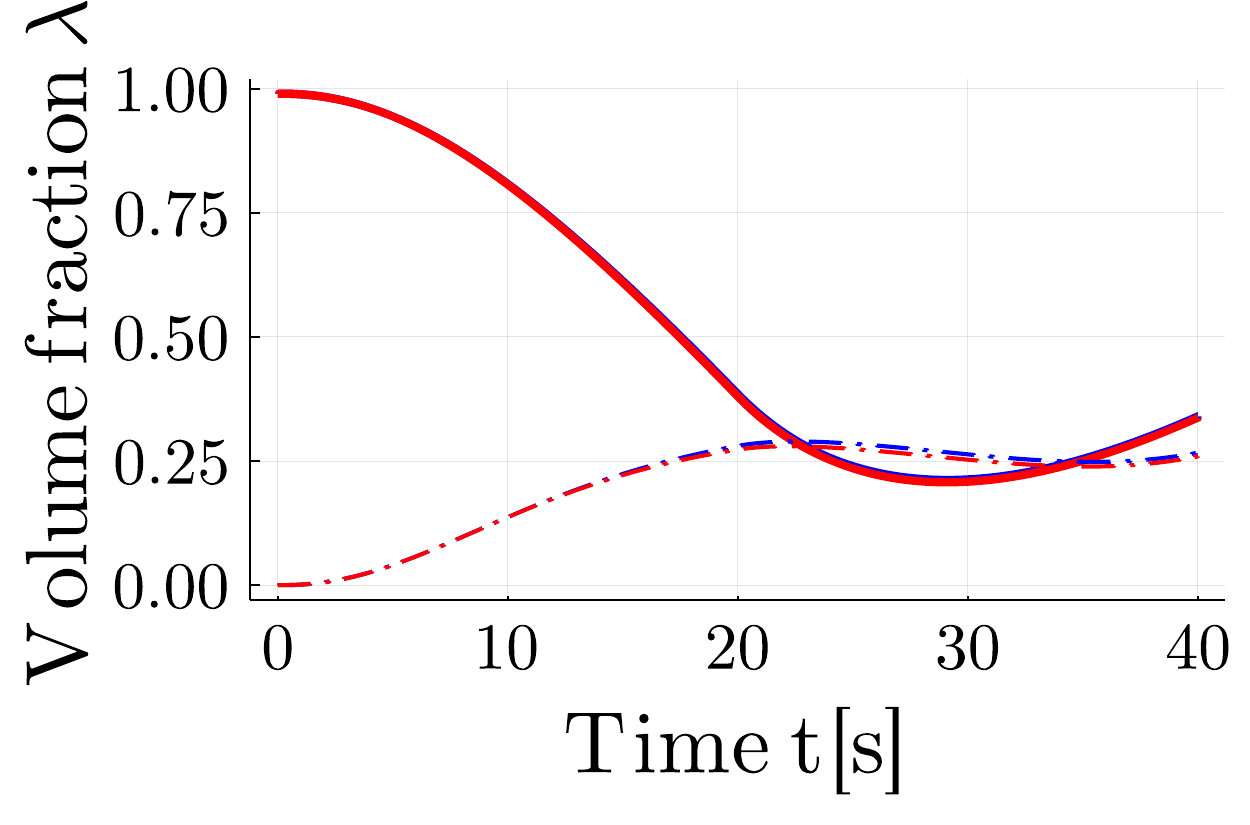}
	\end{subfigure}
	\begin{subfigure}[b]{0.3\textwidth}
		\centering
		\includegraphics[trim=0cm 12.5cm 0cm 0cm,clip,scale=.25]{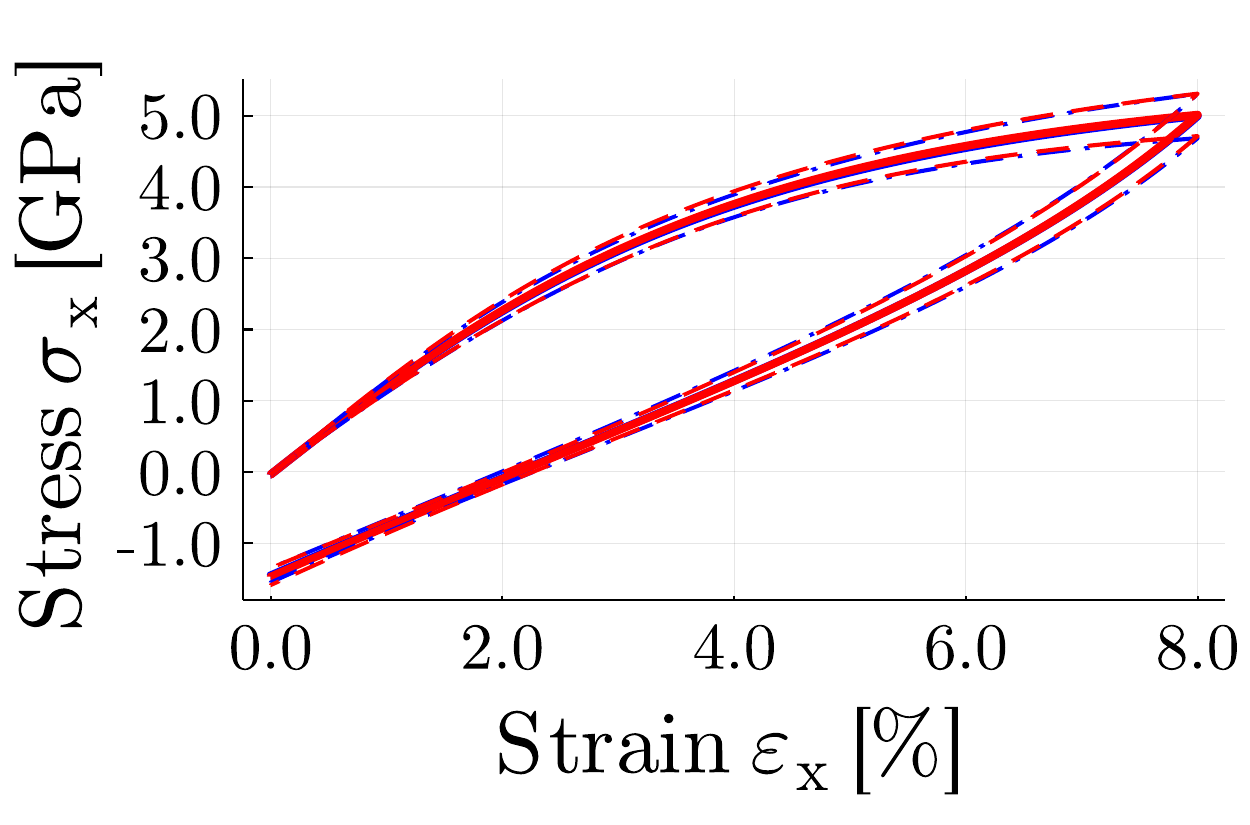}
	\end{subfigure}
	\begin{subfigure}[b]{0.3\textwidth}
		\centering
		\includegraphics[trim=0cm 4cm 0cm 12.5cm,clip,scale=.25]{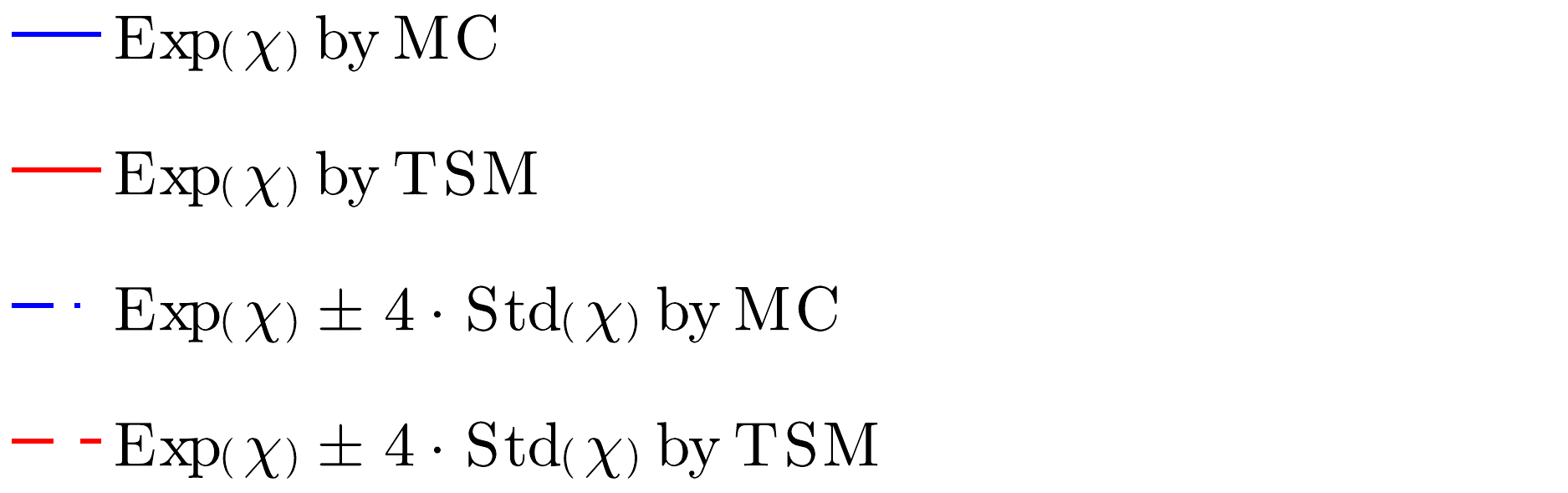}
		\caption{Internal variable $\chi_1$}
		\label{fig:resPhaseTransform2B}
	\end{subfigure}
	\begin{subfigure}[b]{0.3\textwidth}
		\centering
		\includegraphics[trim=4cm 4cm 0cm 12.5cm,clip,scale=.25]{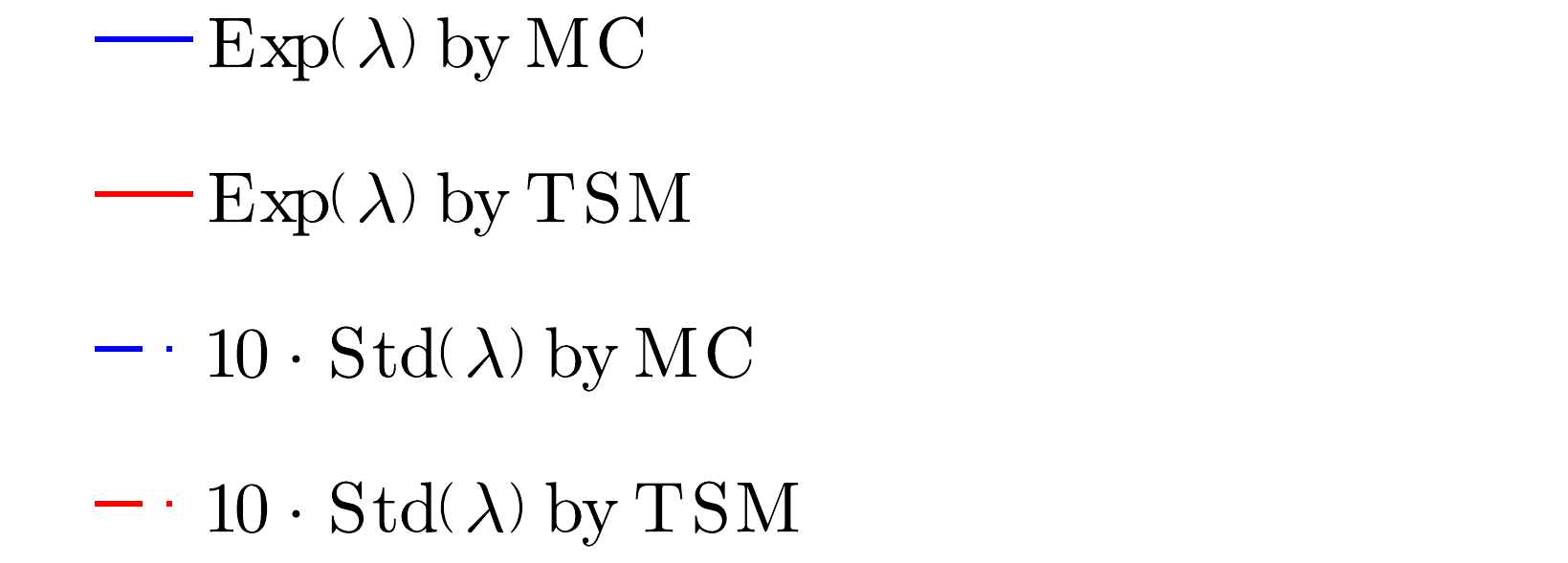}
		\caption{Volume fraction $\lambda_1$}
		\label{fig:resPhaseTransform3B}
	\end{subfigure}
	\begin{subfigure}[b]{0.3\textwidth}
		\centering
		\includegraphics[trim=0cm 4cm 0cm 12.5cm,clip,scale=.25]{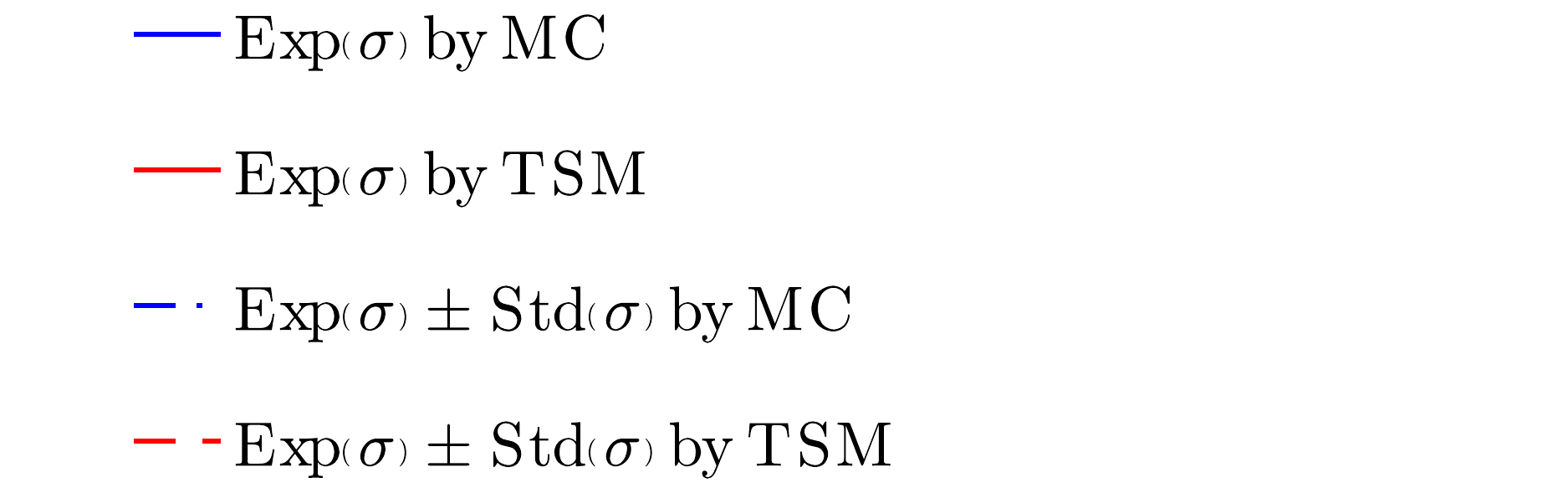}
		\caption{Stress $\sigma_\mathrm{x}$}
		\label{fig:resPhaseTransform1B}
	\end{subfigure}
	\caption{Results for internal variable $\chi_{1}$ over \rev{time} $t$, volume fraction $\lambda_{1}$ over \rev{time} $t$ and stress $\sigma_{x}$ over strain $\epsilon_{x}$ for viscosity parameter $\eta=\SI{2}{GPas}$. Results in blue are obtained by MC, \revv{results} in red by TSM.}
	\label{fig:resPhaseTransformB}
\end{figure}

A phase transformation involving two phases is investigated here.
The first phase resembles austenite with $\langle \lambda \rangle = \SI{70}{GPa}, \langle \mu \rangle = \SI{30}{GPa}$ and zero transformation strains. 
The second phase is martensite with  $\langle \lambda \rangle = \SI{35}{GPa}, \langle \mu \rangle = \SI{15}{GPa}$ and a transformation strain of $\eta = 0.055\cdot\left[ 1.0, -0.45, -0.45, 0.0, 0.0, 0.0 \right]$. The standard deviation of all material parameters is $10\%$ of their mean. In addition, all material parameters are stochastically independent.
An energetically stable start configuration with 99\% austenite is chosen.
Due to the high nonlinearity of the evolution equation \rev{and \revv{the} explicit Euler time integration scheme applied here}, a small time step with $\Delta t = \SI{0.4}{ms}$ is used.
A time-proportional loading with a loading rate of $0.4\%/s$ is applied along the \rev{$x$}-direction. The viscosity parameter is chosen as $\eta = \SI{0.2}{GPa.s}$.
The results from the TSM approach are compared against a reference Monte Carlo (MC) simulation with 1000 iterations in Figure~\ref{fig:resPhaseTransform}.
Due to the increase in loading, the martensite phase \rev{becomes} energetically preferable. Thus a transformation from austenite into martensite takes place.
This can be identified in the plot of the internal variable in Figure~\ref{fig:resPhaseTransform2} and the plot of the volume fraction of austenite $\lambda_{1}$ in Figure~\ref{fig:resPhaseTransform3}. As reference, a $\chi = 5$ equals a volume fraction of $\lambda = 99.3\%$. As the internal variable and volume fraction of both phases are directly linked, i.e., $\chi_2 = -\chi_1$ and $\lambda_2 = 1-\lambda_1$, we only plot the results for the first phase.
Until the transformation is completed at approximately $t = \SI{16}{s}$ and $\varepsilon_\textrm{x} = 6\%$, a nonlinear behavior of the stress with respect to the strain is visible in Figure~\ref{fig:resPhaseTransform1}. Afterwards, a linear elastic behavior between 6\% and 8\% strain is visible. After the load is decreased the same effect takes place in reverse: martensite is transformed back into austenite. At the end of the simulation at $t = \SI{40}{s}$ the transformation is not completed such that due to the transformative strains a negative eigenstress occurs. Due to the dissipative character of the phase transformation, a hysteresis is visible in the stress-strain curve. 
The standard deviation of the stress is nearly proportional to the applied strain yet in a nonlinear \rev{fashion}. Even more interesting is the standard deviation of the internal variable $\chi_{1}$ in Figure~\ref{fig:resPhaseTransform2}. The standard deviation is very low at the majority of the simulation. However, it spikes during the end of the transformation of austenite to martensite and at the beginning of the opposite process. 
This is due to the rapid change of the volume fraction of the phases. Therefore, a high sensitivity with respect to the uncertain material parameters is visible. The only differences between TSM and MC are visible here. The TSM estimates higher but more narrow spikes in the standard deviation. Surprisingly, there is no difference in the standard deviation of the volume fraction in Figure~\ref{fig:resPhaseTransform3} visible. In addition, the expectation and standard deviation of the stress are identical between TSM and MC.
The results of a second load case with $\eta = \SI{2}{GPa.s}$ and unchanged loading are presented in Figure~\ref{fig:resPhaseTransformB}. Due to the higher viscosity, the phase transformation is even more incomplete. Again, the results from TSM and MC for expectation and standard deviation of internal variable and stress are nearly identical. This makes the TSM a valuable tool in the investigation of such nonlinear processes as phase transformations.

\subsection{Elasto-viscoplasticity}
\label{sec:ViscoPlasticity}
Elasto-viscoplasticity is the mathematical description of a material behavior commonly found in certain materials as \rev{some classes} of metals, polymers and soils. It describes rate-dependent inelastic behavior. Plastic deformations occur if the stress surpasses a yield limit and \rev{they} develop afterwards in a viscous manner. 
A material model for elasto-viscoplasticity can be \rev{formulated} using an additive split of the total strains in an elastic and a viscoplastic part as $\bfvarepsilon = \bfvarepsilon^{\textrm{el}} + \bfvarepsilon^{\textrm{vp}}$ \cite{simo_computational_1998}. With this, the stress reads as 
\begin{equation}
    \bfsigma(t) = \dsE \cdot (\bfvarepsilon(t) - \bfvarepsilon^{\textrm{vp}}(t)).    \label{eq:sigmavp}
\end{equation}
The evolution equation for the viscoplastic part of the strain is then given by
\begin{equation}
    \dot{\bfvarepsilon}^{\textrm{vp}}(t) = \frac{1}{\eta} \left( || \text{dev}(\bfsigma(t))|| - \sigma^{\textrm{Y}} \right)_+ \frac{\text{dev}(\bfsigma(t))}{||\text{dev}(\bfsigma(t))||} \label{eq:evpdeq}
\end{equation}
with the scalar yield limit $\sigma^{\textrm{Y}}$ and the Macauley brackets defined as $\left( x \right)_+ = \text{max}(0, x)$. 
Here, the deviatoric part of the stress is calculated by subtracting the hydrostatic stress from the total stress as $\textrm{dev}(\bfsigma) = \bfsigma - \frac{1}{3}\left( \sigma_1+\sigma_2+\sigma_3\right) \rev{\dsI} =  \dsS \cdot \bfsigma$ \revv{where} $\dsS$ denotes the deviator operator in Voigt notation as
\begin{equation}
    \begin{pmatrix}
\frac{2}{3} & -\frac{1}{3} & -\frac{1}{3} & 0 & 0 & 0\\
-\frac{1}{3} & \frac{2}{3} & -\frac{1}{3} & 0 & 0 & 0\\
-\frac{1}{3} & -\frac{1}{3} & \frac{2}{3} & 0 & 0 & 0\\
0& 0 & 0& 1 & 0 & 0 \\
0& 0 & 0& 0 & 1 & 0 \\
0& 0 & 0& 0 & 0 & 1 \\
\end{pmatrix}
\end{equation}
For a deviatoric stress larger then the yield limit, a viscous evolution of the viscoplastic strains takes place. 

\subsubsection{Extended material model}
The material model has three material parameters: the elasticity tensor $\dsE$, the yield limit $\sigma^\textrm{Y}$ and the viscosity parameter $\eta$. Here, we assume uncertainty in the elasticity tensor and the yield limit and present how the extended material model is developed.
We assume that elasticity tensor and yield limit are independent, however any (partial) dependency can be incorporated easily.
\rev{With the approach developed in Section~\ref{sec:DiffEq}} the evolution of the internal variable \rev{is approximated} by a stochastic linear Taylor series of the form (cf. Equation~\eqref{eq:TaylorMultiDim2})
\begin{equation}
    \bfvarepsilon^{\textrm{vp}}(t) = \bfvarepsilon^{\textrm{vp}, (0)}(t) + \bfI^\textrm{E}(t) : \dsD + \bfI^\textrm{Y}(t) \, \tilde{\sigma}^\textrm{Y} \label{eq:vpTSM}
\end{equation}
with the tangents $\bfI^\textrm{E} \in \dsR^{6, 6, 6}$ and $\bfI^\textrm{Y} \in \dsR^6$. \rev{Here, the tensor of random variables $\Phi$ is split up into $\dsD$ and $\tilde{\sigma}^\textrm{Y}$ for readability.}
As before, $\dsD = \dsE - \langle \dsE \rangle$ captures the fluctuations of the elasticity tensor. Fluctuations of the yield limit are captured by $\tilde{\sigma}^\mathrm{Y} = \sigma^\mathrm{Y} - \langle \sigma^\mathrm{Y} \rangle = \sigma^\mathrm{Y} - \sigma^{\mathrm{Y}, 0}$.
Equivalently, the time derivative of Equation~\eqref{eq:vpTSM} reads as
\begin{equation}
    \dot{\bfvarepsilon}^{\textrm{vp}}(t) = \dot{\bfvarepsilon}^{\textrm{vp}, (0)}(t) + \dot{\bfI}^\textrm{E}(t) : \dsD + \dot{\bfI}^\textrm{Y}(t) \, \tilde{\sigma}^\textrm{Y}.
\end{equation}

\rev{The term} $\dot{\bfvarepsilon}^{\textrm{vp}, (0)}(t)$ is given by the evaluation of Equation~\eqref{eq:evpdeq} at $\dsD = \bf0$ and $\tilde{\sigma}^\textrm{Y} = 0$ as
\begin{equation}
    \dot{\bfvarepsilon}^{\textrm{vp}, (0)}(t) = \dot{\bfvarepsilon}^{\textrm{vp}}(t) \Big\vert_{\dsD=\bf0,~\tilde{\sigma}^\textrm{Y}=0} \label{eq:evp0}
\end{equation}

We find the tangent $\bfI^\textrm{E}(t)$ by taking the derivative of the evolution equation (cf. Equation~\eqref{eq:I1}) for each component $\alpha$ as
\begin{align}
    \dot{\bfI}^\textrm{E}_{\alpha st}(t) &= \frac{\mathrm{d}\dot{\bfvarepsilon}^{\textrm{vp}}_\alpha (t)}{\mathrm{d}\dsD_{st}}\Big\rvert_{\dsD = \bf0,\, \tilde{\sigma}^\mathrm{Y}=0} \label{eq:evpIE} \\
    &= \frac{1}{\eta} \bigg( H\left(||\text{dev}(\bfsigma)||-\sigma^\textrm{Y}\right) \frac{\text{dev}(\bfsigma)_d}{||\text{dev}(\bfsigma)||} \frac{\mathrm{d}}{\mathrm{d}\dsD_{st}} (\text{dev}(\bfsigma)_d)\frac{\text{dev}(\bfsigma)_\alpha}{||\text{dev}(\bfsigma)||} \nonumber \\
    &\quad +\left( ||\text{dev}(\bfsigma)|| - \sigma^\textrm{Y}\right)_+ \left( \frac{1}{||\text{dev}(\bfsigma)||} \frac{\mathrm{d}}{\mathrm{d}\dsD_{st}} (\text{dev}(\bfsigma)_\alpha) + \text{dev}(\bfsigma)_\alpha \frac{\mathrm{d}}{\mathrm{d}\dsD_{st}} \left(\frac{1}{|| \text{dev}(\bfsigma) ||} \right)\right) \bigg)\Big\rvert_{\dsD = \bf0,\,\tilde{\sigma}^\mathrm{Y}=0} \nonumber 
\end{align}
where \rev{$H(x)$} denotes the Heaviside step function. Here, the Einstein summation is only carried out over the free index $d$.
It remains to derive
\begin{equation}
    \frac{\mathrm{d}}{\mathrm{d}\dsD_{st}} (\text{dev}(\bfsigma)_d) = \dsS_{ds} (\bfvarepsilon_t - \bfvarepsilon^{\mathrm{vp}}_t) - \dsS_{de} \dsE^{(0)}_{ef} I^\textrm{E}_{fst}
\end{equation}
and
\begin{equation}
    \frac{\mathrm{d}}{\mathrm{d}\dsD_{st}} (||\text{dev}(\bfsigma)||^{-1}) = - ||\text{dev}(\bfsigma)||^{-2} \frac{\text{dev}(\bfsigma)_d}{||\text{dev}(\bfsigma)||} \frac{\mathrm{d}}{\mathrm{d}\dsD_{st}} (\text{dev}(\bfsigma)_d).
\end{equation}
As the indices $s, t$ and $d$ are fixed component indices, no summation is carried out over these indices

For the inclusion of a stochastic yield limit, the tangent $\bfI^\mathrm{Y}(t)$ can be derived as 
\begin{align}
    \dot{\bfI}^\textrm{Y}_\alpha(t) &= \frac{\mathrm{d}\dot{\bfvarepsilon}^{\mathrm{vp}}_\alpha(t)}{\mathrm{d}\tilde{\sigma}^\mathrm{Y}} \Big\rvert_{\dsD = \bf0,\,\tilde{\sigma}^\mathrm{Y}=0} \label{eq:evpIY} \\
    &= \frac{1}{\eta} \bigg( H(||\text{dev}(\bfsigma)|| - \sigma^{\mathrm{Y}}) \left(\frac{\text{dev}(\bfsigma)_d}{||\text{dev}(\bfsigma)||} \frac{\mathrm{d}}{\mathrm{d}\tilde{\sigma}^\mathrm{Y}} (\text{dev}(\sigma)_d) - 1 \right) \frac{\text{dev}(\bfsigma)_\alpha}{||\text{dev}(\bfsigma)||} \nonumber \\
    &\quad + \left( ||\text{dev}(\bfsigma)|| - \sigma^{\mathrm{Y}} \right)_+ \frac{\mathrm{d}}{\mathrm{d}\tilde{\sigma}^\mathrm{Y}} \Big( \frac{\text{dev}(\bfsigma)_\alpha}{||\text{dev}(\bfsigma)||} \Big) \bigg)\Big\rvert_{\dsD = \bf0,\,\tilde{\sigma}^\mathrm{Y}=0} \nonumber
\end{align}
with the quantities
\begin{equation}
\frac{\mathrm{d}}{\mathrm{d}\tilde{\sigma}^\mathrm{Y}} \Big(\text{dev}(\bfsigma)_d \Big) = - \dsS_{db} \dsE^{(0)}_{bc} I^\mathrm{Y}_c,
\end{equation}
\begin{equation}
    \frac{\mathrm{d}}{\mathrm{d}\tilde{\sigma}^\mathrm{Y}} \Big( \frac{\text{dev}(\bfsigma)_\alpha}{||\text{dev}(\bfsigma)||} \Big) = \frac{1}{||\text{dev}(\bfsigma)||} \frac{\mathrm{d}}{\mathrm{d}\tilde{\sigma}^\mathrm{Y}}(\text{dev}(\bfsigma)_\alpha) + \text{dev}(\bfsigma)_\alpha \frac{\mathrm{d}}{\mathrm{d}\tilde{\sigma}^\mathrm{Y}} (|| \text{dev}(\bfsigma)||^{-1} )
\end{equation}
and
\begin{equation}
    \frac{\mathrm{d}}{\mathrm{d}\tilde{\sigma}^\mathrm{Y}} (|| \text{dev}(\bfsigma)||^{-1} ) = - ||\text{dev}(\bfsigma)||^{-2} \frac{\text{dev}(\bfsigma)_d}{||\text{dev}(\bfsigma)||} 
    \frac{\mathrm{d}}{\mathrm{d}\tilde{\sigma}^\mathrm{Y}} \Big(\text{dev}(\bfsigma)_d \Big).
\end{equation}
In these equations, $\alpha$ is a fixed index such that the Einstein summation is only carried out over the remaining latin indices.

The whole extended material model is given by Equations~\eqref{eq:evp0}, \eqref{eq:evpIE} and \eqref{eq:evpIY}.
\rev{The term} $\bfvarepsilon^{\mathrm{vp},0}$ denotes the standard deterministic part of the set of equations. The terms $\bfI^\textrm{E}(t)$ and $\bfI^\textrm{Y}(t)$ track the impact of stochastic fluctuations in the elasticity tensor and the yield limit respectively.

\subsubsection{Probabilistic analysis}
After the time-dependent deterministic terms $\bfvarepsilon^{\textrm{vp}, (0)}(t), \bfI^\textrm{E}(t)$ and $\bfI^\textrm{Y}(t)$ have been calculated, the expectation and variance of internal variable and stress can be determined.

\paragraph{Expectation and variance of the internal variable $\bfvarepsilon^{vp}$}
The expectation is given with the approximation of Equation~\eqref{eq:vpTSM} by
\begin{equation}
    \langle \bfvarepsilon(t) \rangle = \bfvarepsilon^{\textrm{vp}, (0)}(t)
\end{equation}
\rev{Due to the split of the tensor of random quantities in Equation~\eqref{eq:vpTSM} we cannot make direct use of Equation~\eqref{eq:DiffVariance} to calculate the variance. However, we can follow the same derivation. }
\rev{Thus,} the variance of each component $\alpha$ is calculated by
\begin{equation}
    \rev{\textrm{Var}}(\varepsilon_\alpha^{\textrm{vp}}) = \langle \left( \varepsilon_\alpha^{\textrm{vp}} \right)^2 \rangle -\langle \left( \varepsilon_\alpha^{\textrm{vp}} \right) \rangle^2.
\end{equation}
The second moment is given with the approximation in Equation~\eqref{eq:vpTSM} by 
\begin{equation}
    \langle (\varepsilon^{\textrm{vp}}_\alpha(t))^2 \rangle = (\varepsilon_\alpha^{\textrm{vp}, (0)}(t))^2 + I_{\alpha bc}^\mathrm{E}(t) \langle \dsD_{bc} \dsD_{de} \rangle I_{\alpha de}^\mathrm{E}(t) + (I_\alpha^\mathrm{Y}(t))^2 \langle (\tilde{\sigma}^\mathrm{Y})^2 \rangle + 2 I_{\alpha fg}^\mathrm{E}(t) \langle \tilde{\sigma}^\mathrm{Y} \dsD_{fg} \rangle I_\alpha^\mathrm{Y}(t). \label{eq:evpsecmom}
\end{equation}
Here, $\alpha$ is a fixed index and Einstein summation is only carried out over the indices $b,c,d,e,f,g$. 
Despite the inclusion of the stochastic yield limit as second uncertain material parameter, the structure of the second moment remains simple.
\rev{As discussed in Section~\ref{sec:TSM},} the expectations of all stochastic quantities in Equation~\eqref{eq:evpsecmom} are found by sampling. During sampling, any type of correlation between the randomness of $\sigma^\textrm{Y}$ and $\dsE$ can be incorporated. It may be remarked, if the fluctuations in $\sigma^Y$ and $\dsE$ are stochastically independent, the coupling term $\langle \tilde{\sigma}^Y \dsD \rangle$ equals a zero tensor.

\paragraph{Expectation and variance of the stress} 
The stress is given by Equations~\eqref{eq:sigmavp} and \eqref{eq:vpTSM} as
\begin{equation}
    \bfsigma(t) = (\dsE^{(0)} + \dsD) \cdot (\bfvarepsilon(t) - (\bfvarepsilon^{\textrm{vp}, (0)}(t) + \bfI^\mathrm{E}(t) : \dsD + \bfI^\mathrm{Y}(t) \tilde{\sigma}^\mathrm{Y})).
\end{equation} 
\rev{This equation is algebraic, thus we make use of the approach as discussed in Section~\ref{sec:AlgEq}.}
The stress is linearized in each component $\alpha$ as (cf. Equation~\eqref{eq:TaylorMultiDim})
\begin{equation}
    \sigma_\alpha(t) = \sigma_\alpha^{(0)}(t) + T_{\alpha st}^\mathrm{E}(t) \dsD_{st} + T_\alpha^\mathrm{Y} \tilde{\sigma}^\mathrm{Y}_\alpha.
\end{equation}
The expectation for each component $\alpha$ is then given by
\begin{equation}
    \langle \sigma_\alpha(t)  \rangle = \sigma_\alpha^{(0)}(t) = \sigma_\alpha(t) \big\rvert_{\dsD = \bf0, \tilde{\sigma}^\mathrm{Y} = 0} = \dsE^{(0)}_{\alpha b} (\varepsilon_b(t) - \varepsilon_b^{\textrm{vp}, (0)}(t)).
\end{equation}

The tangents $\bfT^\mathrm{E}(t)$ and $T^\mathrm{Y}(t)$ are given by the derivatives of the stress with respect to the random part of the elasticity tensor and the yield stress as
\begin{align}
	T^\mathrm{E}_{\alpha st}(t) &= \frac{\mathrm{d} \sigma_\alpha (t)}{\mathrm{d} \dsD_{st}} \bigg\vert_{\dsD=\bf0, \tilde{\sigma}^\mathrm{Y} =0} \\
	&= \dsI_{\alpha s} \cdot (\varepsilon_t(t) - \varepsilon^{\mathrm{vp}, (0)}_t(t)) - \dsE^{(0)}_{\alpha ,b} \cdot I^\textrm{E}_{bst}(t) \nonumber
\end{align}
and
\begin{align}
	T^\mathrm{Y}_{\alpha }(t) &= \frac{\mathrm{d} \sigma_\alpha (t)}{\mathrm{d} \tilde{\sigma}^\mathrm{Y}} \bigg\vert_{\dsD=\bf0, \tilde{\sigma}^\mathrm{Y}=0} = - \dsE^{(0)}_{ab} \cdot I^\mathrm{Y}_b(t)
\end{align}

The variance of the stress for each component $\alpha$ is given by (cf. Equation~\eqref{eq:AlgVariance})
\begin{equation}
    \text{Var}(\sigma_\alpha) = \langle \sigma_\alpha^2 \rangle - \langle \sigma_\alpha \rangle^2
\end{equation}
with the second moment
\begin{equation}
    \langle \sigma_\alpha^2 \rangle = (\sigma_a^{(0)})^2 + T^\mathrm{E}_{\alpha st}(t) \langle \dsD_{st} \dsD_{uv} \rangle T^\mathrm{E}_{\alpha uv}(t) + T^\mathrm{Y}_\alpha(t) \langle ( \tilde{\sigma}^\mathrm{Y} )^2 \rangle T^\mathrm{Y}_\alpha(t) + 2 T_{\alpha bc}^\mathrm{E}(t) \langle \tilde{\sigma}^\mathrm{Y} \dsD_{bc} \rangle T_\alpha^\mathrm{Y}(t).
\end{equation}
As before, $\alpha$ is a fixed index such that no summation is performed over this index.

\subsubsection{Numerical experiments}
The presented extended material model is compared against reference Monte Carlo simulations.
In total 1000 iterations are used for the Monte Carlo method.
The system is simulated for a total of 100 seconds with a time increment of $\Delta t = \SI{0.05}{s}$.
The material parameters are chosen as the Lam\'e parameter $\langle \lambda \rangle = \SI{12}{G.Pa}, \langle \mu \rangle = \SI{8}{G.Pa}$ and the yield limit $\langle \sigma^Y \rangle = \SI{50}{M.Pa}$. The standard deviation of the Lam\'e parameters is chosen as 10\% of their mean. The standard deviation of the yield limit is chosen as 20\% of its mean. 

Multiple test cases were investigated to compare TSM and reference Monte Carlo simulations.
These differ in the viscosity parameter controlling the evolution of the viscoplastic strains and the assumption whether yield limit and Lam\'e parameter are stochastically (in)dependent. The results for \rev{stochastically} dependent parameters, e.g., all parameters fluctuate simultaneously are presented in Table~\ref{tab:resVP1}. The results for \rev{stochastically} independent yield limit and Lam\'e parameters are presented in Table~\ref{tab:resVP2}.
For both cases, the simulation was performed for different values of the viscosity as $\eta = \SI{20}{GPa.s}, \eta = \SI{80}{GPa.s}$ and $\eta = \SI{200}{GPa.s}$.
The hysteresis behavior of the stress and strain along the $xy$-component is presented along with the viscoplastic strains over time.
The results by the TSM are presented in red; the results by MC in blue. The expectation is indicated by a solid line, the standard deviation by a dashed lines. For the hysteresis curves, the expectation $\pm$ standard deviation of the stress is displayed. The standard deviation of the viscoplastic strains is shown in the same figure as the expectation, but magnified by a factor of 20.
With increasing viscosity parameter the evolution of the viscoplastic strains is slowed down. This leads to a smaller hysteresis and smaller viscoplastic strains. While the standard deviation of the stress is zero for two points of the hysteresis curve for stochastically dependent material parameter, the standard deviation is nearly constant for independent material parameters.
The standard deviation of the viscoplastic strains show a rather complicated behavior: at the beginning of the \rev{decreasing} viscoplastic strains, sharp minima with a standard deviation of zero occur. In comparison, the standard deviation for fully \rev{dependent} material parameters are more complex and harder to interpret as for the independent case. 
It is remarkable how well the TSM is able to approximate expectation and standard deviation even for a very low viscosity parameter which resembles classical plasticity. For nearly all figures, the TSM and MC results are identical. Only for the case of $\eta = \SI{20}{GPa.s}$ and independent material parameters, slight differences at the transition between elasticity and viscoplasticity are visible. This is due to the sharp change in behavior. \rev{Afterwards, a very satisfactory agreement can be observed.} Interestingly, for dependent material parameters no differences between TSM and MC are visible.

\begin{table}
	\centering
	\begin{tabular}{p{2cm}|m{4cm}m{4cm}m{4cm}m{4cm}}
		 & $\eta = \SI{20}{GPa.s}$ & $\eta = \SI{80}{GPa.s}$ & $\eta = \SI{200}{GPa.s}$ & \\[2pt] \hline
		 Stress & \includegraphics[trim=0cm 0cm 9.2cm 0cm, clip, scale=0.25]{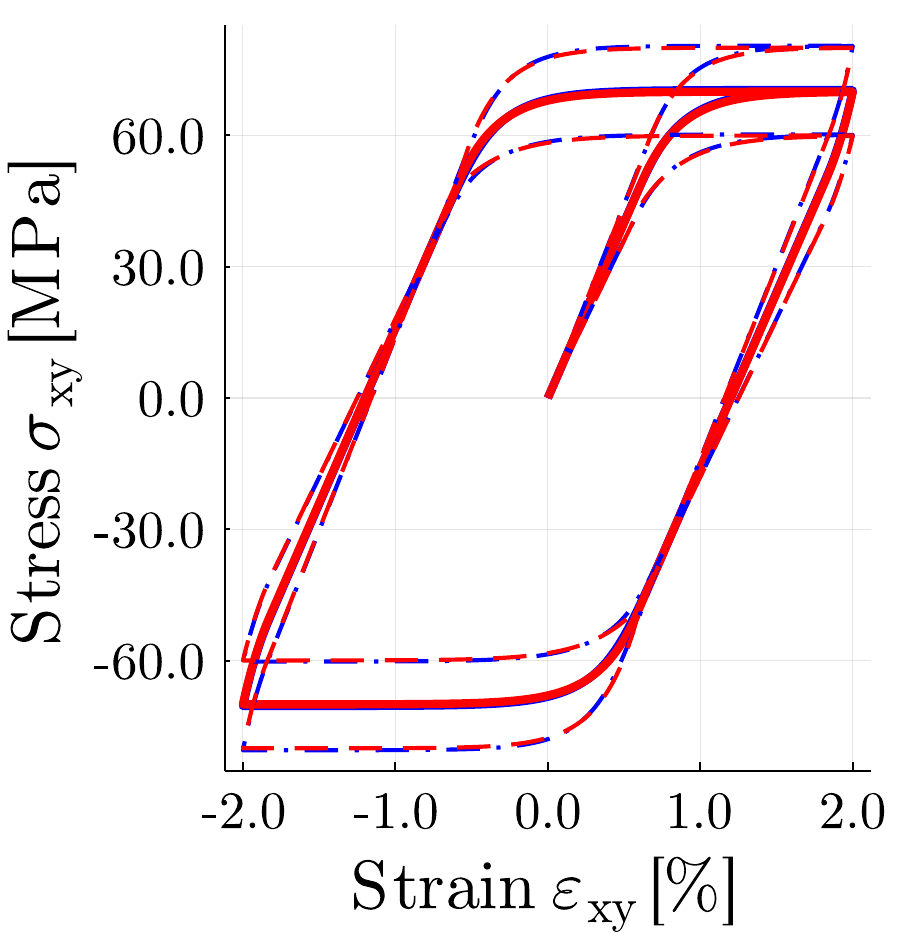} & \includegraphics[trim=0cm 0cm 9.2cm 0cm, clip, scale=0.25]{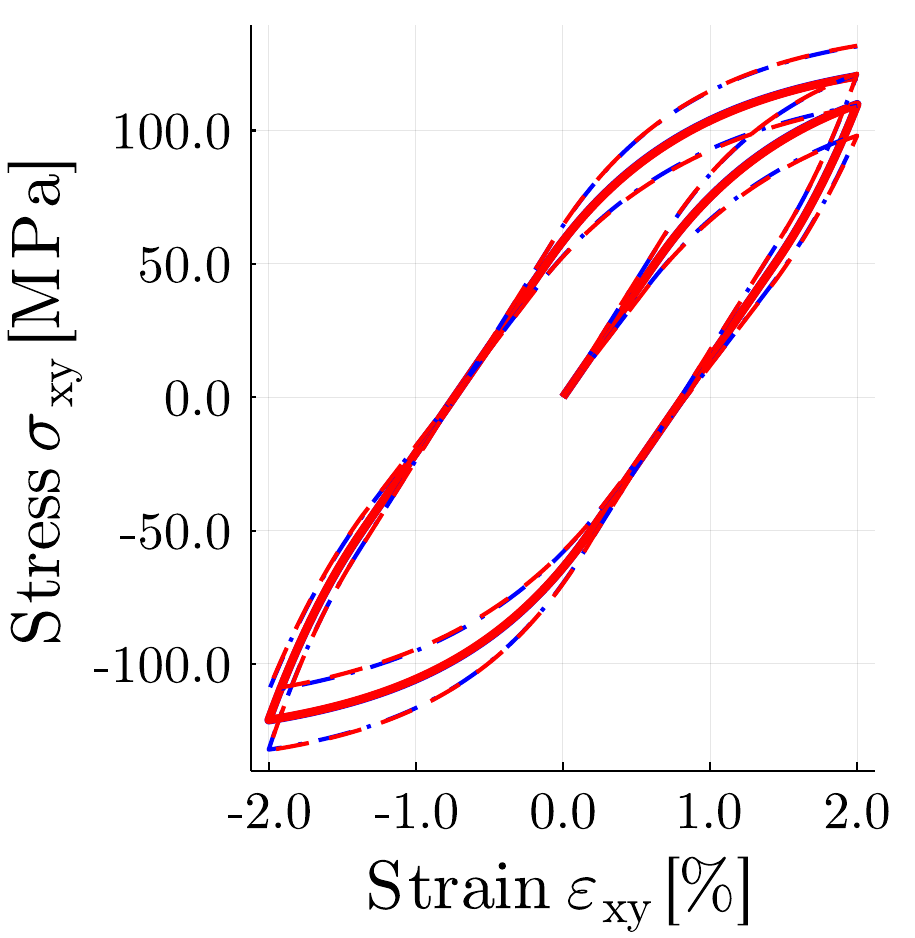} & 
		 \includegraphics[trim=0cm 0cm 9.2cm 0cm, clip, scale=0.25]{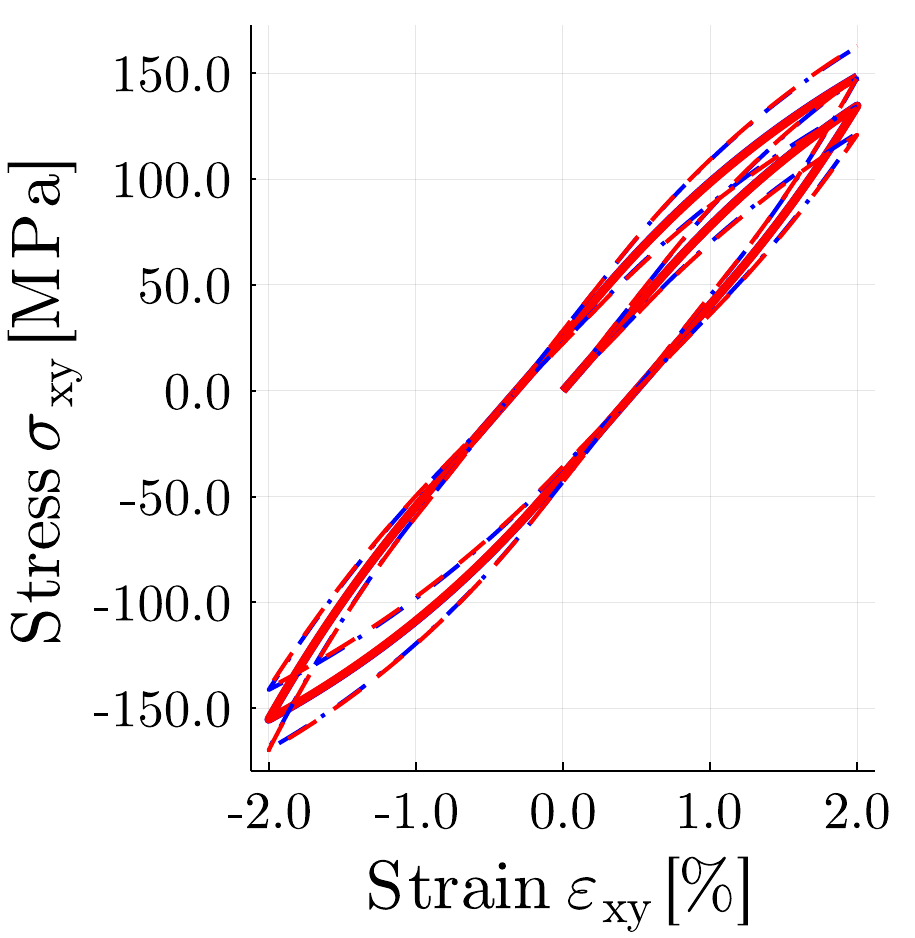} & 
		 \includegraphics[trim=16cm 0cm 0cm 0cm, clip, scale=0.3]{Pictures/ViscoPlasticity/StressXY_A_200E9.pdf} \\
		 Viscous strains & \includegraphics[trim=0cm 0cm 12.5cm 0cm, clip, scale=0.25]{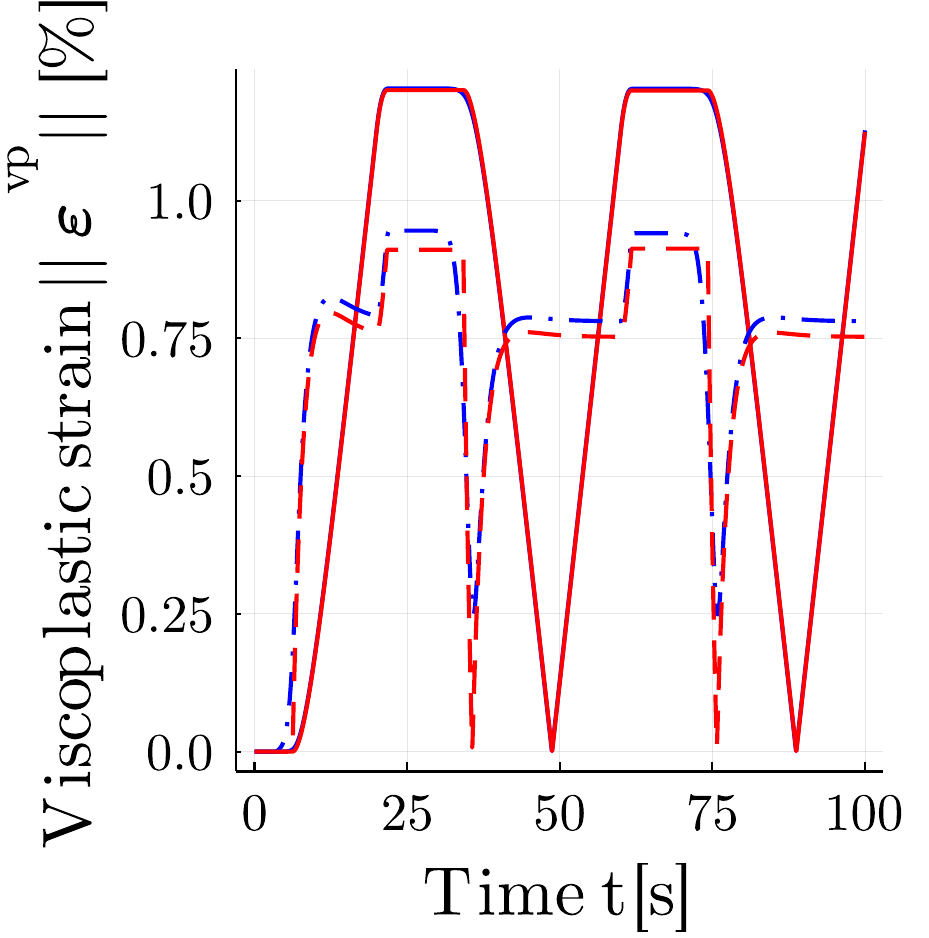} & 
		 \includegraphics[trim=0cm 0cm 12.5cm 0cm, clip, scale=0.25]{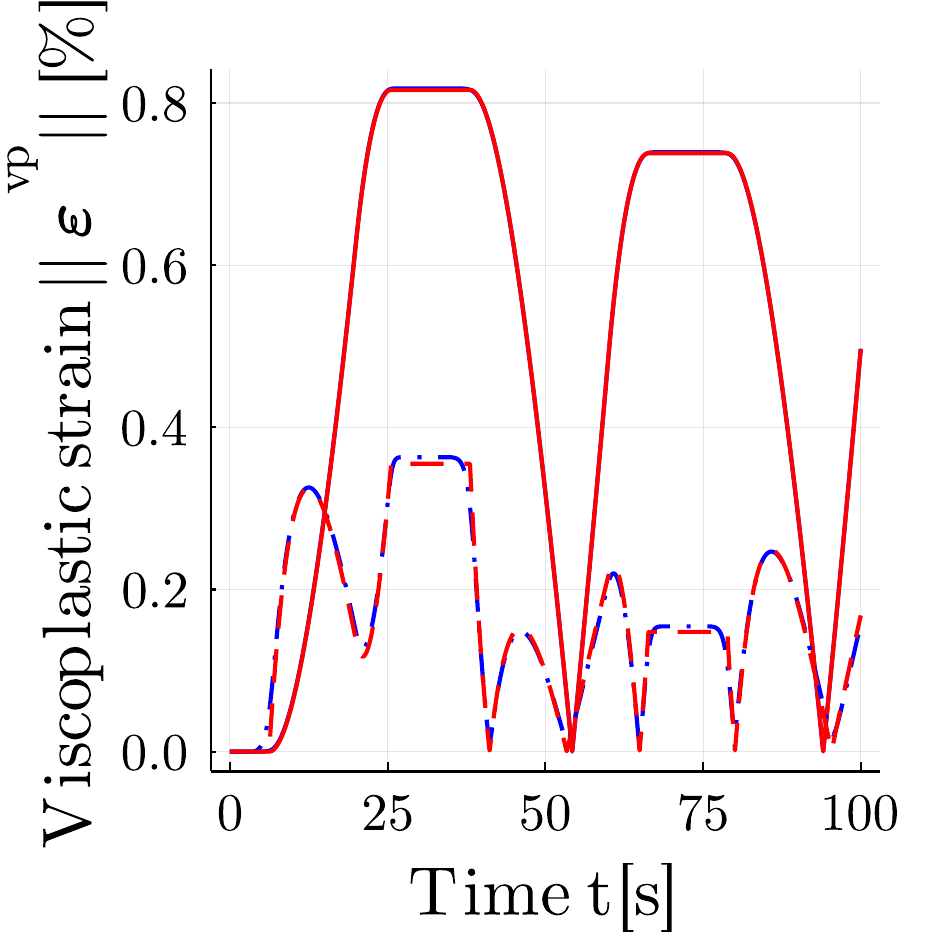} &
		 \includegraphics[trim=0cm 0cm 12.5cm 0cm, clip, scale=0.25]{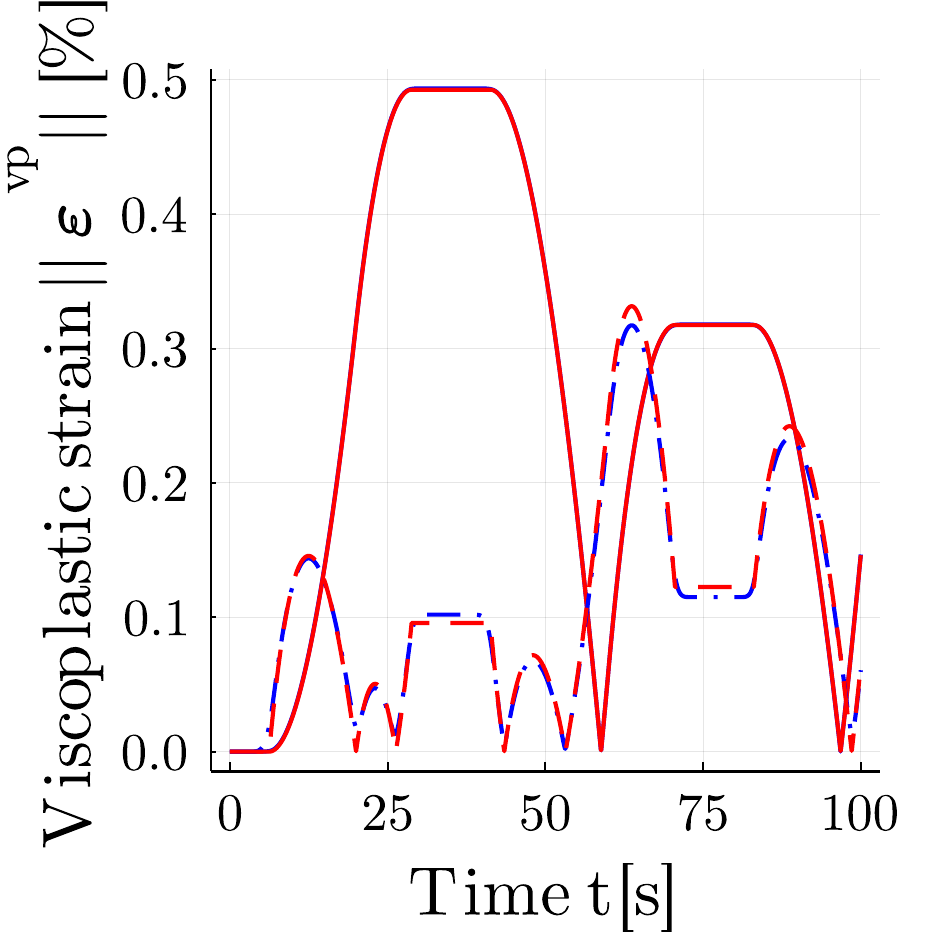} &
		 \includegraphics[trim=16cm 0cm 0cm 0cm, clip, scale=0.25]{Pictures/ViscoPlasticity/Strains_A_200E9_alt.pdf} \\ 
	\end{tabular}
	\caption{Results for the elasto-viscoplasticity model with fully correlated Lam\'e parameters and yield limit. Results obtained \rev{by MC are blue, results in red by TSM.} } 
	\label{tab:resVP1}
\end{table}

\begin{table}
	\centering
	\begin{tabular}{p{2cm}|m{4cm}m{4cm}m{4cm}m{4cm}}
		& $\eta = \SI{20}{GPa.s}$ & $\eta = \SI{80}{GPa.s}$ & $\eta = \SI{200}{GPa.s}$ & \\[2pt] \hline
		Stress & \includegraphics[trim=0cm 0cm 9.2cm 0cm, clip, scale=0.25]{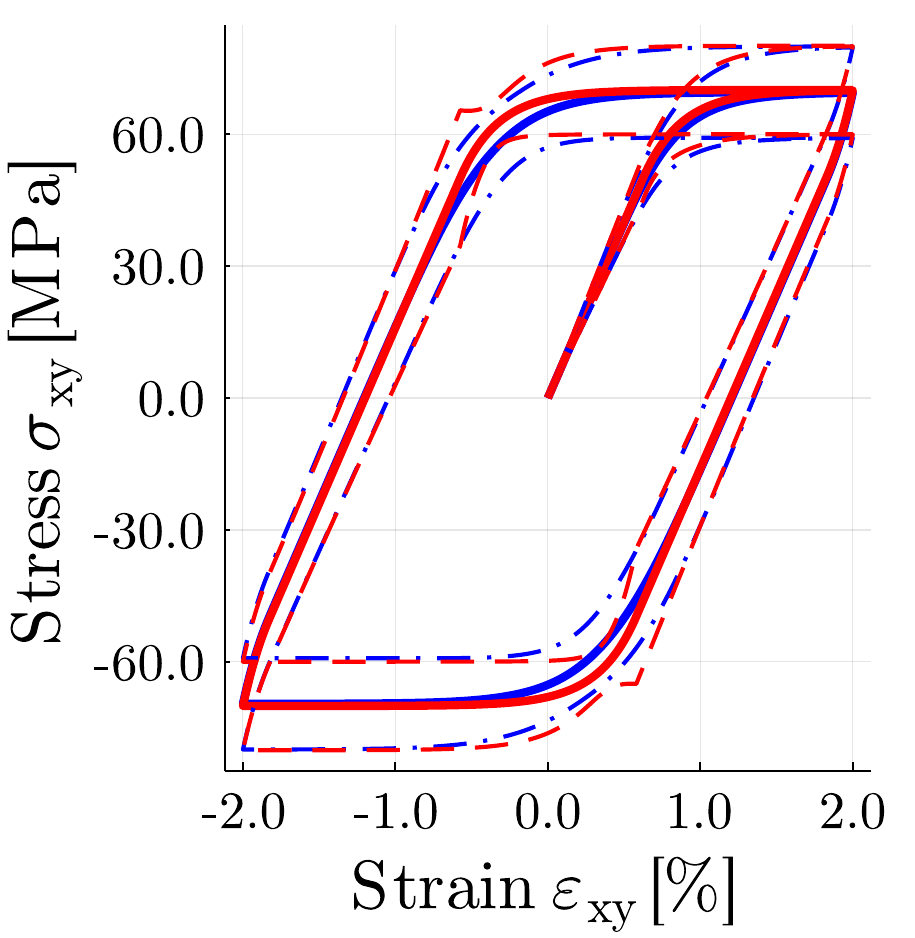} & \includegraphics[trim=0cm 0cm 9.2cm 0cm, clip, scale=0.25]{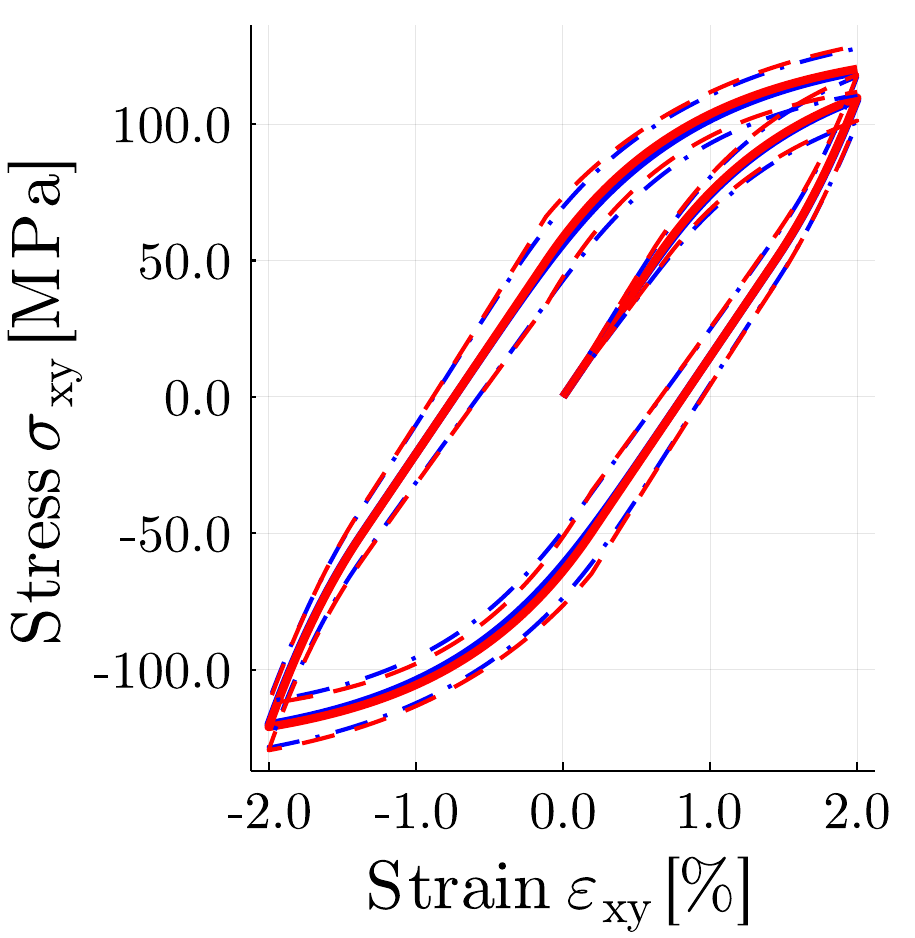} & 
		\includegraphics[trim=0cm 0cm 9.2cm 0cm, clip, scale=0.25]{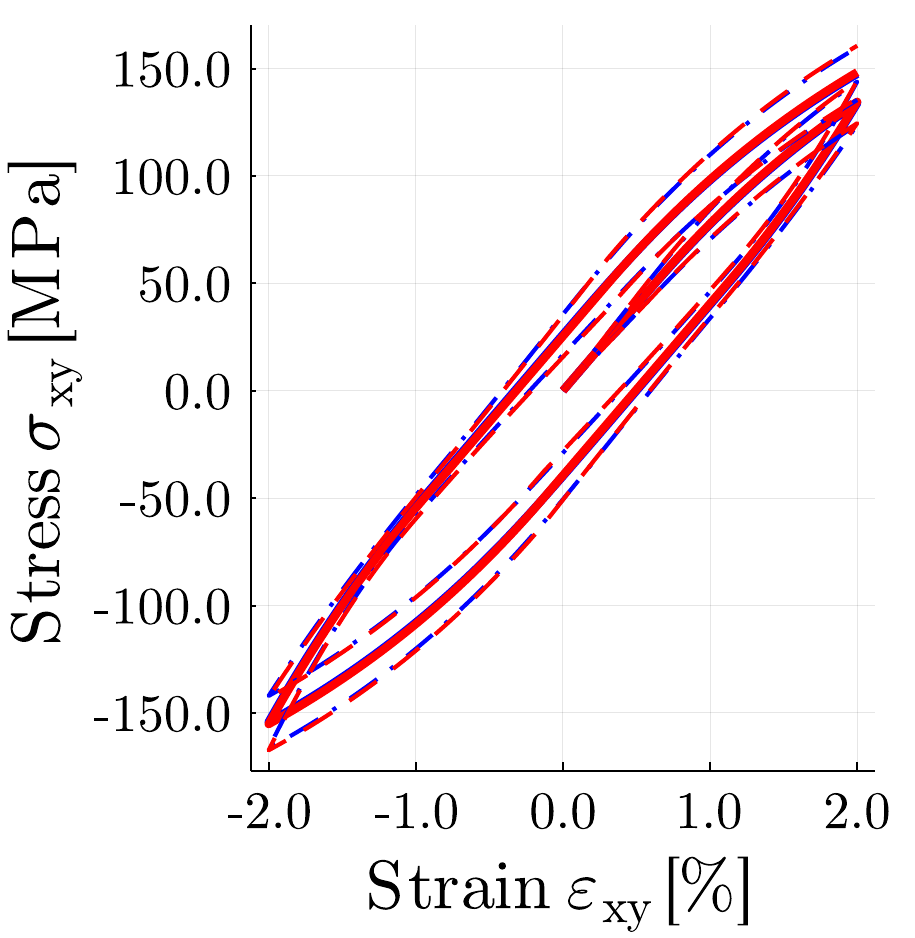} & 
		\includegraphics[trim=16cm 0cm 0cm 0cm, clip, scale=0.3]{Pictures/ViscoPlasticity/StressXY_B_200E9.pdf} \\
		Viscous strains & \includegraphics[trim=0cm 0cm 12.5cm 0cm, clip, scale=0.25]{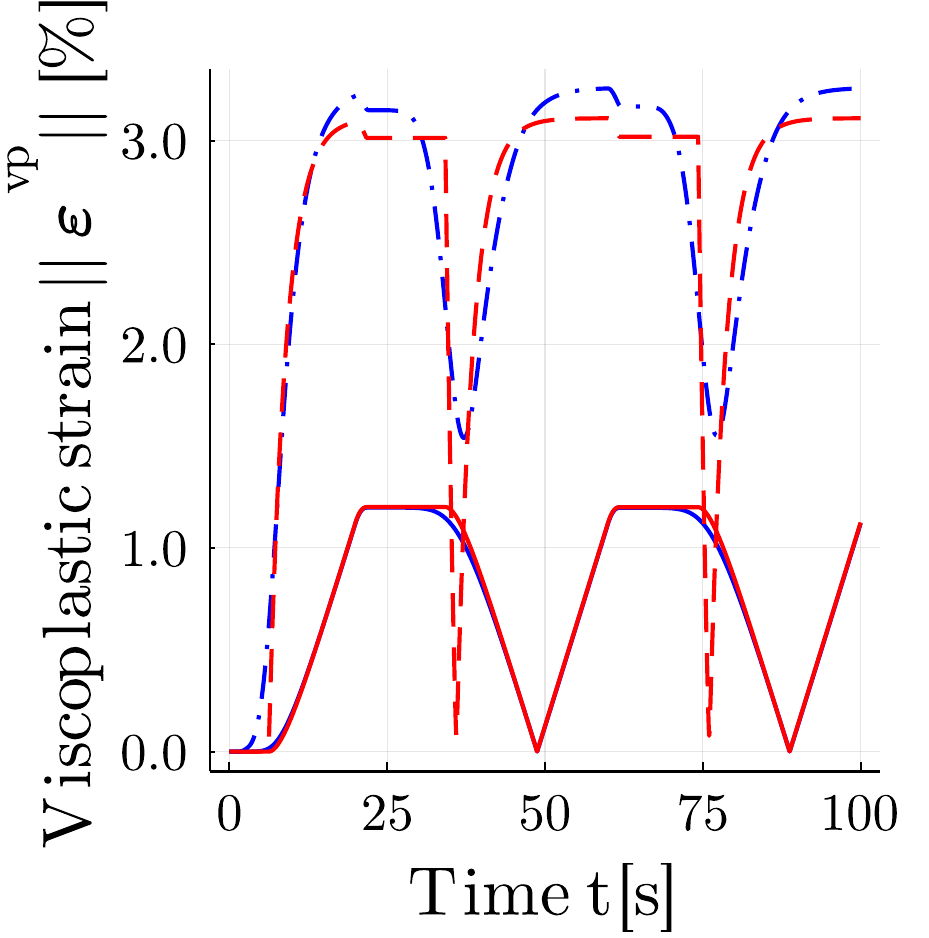} & 
		\includegraphics[trim=0cm 0cm 12.5cm 0cm, clip, scale=0.25]{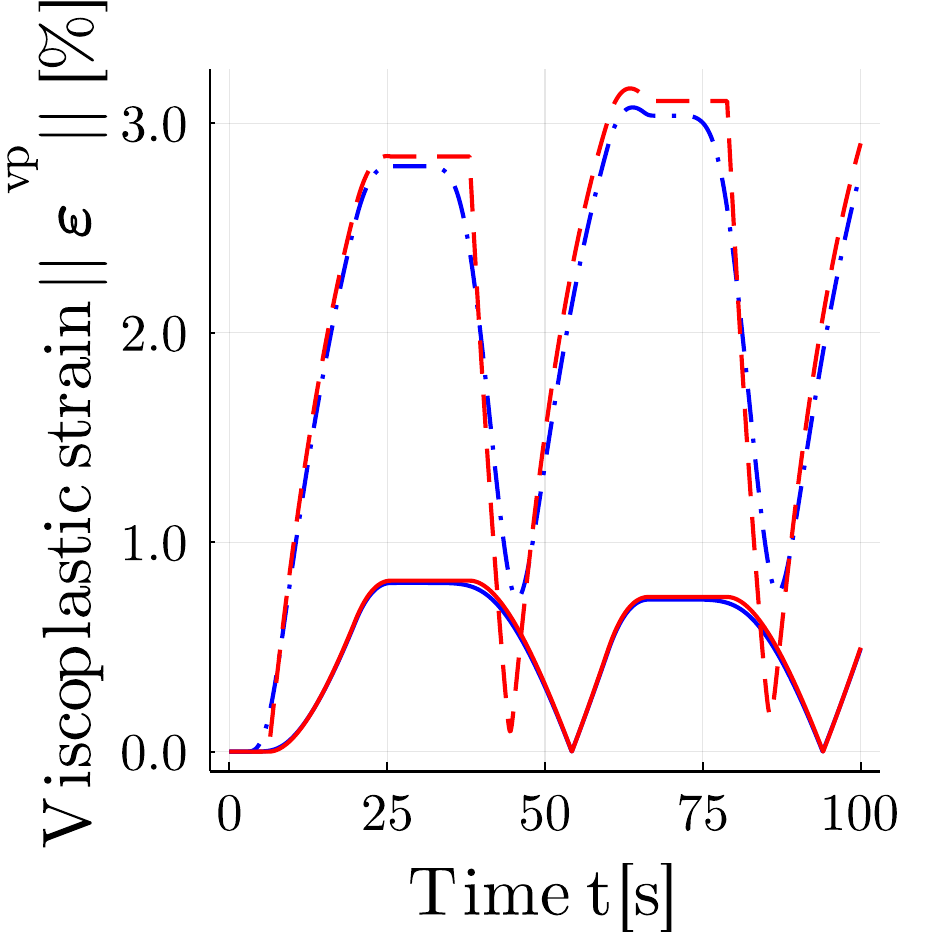} &
		\includegraphics[trim=0cm 0cm 12.5cm 0cm, clip, scale=0.25]{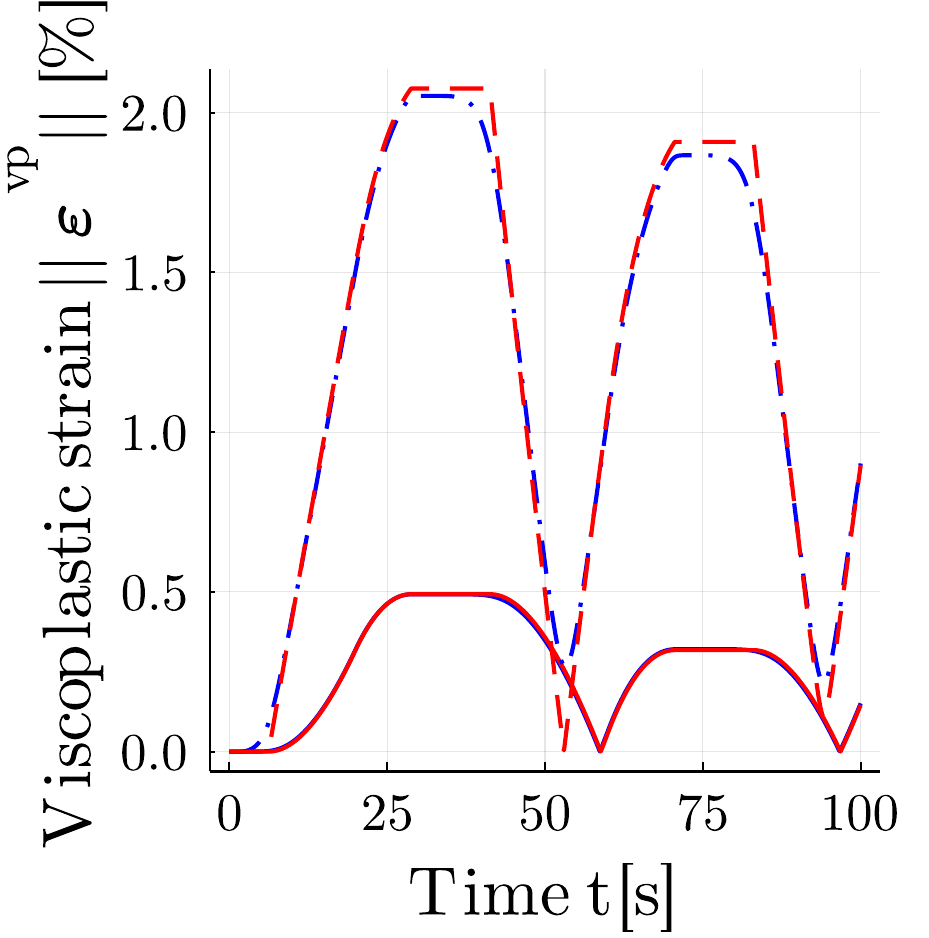} &
		\includegraphics[trim=16cm 0cm 0cm 0cm, clip, scale=0.25]{Pictures/ViscoPlasticity/Strains_B_200E9_alt.pdf} \\ 
	\end{tabular}
	\caption{Results for the elasto-viscoplasticity model with independent Lam\'e parameters and yield limit. Results obtained \rev{by MC are blue, results in red by TSM.}}
	\label{tab:resVP2}
\end{table}

\section{Computational Effort}
\label{sec:ComputationalEffort}
Reducing the computational effort compared to the Monte Carlo method while assuring high accuracy is the main motivation of the presented method. The separation of the material model in deterministic time-dependent and stochastic time-independent terms allows to reduce the number of needed simulations to a single one. This comes at the expense of an extended material model with additional internal variables and additional post-processing calculations. Here, we present the evaluation of the computational effort for all application examples.

The estimation error of the Monte Carlo method decays inversely proportional to the square root of the number of iterations \cite{thomopoulos_essentials_2013}. In all numerical experiments, the number of iterations of the Monte Carlo method is set to 1000 such that the expectation and standard deviation can be treated as converged. It may be remarked that a higher number of Monte Carlo iterations would only shift the analysis in favor of the novel TSM approach.

In Table~\ref{tab:CompEffort}, the runtime of the TSM and the Monte Carlo simulation are presented for all application examples.
Further, the runtime of the TSM is differentiated in the evaluation of the standard material model, the calculation of the additional internal variables and the probabilistic analysis of internal variable and stress respectively. The individual categories do not add up due to additional operations, e.g., saving the state. 
All experiments are performed on an Intel i5-1145G7 processor with a nominal frequency of 2.6\,GHz and 16\,GB of RAM. The implementation was done in Julia \cite{bezanson_julia_2017}.
The runtime differs between the different application examples due to the individual computational complexity. For each case the TSM leads to a significant speed-up. The speed-up is between $\approx 49$ for the damage material model and the elasto-viscoplasticity and 115 for the phase transformation.
It is obvious that the evaluation of the additional internal variables and the post-processing calculations increase the computation time well beyond the evaluation of the standard deterministic material model. However, with the combined computation time of only several ($\approx 8 - 20$) deterministic simulations, all stochastic information could be extracted.
\rev{For such a small number of deterministic simulations, the Monte Carlo simulation would not have been converged and huge errors are present. In contrast, TSM delivers almost exact results.}
The usage of AceGen for the derivation of the tangent of the phase transformation material model leads to an impressively low computation time. 
No additional speed-up results for the other material models because the analytical equations for the tangents are sufficiently small and no common calculations can be exploited. The same \rev{is valid} for the calculation of expectation and variance of internal variable and stress.

\begin{table}[]
	\centering
	\begin{tabular}{lp{4cm}|ccc}
		& & Damage & Phase transformation & Elasto-viscoplasticity \\ \hline
		\bf{MC} & & \bf{39.7\,s} & \bf{68\,min} & \bf{14.4\,s} \\
		\bf{TSM} & & \bf{810\,ms} & \bf{35.3\,s} & \bf{295\,ms}\\
		& Mean evaluation & 15\,ms & 4\,s & 15\,ms \\
		& Tangent evaluation & 85\,ms & 5.3\,s & 140\,ms \\
		& Expectation and variance of internal variable & 185\,ms & 2.4\,s & 50\,ms \\
		& Expectation and variance of stress & 495\,ms & 15.5\,s & 50\,ms \\ \hline
		Speed-up & & 49.0 & 115.6 & 48.8 \\
	\end{tabular}
	\caption{Computational effort}
	\label{tab:CompEffort}
\end{table}

\section{Conclusion}
\label{sec:Conclusion}
A new paradigm for the inclusion of stochasticity in engineering simulations is presented in this work.
The method is based on extended material models in which additional internal variables are used to track the effect of material parameter fluctuations. The extended material models are easily derived from standard material models.
This approach requires only one deterministic simulation with an increased set of internal variables. 
During post-processing, time-dependent deterministic terms and \rev{time-independent stochastic} terms are combined to approximate the evolution of expectation and standard deviation of the internal variables and stress. The approach is presented exemplary for three different material models: viscous damage, viscous phase transformations and elasto-viscoplasticity. 
Compared to reference Monte Carlo, simulations the method is able to approximate expectation and standard deviation of internal variable and stress for a variety of different load cases remarkably well. In many results, this method and the reference are identical.
The computational speed-up depends on the investigated material model and is well over one order of magnitude. It remains to transfer the approach to Finite Element simulations to enable large scale uncertainty quantification of engineering structures.

\section*{Acknowledgment}
This work has been supported by the German Research Foundation (DFG) 
within the framework of the international research training group IRTG 
2657 "Computational Mechanics Techniques in High Dimensions" (Reference: 
GRK 2657/1, Project number 433082294).

\appendix

\rev{
\section{Derivation of the material model for viscous damage}
\label{sec:derivDamage}
The material model for viscous damage is based on the minimum of the dissipation potential \cite{junker_extended_2021}.
The damage evolution is tracked by an internal variable $d(t)$ with an accompanying damage function $f(d(t)) = \exp(-d(t)) \in (0,1]$.
The mechanical part of the Helmholtz free energy is modeled as 
\begin{equation}
	\Psi_d(t) = \frac{1}{2} f(d) \bfvarepsilon \cdot \dsE \cdot \bfvarepsilon
\end{equation}
with the elasticity tensor $\dsE$ and the strains $\bfvarepsilon$. We denote the Helmholtz free energy in the undamaged state ($d = 0$) as $\Psi_0$.
We choose a rate-independent dissipation function
\begin{equation}
	\Delta_d^\textrm{diss} = \frac{1}{2} \eta \dot{d}^2
\end{equation}
\revv{with the viscosity $\eta$}.
The stationarity condition of the Hamilton functional with respect to $d$ reads locally as 
\begin{equation}
	\pf{\Psi_d}{d} + \pf{\Delta_d^\textrm{diss}}{\dot{d}} = 0.
\end{equation}
Thus, the evolution equation
\begin{equation}
	\dot{d}(t) = \frac{1}{\eta} \exp(-d) \Psi_0(t)
\end{equation}
can be recovered.
The \revv{stress is given by}
\begin{equation}
	\bfsigma = \revv{\pf{\Psi}{\bfvarepsilon} = } f(d) \, \dsE \cdot \bfvarepsilon.
\end{equation}
}

\section{Derivation of the material model for phase transformation}
\label{sec:derivPT}
\revv{For the material model for phase transformations, we have to define the Helmholtz free energy, the dissipation functional and the constraints. The Helmholtz free energy is given by Equation~\eqref{eq:PTHelmholtz} and recalled here as
\begin{equation}
	\Psi_\lambda = \frac{1}{2} (\bfvarepsilon - \bar{\bfeta}) \cdot \bar{\dsE} \cdot (\bfvarepsilon - \bar{\bfeta}) + \Lambda \sum_i \left( \frac{1}{\lambda_i^2(\lambda_i-1)^2} \right)
\end{equation}
We assume a rate-dependent evolution of the internal variable and thus propose the dissipation function
\begin{equation}
\Delta_\lambda^{\textrm{diss}} = \frac{1}{2} \eta \sum_i \dot{\lambda}_i^2.
\end{equation}	
The constraint is given by the volume conservation
\begin{equation}
\sum_i \dot{\lambda}_i = 0.
\end{equation}
Then, the stationarity condition of the Hamilton functional \cite{junker2021extended} demands
\begin{equation}
	\pf{\Psi_\lambda}{\lambda_i} + \pf{\Delta_\lambda^\textrm{diss}}{\dot{\lambda}_i} + \gamma = 0
\end{equation}
with the Lagrange parameter $\gamma$.
It is found as
\begin{equation}
	\gamma = - \frac{1}{n} \sum_i \frac{\partial \Psi}{\partial \lambda_i}.
\end{equation}
Employing the relationship $\dot{\lambda}_i = \frac{\mathrm{d} \lambda_i}{\mathrm{d}t} = \frac{\partial \lambda_i}{\partial \chi_i} \frac{\partial \chi_i}{\partial t}$ we find the evolution equation for the volume fraction variable $\chi_i$ as
\begin{equation}
\dot{\chi}_i = \left( \frac{\partial \lambda_i}{\partial \chi_i} \right)^{-1} \frac{1}{\eta} \left( - \frac{\partial \Psi}{\partial\lambda_i} +\frac{1}{n} \sum_j \frac{\partial \Psi}{\partial \lambda_j} \right). 
\end{equation}
}

\section{Tangent of extended material model for phase transformations}
\label{sec:tangentEMM}
To arrive at an differential equation for the time-dependent variable $\bfI_{ij}$ the following derivative  
\begin{align}
\dot{\bfI}_{ij, st} &= \frac{\mathrm{d}\dot{\chi}_i}{\mathrm{d}\dsD_{j, st}} \Bigg\vert_{\dsD = \bf0} = \frac{\mathrm{d}}{\mathrm{d}\dsD_{j, st}} \Bigg( \left( \pf{\lambda_i}{\chi_i} \right)^{-1} \Bigg)  \frac{1}{\eta} \left( - \pf{\Psi}{\lambda_i} + \frac{1}{n} \sum_k \pf{\Psi}{\lambda_k}\right) \nonumber \label{eq:derchi} \\
&\quad \quad + \left( \pf{\lambda_i}{\chi_i} \right)^{-1} \frac{1}{\eta} \Bigg( \frac{\mathrm{d}}{\mathrm{d}\dsD_{j,st}} \left(- \pf{\Psi}{\lambda_i}\right) + \frac{1}{n} \sum_k \frac{\mathrm{d}}{\mathrm{d}\dsD_{j,st}} \left( \pf{\Psi}{\lambda_k} \right) \Bigg) \Bigg\vert_{\dsD = \bf0} \nonumber
\end{align}
is needed. \rev{The indices $s$ and $t$ denote components of the tensor $\dot{\bfI}$, whereas $i$ and $j$ denote the index of the phase and the uncertain elasticity tensor respectively.}
A symbolic derivation of the derivative is presented below. The equation is split up in subexpressions to increase the clarity.
The indices $i$ and $j$ refer to the derivative of the $i$th phase with respect to the uncertain elasticity tensor of the $j$th phase.
All other indices are used according the rules of the index notation.

This equation for the derivative consists of multiple derivatives itself which are derived below.
The first derivative is found as
\begin{equation}
\frac{\mathrm{d}}{\mathrm{d}\dsD_{j,st}} \left( \frac{\mathrm{d}\lambda_i}{\mathrm{d}\chi_i}\right)^{-1}  = - \left( \frac{\mathrm{d}\lambda_i}{\mathrm{d}\chi_i}\right)^{-2} \frac{\mathrm{d}\lambda'_i}{\mathrm{d}\dsD_{j, st}}
\end{equation} 
where
\begin{align}
\frac{\mathrm{d}\lambda'_i}{\mathrm{d}\dsD_{j, st}} = 2 \lambda_i \frac{\mathrm{d}\lambda_i}{\mathrm{d}\dsD_{j, st}} \exp(-\chi_i) - \lambda_i^2 \exp(-\chi_i) I_{ij, st}
\end{align}
and
\begin{equation}
\frac{\mathrm{d}}{\mathrm{d}\dsD_j} \left(2 \lambda_i -1 \right) = 2 \frac{\mathrm{d}\lambda_i}{\mathrm{d}\dsD_j}.
\end{equation}
\rev{In Equation~\eqref{eq:dlambda_dD}, it was shown that
	\begin{equation}
		\frac{\mathrm{d}\lambda_i}{\mathrm{d}\dsD_j} = (\lambda_i^{(0)})^2 \exp(-\chi_i^{(0)}) \bfI_{ij}
	\end{equation}  holds.
}
\rev{
For clarity in the following presentation the Helmholtz energy is separated into two parts
\begin{equation}
    \Psi_\lambda = \bar{\Psi} + \Psi^r
\end{equation}
where 
\begin{equation}
	\bar{\Psi} = \frac{1}{2} (\bfvarepsilon - \bar{\bfeta}) \cdot \bar{\dsE} \cdot (\bfvarepsilon - \bar{\bfeta}) \quad \textrm{and} \quad 
	\Psi^r = \Lambda \sum_i \left( \frac{1}{\lambda_i^2(\lambda_i-1)^2} \right).
\end{equation}
Accordingly, the driving forces given as the derivative of the Helmholtz energy can be calculated as
\begin{align}
    \pf{\bar{\Psi}}{\lambda_i} &= - \frac{1}{2} \left( \bfeta_i \cdot \bar{\dsE} \cdot (\bfvarepsilon - \bar{\bfeta}) + (\bfvarepsilon - \bar{\bfeta}) \cdot \bar{\dsE} \cdot \dsE_i^{-1} \cdot \bar{\dsE} \cdot (\bfvarepsilon - \bar{\bfeta}) + (\bfvarepsilon - \bar{\bfeta}) \cdot \bar{\dsE} \cdot \bfeta_i \right) \nonumber \\
    &= - \eta_i \cdot \bar{\dsE} \cdot (\bfvarepsilon - \bar{\bfeta}) - \frac{1}{2} (\bfvarepsilon - \bar{\bfeta}) \cdot \bar{\dsE} \cdot \dsE_i^{-1} \cdot \bar{\dsE} \cdot (\bfvarepsilon - \bar{\bfeta})
\end{align}
and
\begin{align}
    \pf{\Psi^r}{\lambda_i} &= \Lambda (-2 \lambda_j^{-3} (1-\lambda_j)^{-2} + \lambda_j^{-2} (2 (1-\lambda_j)^{-3}) \nonumber \\
    &= 2 \Lambda \left( \frac{2\lambda_j - 1}{\lambda_j^3 (1-\lambda_j)^3} \right).
\end{align}
The derivative of the driving force results as
\begin{equation}
\frac{\mathrm{d}}{\mathrm{d}\dsD_{j, st}} \pf{\Psi_\lambda}{\lambda_i} = \frac{\mathrm{d}}{\mathrm{d}\dsD_{j, st}} \frac{\partial \bar{\Psi}}{\partial \lambda_i} + \frac{\mathrm{d}}{\mathrm{d}\dsD_{j, st}} \pf{\bar{\Psi^r}}{\lambda_i}.
\end{equation}
}
The first part of the preceding equation is given as
\begin{align}
    \frac{\mathrm{d}}{\mathrm{d}\dsD_j} \left( \pf{\bar{\Psi}}{\lambda_i} \right) &=  - \eta_{i, a} \frac{\mathrm{d}}{\mathrm{d}\dsD_{j, st}} \left( \bar{\dsE}_{ab} \right) \cdot (\varepsilon_b - \bar{\eta}_b) + \eta_{i, c} \cdot \bar{\dsE}_{cd} \cdot \sum_i \frac{\mathrm{d}}{\mathrm{d}\dsD_{j, st}} \left( \lambda_i \right) \eta_{i, d}  \\ 
    &\quad - \frac{1}{2} (\varepsilon_e - \bar{\eta}_e) \left( \frac{\mathrm{d}}{\mathrm{d}\dsD_{j, st}} \left( \bar{\dsE}_{ef} \right) \dsE_{i, fg}^{-1} \bar{\dsE}_{gh} + \bar{\dsE}_{kl} \frac{\mathrm{d}}{\mathrm{d}\dsD_{j, st}} \left( \dsE_{i, lm}^{-1} \right) \bar{\dsE}_{mh} + \bar{\dsE}_{no} \dsE_{i, op}^{-1} \frac{\mathrm{d}}{\mathrm{d}\dsD_{j, st}} \left( \bar{\dsE}_{ph} \right) \right) (\varepsilon_h -\bar{\eta}_h)\nonumber
\end{align}
with the derivatives
\begin{align}
    \frac{\mathrm{d}\bar{\dsE}_{ab}}{\mathrm{d}\dsD_{j,st}} &= - \bar{\dsE}_{ag} \frac{\mathrm{d}\left( \sum_i \lambda_i \dsE_i^{-1} \right)_{gh}}{\mathrm{d}\dsD_{j, st}} \bar{\dsE}_{hb},
\end{align}
and
\begin{align}
    \frac{\mathrm{d}\left( \sum_i \lambda_i \dsE_i^{-1} \right)_{gh}}{\mathrm{d}\dsD_{j, st}} &= \sum_i \left( \frac{\mathrm{d}\lambda_i}{\mathrm{d}\dsD_{j, st}} \left( \dsE^{-1} \right)_{gh} + \lambda_i \frac{\mathrm{d}\left( \dsE^{-1} \right)_{gh} }{\mathrm{d}\dsD_{j, st}} \right). \label{eq:derivativechi1}
\end{align}
Furthermore, it can be found that
\begin{align}
    \frac{\mathrm{d}\left(\dsE^{-1} \right)_{gh} }{\mathrm{d}\dsD_{j, st}} = - \left( \dsE^{-1} \right)_{gs} \left(\dsE^{-1} \right)_{th} 
\end{align}
with the Dirac delta function $\delta_{ij}$.
The second part of the derivative of the driving force with respect to the uncertain part of the material tensor $\dsD$ is given as
\begin{align}
\frac{\mathrm{d}}{\mathrm{d}\dsD_j} \left( \frac{\mathrm{d}\Psi^r}{\mathrm{d}\lambda_i} \right) &= \frac{\mathrm{d}}{\mathrm{d}\dsD_j} \left( 2 \Lambda \left( \frac{2 \lambda_i - 1}{\lambda_i^3 (1-\lambda_i)^3} \right) \right) \\ 
&= 2 \Lambda \left( \frac{\left( \frac{\mathrm{d}}{\mathrm{d}\dsD_j} \left( 2 \lambda_i - 1 \right) \right) \lambda_i^3 (1-\lambda_i)^3 - (2 \lambda_i - 1) \left( \frac{\mathrm{d}}{\mathrm{d}\dsD_j} \left( \lambda_i^3 (1-\lambda_i)^3 \right) \right) }{(\lambda_i^3 (1-\lambda_i)^3)^2 } \right) \nonumber
\end{align}
with the derivative
\begin{equation}
\frac{\mathrm{d}}{\mathrm{d}\dsD_j} \left( \lambda_i^3 (1-\lambda_i)^3 \right) = \left( 3 \lambda_i^2 (1-\lambda_i)^3 - 3 \lambda_i^3 (1-\lambda_i)^2 \right) \frac{\mathrm{d}\lambda_i}{\mathrm{d}\dsD_j}.
\end{equation}

With these equations above, $\dot{\bfI}_{ij}$ can be calculated. All equations are evaluated at $\dsD_j = \boldface{0} \quad \forall j$.

\section{Tangent of the stress for phase transformations}
\label{sec:tangentStress}
The linearization of the stress requires the definition of the tangents
\begin{equation}
    \frac{\mathrm{d}\sigma_\alpha}{\mathrm{d}\dsD_j} \Big\rvert_{\dsD_i = \boldface{0}}.
\end{equation}
This is done componentwise, i.e., $\alpha$ is a fixed index over which no summation is performed.
\begin{equation}
    \frac{\mathrm{d}\sigma_\alpha}{\mathrm{d}\dsD_j} \Big\rvert_{\dsD_i = \boldface{0}} = \frac{\mathrm{d}}{\mathrm{d}\dsD_{j, st}}\left( \bar{\dsE}_{\alpha b} \right) \left(\varepsilon_b - \bar{\eta}_b \right) - \bar{\dsE}_{\alpha b} \sum_i \frac{\mathrm{d}}{\mathrm{d}\dsD_{j, st}} \left( \lambda_i \right) \eta_{i, b}.
\end{equation}
All necessary derivatives are already introduced above.

\addcontentsline{toc}{chapter}{Bibliography}
\bibliographystyle{plain}
\bibliography{bib_short}

\end{document}

%% file: LiteratureReview.pdf_tex
%% Creator: Inkscape 1.0.2-2 (e86c870879, 2021-01-15), www.inkscape.org
%% Accompanies image file 'LiteratureReview.pdf' (pdf, eps, ps)
\begingroup%
  \makeatletter%
  \providecommand\color[2][]{%
    \errmessage{(Inkscape) Color is used for the text in Inkscape, but the package 'color.sty' is not loaded}%
    \renewcommand\color[2][]{}%
  }%
  \providecommand\transparent[1]{%
    \errmessage{(Inkscape) Transparency is used (non-zero) for the text in Inkscape, but the package 'transparent.sty' is not loaded}%
    \renewcommand\transparent[1]{}%
  }%
  \providecommand\rotatebox[2]{#2}%
  \newcommand*\fsize{\dimexpr\f@size pt\relax}%
  \newcommand*\lineheight[1]{\fontsize{\fsize}{#1\fsize}\selectfont}%
  \ifx\svgwidth\undefined%
    \setlength{\unitlength}{457.12842467bp}%
    \ifx\svgscale\undefined%
      \relax%
    \else%
      \setlength{\unitlength}{\unitlength * \real{\svgscale}}%
    \fi%
  \else%
    \setlength{\unitlength}{\svgwidth}%
  \fi%
  \global\let\svgwidth\undefined%
  \global\let\svgscale\undefined%
  \makeatother%
  \begin{picture}(1,0.52913525)%
    \lineheight{1}%
    \setlength\tabcolsep{0pt}%
    \put(0,0){\includegraphics[width=\unitlength,page=1]{LiteratureReview.pdf}}%
    \put(0.7699463,0.07108737){\color[rgb]{0,0,0}\makebox(0,0)[lt]{\lineheight{1.25}\smash{\begin{tabular}[t]{l}accuracy of\\material model\end{tabular}}}}%
    \put(0.47228051,0.00546037){\makebox(0,0)[lt]{\lineheight{0}\smash{\begin{tabular}[t]{l}physically correct\end{tabular}}}}%
    \put(0.20625653,0.00546037){\makebox(0,0)[lt]{\lineheight{0}\smash{\begin{tabular}[t]{l}simplified\end{tabular}}}}%
    \put(-0,0.50586669){\makebox(0,0)[lt]{\lineheight{0}\smash{\begin{tabular}[t]{l}computation speed\\ \\ \\\\\end{tabular}}}}%
    \put(0.04336072,0.39453505){\makebox(0,0)[lt]{\lineheight{0}\smash{\begin{tabular}[t]{l}high\end{tabular}}}}%
    \put(0.043271,0.12733913){\makebox(0,0)[lt]{\lineheight{0}\smash{\begin{tabular}[t]{l}low\end{tabular}}}}%
    \put(0,0){\includegraphics[width=\unitlength,page=2]{LiteratureReview.pdf}}%
    \put(0.19922502,0.3835379){\makebox(0,0)[lt]{\lineheight{0}\smash{\begin{tabular}[t]{l}Perturbation\end{tabular}}}}%
    \put(0,0){\includegraphics[width=\unitlength,page=3]{LiteratureReview.pdf}}%
    \put(0.29628551,0.24687411){\makebox(0,0)[lt]{\lineheight{0}\smash{\begin{tabular}[t]{l}Stochastic Collocation\end{tabular}}}}%
    \put(0,0){\includegraphics[width=\unitlength,page=4]{LiteratureReview.pdf}}%
    \put(0.31503611,0.11444812){\makebox(0,0)[lt]{\lineheight{0}\smash{\begin{tabular}[t]{l}Sampling Method\end{tabular}}}}%
    \put(0,0){\includegraphics[width=\unitlength,page=5]{LiteratureReview.pdf}}%
    \put(0.44415459,0.39570691){\makebox(0,0)[lt]{\smash{\begin{tabular}[t]{l}Time-separated\\stochastic mechanics\\\end{tabular}}}}%
  \end{picture}%
\endgroup%